\documentclass[twoside]{IEEEtran}

\usepackage{amsmath}
\usepackage{amssymb}
\usepackage{amsthm}
\usepackage{enumerate}

\usepackage{epsfig}
\usepackage{subfigure}
\usepackage{psfrag}
\usepackage{cite}
\usepackage{stfloats}
\usepackage{hyperref}
\hypersetup{colorlinks=true, linkcolor=blue}

\graphicspath{{figures/}}

\psfrag{Rd=Cmarton}{\small{$R(d)=$Marton's converse \eqref{eq_Cmarton}}}
\psfrag{Rd}{\small{$R(d)$}}
\psfrag{Rnde}{\small{$R(n, d, \epsilon)$}}
\psfrag{n}{\small{$n$}}
\psfrag{R}{\small{$R$}}
\psfrag{V(d)}{\footnotesize{$V(d)$}}
\psfrag{d}{\footnotesize{$d$}}
\psfrag{Cmarton}{\small{Marton's converse \eqref{eq_Cmarton}}}
\psfrag{Ashannon}{\small{Shannon's achievability \eqref{eq_Ashannon}}}

\psfrag{dinf}{\small{$d_\infty$}}
\psfrag{dn}{\small{$d_n$}}
\psfrag{Rinf}{\small{$R_\infty$}}
\psfrag{angle}{\small{$\alpha_n$}}
\psfrag{ABMS}{\small{Achievability \eqref{eq_ABMS}}}
\psfrag{CBMS}{\small{Converse \eqref{eq_CBMS}}}
\psfrag{CGenBMS}{\small{Converse \eqref{eq_CgBMS}}}
\psfrag{NBMS}{\small{Approximation \eqref{eq_2order}}}

\psfrag{AEBMS}{\small{Achievability \eqref{eq_AEBMS}}}
\psfrag{CEBMS}{\small{Converse \eqref{eq_CEBMS}}}
\psfrag{NEBMS}{\small{Approximation \eqref{eq_2orderEBMS}}}

\psfrag{ABES}{\small{Achievability \eqref{eq_ABES}}}
\psfrag{CBES}{\small{Converse \eqref{eq_CBES}}}
\psfrag{NBES}{\small{Approximation \eqref{eq_2order}}}

\psfrag{AGMS}{\small{Achievability \eqref{eq_AGMS}}}
\psfrag{ARogers}{\small{Achievability \eqref{eq_AGMSrogers}}}
\psfrag{CGMS}{\small{Converse \eqref{eq_CGMS}}}
\psfrag{CGenGMS}{\small{Converse \eqref{eq_CgGMS}}}
\psfrag{NGMS}{\small{Approximation \eqref{eq_2order}}}

\newtheorem{thm}{Theorem}
\newtheorem{cor}[thm]{Corollary}
\newtheorem{lemma}{Lemma}

\newtheorem{defn}{Definition}

\theoremstyle{remark}
\newtheorem{remark}{Remark}
\newtheorem{property}{Property}
\newtheorem{example}{Example}

\newcommand{\bigo}[1]{O\left(#1\right)}
\newcommand{\smallo}[1]{o\left(#1\right)}
\newcommand{\Qinv}[1]{Q^{-1}\left(#1\right)}
\newcommand{\Var}[1]{{\rm Var}\left[#1\right]}
\newcommand{\E}[1]{\mathbb E\left[#1\right]}
\newcommand{\Prob}[1]{\mathbb P\left[#1\right]}
\newcommand{\beq}{\begin{equation}}
\newcommand{\eeq}{\end{equation}}

\def\binosum#1#2{\left\langle{#1\atopwithdelims..#2}\right\rangle}
\def\hammingsum#1#2{S_{#2}}
\title{Fixed-length lossy compression\\ in the finite blocklength regime}

\author{
Victoria Kostina,  \IEEEmembership{Student Member, IEEE}, \and Sergio Verd\'{u}, \IEEEmembership{Fellow, IEEE}
\\ Dept. of Electrical Engineering, Princeton University, NJ, 08544, USA.\\
 Email: vkostina@princeton.edu, verdu@princeton.edu.
 \thanks{
This research was supported in part by NSF under grants CCF-1016625 and CCF 09-39370. The first author was supported in part by the Natural Sciences and Engineering Research Council of Canada.}
}

\begin{document}

\maketitle
\begin{abstract}
This paper studies the minimum achievable source coding rate as a function of blocklength $n$ and probability $\epsilon$ that the distortion exceeds a given level $d$. Tight general achievability and converse bounds are derived that hold at arbitrary fixed blocklength. For stationary memoryless sources with separable distortion, the minimum rate achievable is shown to be closely approximated by $R(d) + \sqrt{\frac{V(d)}{n}} \Qinv{\epsilon}$, where $R(d)$ is the rate-distortion function, $V(d)$ is the {\it rate dispersion}, a characteristic of the source which measures its stochastic variability, and $\Qinv{\cdot}$ is the inverse of the standard Gaussian complementary cdf.
\end{abstract}

\begin{IEEEkeywords}
achievability, converse, finite blocklength regime, lossy source coding, memoryless sources, rate-distortion, Shannon theory.
\end{IEEEkeywords}

\section{Introduction}
\label{sec:intro}
The rate-distortion function characterizes the minimal source coding rate compatible with a given distortion level, either in average or excess distortion sense, provided that the blocklength is permitted to grow without limit. However, in some applications relatively short blocklengths are common both due to delay and complexity constraints. It is therefore of critical practical interest to assess the unavoidable penalty over the rate-distortion function required to sustain the desired fidelity at a given fixed blocklength. Neither the lossy source coding theorem nor the reliability function, which gives the asymptotic exponential decay of the probability of exceeding a given distortion level when compressing at a fixed rate, provide an answer to that question.

This paper presents new achievability and converse bounds to the minimum sustainable rate as a function of blocklength and excess probability, valid for general sources and general distortion measures. In addition, for stationary memoryless sources with separable (i.e., additive, or per-letter) distortion, we show that the finite blocklength coding rate is well approximated by
\begin{equation}
\label{eq_2orderIntro}
R(n,d,\epsilon) \approx R(d) + \sqrt{\frac{V(d)}{n}}\Qinv{\epsilon},
\end{equation}
where $n$ is the blocklength, $\epsilon$ is the probability that the distortion incurred by the reproduction exceeds $d$, and $V(d)$ is the {\it rate-dispersion function}.
The evaluation of the new bounds is detailed for:
 \begin{itemize}
 \item  the stationary discrete memoryless source (DMS) with symbol error rate distortion;
 \item  the stationary Gaussian memoryless source (GMS) with mean-square error distortion;
 \item  the stationary binary memoryless source when the compressor observes it through the binary erasure channel (BES), and the distortion measure is bit error rate.
 \end{itemize}
 In the most basic special case, namely that of the equiprobable source with symbol error rate distortion, the rate-dispersion function is zero, and the finite blocklength coding rate is approximated by
\begin{equation}
\label{eq_2orderNonRedundant}
R(n,d,\epsilon) = R(d) + \frac 1 2 \frac{\log n}{n} + \bigo{\frac 1 n}
\end{equation}

Section \ref{sec:nonasymptotic} sets up the problem, introduces the definitions of the fundamental finite blocklengths limits and presents the basic notation and properties of the information density and related quantities used throughout the paper. Section \ref{sec:prior} reviews the few existing finite blocklength achievability and converse bounds for lossy compression, as well as various relevant  asymptotic refinements of Shannon's lossy source coding theorem. Section \ref{sec:new} shows the new general upper and lower bounds to the minimum rate at a given blocklength. Section \ref{sec:2order} studies the asymptotic behavior of the bounds using Gaussian approximation analysis. Sections \ref{sec:BMS}, \ref{sec:DMS}, \ref{sec:BES} and \ref{sec:GMS} focus on the binary memoryless source (BMS)\footnote{Although the results in Section \ref{sec:BMS} are a special case of those in Section \ref{sec:DMS}, it is enlightening to specialize our results to the simplest possible setting.}, DMS, BES and GMS, respectively.

\section{Preliminaries}
\label{sec:nonasymptotic}
\subsection{Operational definitions}
In fixed-length lossy compression, the output of a general source with alphabet $A$ and source distribution $P_X$ is mapped to one of the $M$ codewords from the reproduction alphabet $B$. A lossy code is a (possibly randomized) pair of mappings $\mathsf f \colon A \mapsto \{1, \ldots, M\}$ and $\mathsf c \colon \{1, \ldots, M\} \mapsto B$. A distortion measure $d \colon A \times B \mapsto [0, +\infty]$ is used to quantify the performance of a lossy code. Given decoder $\mathsf c$, the best encoder simply maps the source output to the closest (in the sense of the distortion measure) codeword, i.e. $\mathsf f(x) = \arg\min_m d(x, \mathsf c(m))$.  The average distortion over the source statistics is a popular performance criterion. A stronger criterion is also used, namely, the probability of exceeding a given distortion level (called {\it excess-distortion probability}). The following definitions abide by the excess distortion criterion.
\begin{defn}
An $(M, d, \epsilon)$ code for $\{A, \ B,\ P_X,\ d \colon A \times B \mapsto  [0, +\infty]\}$ is a code with $|\mathsf f| = M$ such that $\mathbb P \left[ d\left( X, \mathsf c(\mathsf f(X))\right) > d\right] \leq \epsilon$.

The minimum achievable code size at excess-distortion probability $\epsilon$ and distortion $d$ is defined by
\begin{equation}
 M^\star(d,\epsilon) = \min \left\{ M\colon \ \exists (M, d, \epsilon) \mbox{ code} \right\}
\end{equation}
\label{defn:(M, d, eps)}
\end{defn}
Note that the special case $d = 0$ and $d(x, y) = 1 \left\{ x \neq y \right\}$ corresponds to almost-lossless compression.
\begin{defn}
In the conventional fixed-to-fixed (or block) setting in which $A$ and $B$ are the $n-$fold Cartesian products of alphabets $\mathcal A$ and $\mathcal B$, an $(M, d, \epsilon)$ code for $\{\mathcal A^n, \ \mathcal B^n,\ P_{X^n},\ d^n\colon \mathcal A^n \times \mathcal B^n \mapsto  [0, +\infty]\}$ is called an $(n, M, d, \epsilon)$ code.

Fix $\epsilon$, $d$ and blocklength $n$. The minimum achievable code size and the finite blocklength rate-distortion function (excess distortion) are defined by, respectively
\begin{align}
M^\star(n,d,\epsilon) &= \min \left\{ M\colon \ \exists (n, M, d, \epsilon) \mbox{ code} \right\}\\
R(n, d, \epsilon) &= \frac 1 n \log M^\star(n, d, \epsilon)
\end{align}
\end{defn}

Alternatively, using an average distortion criterion, we employ the following notations.
\begin{defn}
An $\langle M, d \rangle$ code for $\{A, \ B,\ P_X,\ d\colon A \times B \mapsto [0, +\infty]\}$ is a code with $|\mathsf f| = M$ such that $\mathbb E \left[ d\left( X, \mathsf c(\mathsf f(X))\right) \right] \leq d$. The minimum achievable code size at average distortion $d$ is defined by
\begin{equation}
 M^\star(d) = \min \left\{ M\colon \ \exists \langle M, d\rangle \mbox{ code} \right\}
\end{equation}

\end{defn}

\begin{defn}
 If $A$ and $B$ are the $n-$fold Cartesian products of alphabets $\mathcal A$ and $\mathcal B$, an $\langle M, d \rangle$ code for $\{\mathcal A^n, \ \mathcal B^n,\ P_{X^n},\ d^n\colon \mathcal A^n \times \mathcal B^n \mapsto [0, +\infty]\}$ is called an $\langle n, M, d \rangle$ code.

Fix $d$ and blocklength $n$. The minimum achievable code size and the finite blocklength rate-distortion function (average distortion) are defined by, respectively
\begin{align}
M^\star(n,d) &= \min \left\{ M\colon \ \exists \langle n, M, d\rangle \mbox{ code} \right\}\\
R(n,d) &= \frac{\log M^\star(n, d)}{n}
\end{align}
\end{defn}

In the limit of long blocklengths, the minimum achievable rate is characterized by the rate-distortion function \cite{shannon1948mathematical}\cite{shannon1959coding}.

\begin{defn}
 The rate-distortion function is defined as
\begin{equation}
R(d) = \limsup_{n \to \infty} R(n, d)
\end{equation}
\end{defn}

In a similar manner, one can define the distortion-rate functions $D(n, R, \epsilon)$,  $D(n,R)$ and $D(R)$.

In the review of prior work in Section \ref{sec:prior} we will use the following concepts related to variable-length coding. A variable-length code is a pair of mappings $\mathsf f\colon A \mapsto \{0, 1\}^\star$ and $\mathsf c\colon \{0, 1\}^\star \mapsto B$, where $\{0,1\}^\star$ is the set of all possibly empty binary strings. It is said to operate at distortion level $d$ if $\Prob{d(X, \mathsf c(\mathsf f(X))) \leq d} = 1$. For a given code $(\mathsf f, \mathsf c)$ operating at distortion $d$, the length of the binary codeword assigned to $x \in A$ is denoted by $\ell(x) = \text{length of } \mathsf f(x)$. 

\subsection{Tilted information}
\label{ssec:dtilted}
Denote by
\begin{equation}
\imath_{X; Y}(x; y) = \log \frac{dP_{XY}}{d(P_X \times P_Y)} (x,y) \label{eq_ixy}
\end{equation}
the information density of the joint distribution $P_{XY}$ at $(x, y) \in A \times B$. Further, for a discrete random variable $X$, the information in outcome $x$ is denoted by
\begin{equation}
 \imath_X(x) = \log \frac 1 {P_X(x)}\label{eq_ix}
\end{equation}
Under appropriate conditions, the number of bits that it takes to represent $x$ divided by $\imath_X (x)$ converges to $1$ as these quantities
go to infinity. Note that if $X$ is discrete, then $\imath_{X;X} (x;x) = \imath_X (x)$.

For a given $P_X$ and distortion measure, denote
\begin{equation}
\mathbb R_X(d) =  \inf_{\substack{P_{Y|X}\colon\\ \E{d(X, Y)} \leq d}} I(X; Y) \label{eq_RR(d)}
\end{equation}
We impose the following basic restrictions on the source and the distortion measure.
 \begin{enumerate}[(a)]
  \item  \label{item:a}
 $\mathbb R_X(d)$ is finite for some $d$, i.e. $ d_{\min} < \infty$, where
\begin{equation}
 d_{\min} = \inf \left\{ d\colon ~ \mathbb R_X(d) < \infty \right\} \label{eq_dmin}
\end{equation}
 \item \label{item:c} The distortion measure is such that there exists a finite set $E \subset B$ such that
\begin{equation}
 \E{ \min_{y \in E} d(X, y)} < \infty \label{eq_dcsiszar}
\end{equation}
\item \label{item:d} The infimum in \eqref{eq_RR(d)} is achieved by a unique $P_{Y|X}^\star$, and distortion measure is finite-valued. \footnote{Restriction \eqref{item:d} is imposed for clarity of presentation. We will show in Section \ref{sec:2order} that it can be dispensed with. }
\end{enumerate}

The counterpart of \eqref{eq_ix} in lossy data compression, which roughly corresponds to the number of bits one needs to spend to encode $x$ within distortion $d$, is the following.
\begin{defn}[$d-$tilted information] \label{defn:id}
For $d > d_{\min}$, the $d-$tilted information in $x$ is defined as
 \begin{equation}
\jmath_{X}(x, d) =\log \frac 1 {\E{ \exp\left\{ \lambda^\star d - \lambda^\star d(x, Y^\star)\right\}}} \label{eq_idball}
\end{equation}
where the expectation is with respect to the unconditional distribution\footnote{Henceforth, $Y^\star$ denotes the rate-distortion-achieving reproduction random variable at distortion $d$, i.e. its distribution $P_Y^\star$ is the marginal of $P_{Y|X}^\star P_X$, where $P_{Y|X}^\star$ achieves the infimum in \eqref{eq_RR(d)}. } of $Y^\star$, and
\begin{equation}
 \lambda^\star = -\mathbb R_X^\prime (d) \label{eq_lambdastar}
\end{equation}
\end{defn}
It can be shown that \eqref{item:d} guarantees differentiability of $\mathbb R_X (d)$, thus $\eqref{eq_idball}$ is well defined. A measure-theoretic proof of the following properties can be found in \cite[Lemma 1.4]{csiszar1974extremum}.
\begin{property}
For $P_{Y}^\star$-almost every $y$,
\begin{equation}
 \jmath_{X}(x, d) = \imath_{X; Y^\star}(x; y) + \lambda^\star d(x, y) - \lambda^\star d \label{eq_iddensity}
\end{equation}
hence the name we adopted in Definition \ref{defn:id}, and 
\begin{equation}
\mathbb R_X(d) = \E{ \jmath_{X}(X, d)} \label{eq_Ejd}
\end{equation}
\end{property}
\begin{property}
For all $y \in B$,
\begin{equation}
\E{ \exp\left\{ \lambda^\star d - \lambda^\star d(X, y) + \jmath_X(X, d)\right\} } \leq 1 \label{eq_csiszar}
\end{equation}
with equality for $P_{Y}^\star$-almost every $y$.
\end{property}

\begin{remark}
While Definition \ref{defn:id} does not cover the case $d = d_{\min}$, for discrete random variables with $d(x, y) = 1 \left\{ x \neq y \right\}$ it is natural to define $0$-tilted information as
\begin{equation}
 \jmath_X(x, 0) = \imath_X(x)
\end{equation}
\end{remark}

\begin{example}
For the BMS with bias $p \leq \frac 1 2$ and bit error rate distortion, 
\begin{equation}
 \jmath_{X^n}(x^n, d) = \imath_{X^n}(x^n) - n h(d) \label{eq_jBMS}
\end{equation}
if $0 \leq d < p$, and $0$ if $d \geq p$. 
\end{example}
\begin{example}
For the GMS with variance $\sigma^2$ and mean-square error distortion,\footnote{We denote the Euclidean norm by $|\cdot|$, i.e. $|x^n|^2 = x_1^2 + \ldots +x_n^2$.}
\begin{equation}
 \jmath_{X^n}(x^n, d) = \frac n 2 \log \frac {\sigma^2}{d} +\left( \frac{|x^n|^2}{\sigma^2} - n \right) \frac {\log e}2  \label{eq_jGMS}
\end{equation}
if $0 < d < \sigma^2$, and $0$ if $d \geq \sigma^2$. 
\end{example}

The distortion $d$-ball around $x$ is denoted by
\begin{equation}
 B_d(x) =  \{y \in B\colon ~ d(x, y) \leq d\}
\end{equation}
Tilted information is closely related to the (unconditional) probability that $Y^\star$ falls within distortion $d$ from $X$. Indeed, since $\lambda^\star > 0$,  for an arbitrary $P_Y$ we have by Markov's inequality,
\begin{align}
P_Y(B_d(x)) &= \Prob{d(x, Y) \leq d}\\
&\leq \E{ \exp\left\{\lambda^\star d - \lambda^\star  d(x, Y)\right\}}  \label{eq_Pdball_Markov}
\end{align}
where the probability measure is generated by the unconditional distribution of $Y$. Thus
\begin{equation}
 \log \frac 1 {P_Y^\star(B_d(x))} \geq  \jmath_X(x, d)\label{eq_logPdball_Markov}
\end{equation}
As we will see in Theorem \ref{thm:AEP}, under certain regularity conditions the equality in \eqref{eq_logPdball_Markov} can be closely approached.

\subsection{Generalized tilted information}
\label{ssec:dtiltedgen}
Often it is more convenient \cite{blahut1972computation} to fix $P_Y$ defined on $B$ and to consider, in lieu of \eqref{eq_RR(d)}, the following optimization problem:
\begin{equation}
 \mathbb R_{X, Y} (d)  = \min_{\substack{P_{Z|X}\colon\\ \E{d(X; Z)} \leq d}} D(P_{Z|X}\| P_{Y} | P_X)    \label{eq_RR(d)gen}
\end{equation}
In parallel with Definition \ref{defn:id}, define for any $\lambda \geq 0$
\begin{equation}
 \Lambda_{Y}(x, \lambda) = \log \frac 1 {\E{\exp\left( \lambda d -\lambda d(x, Y) \right) }} \label{eq_jlambdagen}
\end{equation}


As long as
$ d > d_{\min\mid X, Y}$, 
where
\begin{equation}
 d_{\min\mid X, Y} = \inf \left\{ d\colon ~ \mathbb R_{X, Y}(d) < \infty \right\} \label{eq_dminXY}
\end{equation} 
the minimum in \eqref{eq_RR(d)gen} is always achieved by a $P_{Z^\star|X}$ that satisfies  \cite{csiszar1974extremum}
\begin{align}
 &~\log \frac{dP_{Z^\star|X}(y|x)}{dP_{Y}(y)}\notag \\
 =&~ {\log \frac {\exp\left( - \lambda^\star_{X, Y} d(x, y) \right)} {\E{\exp\left(-\lambda^\star_{X, Y} d(x, Y)\right) }}}\label{eq_Zstardensity}\\
 =&~  \Lambda_{Y}(x, \lambda^\star_{X, Y})  - \lambda^\star_{X, Y} d(x, y) + \lambda^\star_{X, Y} d \label{eq_jlambdagendensity}
\end{align}
where
\begin{equation}
 \lambda^\star_{X, Y} =  -\mathbb R_{X, Y}^\prime (d)
\end{equation}


\section{Prior work}
\label{sec:prior}
In this section, we summarize the main available bounds on the fixed-blocklength fundamental limits of lossy compression and we review the main relevant asymptotic refinements to Shannon's lossy source coding theorem.

\subsection{Achievability bounds}
Returning to the general setup of Definition \ref{defn:(M, d, eps)}, the basic general achievability result can be distilled \cite{verdu2009notes} from Shannon's coding theorem for memoryless sources:
\begin{thm}[Achievability, \cite{shannon1959coding}\cite{verdu2009notes}]
\label{thm:Ashannon}
Fix $P_{X}$, a positive integer $M$ and $d \geq d_{\min}$. There exists an $(M,d,\epsilon)$ code such that
\begin{align}
\label{eq_Ashannon} \epsilon
&\leq
\inf_{P_{Y |X } }\bigg\{ \mathbb{P}\left[ d\left(X, Y
\right) > d \right] \notag\\
&+ \inf_{\gamma > 0} \left\{ \mathbb{P}\left[
\imath_{X; Y}\left(X; Y \right) > \log M - \gamma \right] +
e^{-\exp(\gamma) } \right\} \bigg\}
\end{align}
\end{thm}

Theorem \ref{thm:Ashannon} is the most general existing achievability result (i.e. existence result of a code with a guaranteed upper bound on error probability). In particular, it allows us to deduce that for stationary memoryless sources with separable distortion measure, i.e. when $P_{X^n}  = P_{\mathsf X} \times \ldots \times P_{\mathsf X}$, $d(x^n, y^n) = \frac 1 n \sum_{i = 1}^n d(x_i, y_i)$, it holds that
\begin{align}
 \limsup_{n \to \infty} R(n, d) &\leq \mathbb R_{\mathsf X}(d) \label{eq_AlimR(n,d)}\\
  \limsup_{n \to \infty} R(n, d, \epsilon) &\leq \mathbb R_{\mathsf X}(d)  \label{eq_AlimR(n,d,eps)}
\end{align}
where $\mathbb R_{\mathsf X}(d)$ is defined in \eqref{eq_RR(d)},  and $0 < \epsilon < 1$.

For three particular setups of i.i.d. sources with separable distortion measure, we can cite the achievability bounds of Goblick \cite{goblick1962coding} (fixed-rate compression of a finite alphabet source), Pinkston \cite{pinkston1967encoding} (variable-rate compression of a finite-alphabet source) and Sakrison \cite{sakrison1968geometric} (variable-rate compression of a Gaussian source with mean-square error distortion). Sakrison's achievability bound is summarized below as the least cumbersome of the aforementioned:

\begin{thm}[Achievability, \cite{sakrison1968geometric}]
Fix blocklength $n$, and let $X^n$ be a Gaussian vector with independent components of variance $\sigma^2$. There exists a variable-length code achieving average mean-square error $d$ such that
\begin{align}
\E{\ell(X^n) } &\leq - \frac {n-1} {2} \log \left( \frac d {\sigma^2} - \frac 1 {1.2 n} \right) +  \frac{1}{2} \log n \notag\\
&+ \log 4 \pi + \frac{2}{3} \log e + \frac{5 \log e}{12(n+1) }
\end{align}
\end{thm}

\subsection{Converse bounds}
The basic converse used in conjunction with \eqref{eq_Ashannon} to prove the rate-distortion fundamental limit with average distortion is the following simple result, which follows immediately from the data processing lemma for mutual information:

\begin{thm}[Converse, \cite{shannon1959coding}]
\label{thm:Cshannon}
Fix $P_{X}$, integer $M$ and $d \geq d_{\min}$. Any $\langle M, d \rangle$ code must satisfy
\begin{equation}
\mathbb R_X(d) \leq \log M \label{eq_Cshannon}
\end{equation}
where $\mathbb R_X(d)$ is defined in \eqref{eq_RR(d)}.
\end{thm}
Shannon \cite{shannon1959coding} showed that in the case of stationary memoryless sources with separable distortion, $\mathbb R_{X^n}(d) = n \mathbb R_{\mathsf X}(d)$. Using Theorem \ref{thm:Cshannon}, it follows that  for such sources,
\begin{equation}
\mathbb R_{\mathsf X}(d) \leq R(n, d)
\end{equation}
for any blocklength $n$ and any $d > d_{\min}$, which together with \eqref{eq_AlimR(n,d)} gives
\begin{equation}
R(d) = \mathbb R_{\mathsf X}(d)
\end{equation}

The strong converse for lossy source coding \cite{korner1973coding,kieffer1991strong} states that if the compression rate $R$ is fixed and $R < \mathbb R_{\mathsf X}(d)$, then $\epsilon \to 1$ as $n \to \infty$, which together with \eqref{eq_AlimR(n,d,eps)} yields that for i.i.d. sources with separable distortion and any $0 < \epsilon < 1$,
\begin{equation}
 \limsup_{n \to \infty} R(n, d, \epsilon) = \mathbb R_{\mathsf X}(d) = R(d)
\end{equation}

For prefix-free variable-length lossy compression, the key non-asymptotic converse was obtained by Kontoyiannis \cite{kontoyiannis2000pointwise} (see also \cite{barron1985logically} for a lossless compression counterpart).

 \begin{thm}[Converse, \cite{kontoyiannis2000pointwise}]
Assume that the infimum in the right side of \eqref{eq_RR(d)}  is achieved by some conditional distribution $P_{Y|X}^\star$. If a prefix-free variable-length code for $P_X$ operates at distortion level $d$, then for any $\gamma > 0$,
\begin{equation}
\Prob{\ell(X) \leq \jmath_X(X, d) - \gamma} \leq 2^{-\gamma}
\end{equation}
\end{thm}
 For DMS with finite alphabet and bounded separable distortion measure, a finite blocklength converse can be distilled from Marton's fixed-rate lossy compression error exponent \cite{marton1974error}:

\begin{thm}[Converse, \cite{marton1974error}]
\label{thm:Cmarton}
Consider a DMS with finite input and reproduction alphabets, source distribution $P$ and separable distortion measure with $\max_{\mathsf x}\min_{\mathsf y} d(\mathsf x, \mathsf y) = 0$, $\Delta_{\max} = \max_{\mathsf x, \mathsf y} d(\mathsf x, \mathsf y) < +\infty$.  Fix $0 < d < \Delta_{\max}$. Let the corresponding rate-distortion and distortion-rate functions be denoted by $R_P(d)$ and $D_P(R)$, respectively. Fix an arbitrary $(n, M, d, \epsilon)$ code.
\begin{itemize}
\item If the code rate $R = \frac{\log M}{n}$ satisfies
\begin{equation}
R < R_{P} (d),
\end{equation}
then the excess-distortion probability is bounded away from zero:
\begin{equation}
\epsilon \geq \frac {D_{P}(R) - d}{\Delta_{\max} - d},
\end{equation}
\item If $R$ satisfies
\begin{align}
\label{eq_martonRegion}
R_{P}(d) < R < \max_{Q} R_Q(d),
\end{align}
where the maximization is over the set of all probability distributions on $\mathcal A$, then
\begin{align}
\label{eq_Cmarton}
\epsilon &\geq \sup_{\delta >0, Q} \left( \frac {D_{Q}(R) - d}{\Delta_{\max} - d} - Q^n({G_{\delta,n}}^c)
\right) \notag\\
&\cdot \exp \left( - n \left( D(Q\|P) + \delta \right)\right),
\end{align}
where the supremization is over all probability distributions on
$\mathcal A$ satisfying $R_Q(d) > R$, and
\begin{align}
G_{\delta, n} &= \left\{ x^n \in \mathcal A^n\colon \  \frac 1 n \log
\frac{Q^n(x^n)}{P^n(x^n)} \leq D(Q\|P) + \delta
 \right\} \notag
\end{align}
\end{itemize}
\end{thm}

 It turns out that the converse in Theorem \ref{thm:Cmarton} results in rather loose lower bounds on $R(n, d, \epsilon)$ unless $n$ is very large, in which case the rate-distortion function already gives a tight lower bound. Generalizations of the error exponent results in \cite{marton1974error} are found in \cite{viterbi1979principles,ihara2000error,han2000reliability,iriyama2001error,iriyama2002probability}.

\subsection{Gaussian Asymptotic Approximation}
\label{ssec:priorAEP}
The ``lossy asymptotic equipartition property (AEP)''  \cite{kieffer1991sample}, which leads to strong achievability and converse bounds for variable-rate quantization, is concerned with the almost sure asymptotic behavior of the distortion $d-$balls. Second-order refinements of the ``lossy AEP'' were studied in  \cite{yang1999redundancy,dembo1999asymptotics,kontoyiannis2000pointwise}.\footnote{The result of Theorem \ref{thm:AEP} was pointed out in  \cite[Proposition 3]{kontoyiannis2000pointwise} as a simple corollary to the analyses in \cite{dembo1999asymptotics,yang1999redundancy}. See \cite{dembo2001source} for a generalization to $\alpha$-mixing sources.}
\begin{thm} [``Lossy AEP'']
For memoryless sources with separable distortion measure satisfying the regularity restrictions \eqref{item:first}--\eqref{item:last} in Section \ref{sec:2order},
\begin{equation*}
\log \frac 1 {P_{Y^n}^\star(B_d(X^n))}=  \sum_{i=1}^n \jmath_{\mathsf X}(X_i,d) + \frac 1 2 \log n + \bigo{\log \log n}
\end{equation*}
almost surely. 
\label{thm:AEP}
\end{thm}
\begin{remark}
Note the different behavior of almost lossless data compression:
\begin{equation}
\log \frac{1}{P_{Y^n}^\star(B_0(X^n))} = \log \frac 1 {P_{X^n}(X^n)} = \sum_{i = 1}^n \imath_{\mathsf X}(X_i) \label{eq_AEPlossless}
\end{equation}

\end{remark}

Kontoyiannis  \cite{kontoyiannis2000pointwise} pioneered the second-order refinement of the variable-length rate-distortion function showing that for memoryless sources with separable distortion measures the optimum prefix-free description length at distortion level $d$ satisfies
  \begin{equation}
\ell^\star(X^n) = nR(d) + \sqrt{n} G_n + \bigo{\log n}  \ a.s.
\label{eq_kontoyiannisIntro}
\end{equation}
where $G_n$ converges in distribution to a Gaussian random variable with zero mean and variance equal to the rate-dispersion function defined in Section \ref{sec:2order}.

\subsection{Asymptotics of redundancy}
Considerable attention has been paid to the asymptotic behavior of the redundancy, i.e. the difference between the average distortion $D(n, R)$ of the best $n-$dimensional quantizer and the distortion-rate function $D(R)$. For finite-alphabet i.i.d. sources, Pilc \cite{pilc1967coding} strengthened the positive lossy source coding theorem by showing that
\begin{equation}
D(n, R) - D(R) \leq  -\frac{\partial D(R)}{\partial R}\frac{\log n}{2n}  + \smallo{\frac{\log n}{n}}  \label{eq_Aredundancy}
\end{equation}
 Zhang, Yang and Wei \cite{zhang1997redundancy} proved a converse to \eqref{eq_Aredundancy}, thereby showing that for memoryless sources with finite alphabet,
 \begin{equation}
 D(n, R) - D(R) = -\frac{\partial D(R)}{\partial R}\frac{\log n}{2n} + \smallo{\frac{\log n}{n}}
 \end{equation}
Using a geometric approach akin to that of Sakrison \cite{sakrison1968geometric}, Wyner \cite{wyner1968communication} showed that \eqref{eq_Aredundancy} also holds for stationary Gaussian sources with mean-square error distortion, while Yang and Zhang \cite{yang1999redundancy} extended  \eqref{eq_Aredundancy} to abstract alphabets. Note that as the average overhead over the distortion-rate function is dwarfed by its standard deviation, the analyses of \cite{pilc1967coding,wyner1968communication,zhang1997redundancy,yang1999redundancy}  are bound to be overly optimistic since they neglect the stochastic variability of the distortion.

\section{New finite blocklength bounds}
\label{sec:new}
In this section we give achievability and converse results for any source and any distortion measure according to the setup of Section \ref{sec:nonasymptotic}. When we apply these results in Sections \ref{sec:2order} - \ref{sec:GMS}, the source $X$ becomes an $n-$tuple $(X_1, \ldots, X_n)$.

\subsection{Converse bounds}
Our first result is a general converse bound.
\begin{thm}[Converse]
\label{thm:Cg}
Assume the basic conditions  \eqref{item:a}--\eqref{item:d} in Section \ref{sec:nonasymptotic} are met. Fix $d > d_{\min}$. Any $(M,d,\epsilon)$ code must satisfy
\begin{equation}
\epsilon \geq \sup_{\gamma \geq 0} \left\{ \Prob{\jmath_X(X,d) \geq \log M + \gamma } - \exp(-\gamma) \right\}\label{eq_Cg}
\end{equation}
\end{thm}
\begin{proof}
Let the encoder and decoder be the random transformations $P_{Z|X}$ and $P_{Y|Z}$, where $Z$ takes values in $\{1, \ldots, M\}$. Let $Q_Z$ be equiprobable on $\{1, \ldots, M\}$, and let $Q_Y$ denote the marginal of $P_{Y|Z} Q_Z $.
We have\footnote{We write summations over alphabets for simplicity. All our results in Sections \ref{sec:new} and \ref{sec:2order} hold for arbitrary probability spaces.}, for any $\gamma \geq 0$
\begin{align}
&~ \Prob{ \jmath_X(X,d) \geq \log M + \gamma} \\
=&~ \Prob{ \jmath_X(X,d) \geq \log M + \gamma, d(X, Y) > d} \notag\\
+&~ \Prob{ \jmath_X(X,d) \geq \log M + \gamma, d(X, Y) \leq d}\\
\leq&~ \epsilon + \sum_{x \in A} P_X(x) \sum_{z = 1}^M P_{Z|X}(z|x) \notag\\
\cdot&~\sum_{y \in B_d(x)} P_{Y|Z}(y|z)  1\left\{ M \leq \exp\left(\jmath_X(x,d) -\gamma \right) \right\}  \\
\leq&~ \epsilon + \exp\left(-\gamma\right) \sum_{x \in A} P_X(x) \exp\left( \jmath_X(x,d)\right) \notag\\
\cdot&~ \sum_{z = 1}^M \frac 1 {M}   \sum_{y \in B_d(x)} P_{Y|Z}(y|z)   \label{eq_-Ca}\\
=&~ \epsilon + \exp\left(-\gamma\right)  \sum_{x \in A} P_X(x) \exp\left( \jmath_X(x,d)\right) Q_Y(B_d(x))    \\
\leq&~  \epsilon + \exp\left(-\gamma\right)   \sum_{y \in B} Q_Y(y) \notag\\
\cdot&~\sum_{x \in A} P_X(x) \exp\left( \lambda^\star d - \lambda^\star d(x, y) + \jmath_X(x,d)\right) \label{eq_-Ca1}\\
\leq&~ \epsilon + \exp\left(-\gamma\right) \label{eq_-Cb}
\end{align}
where
\begin{itemize}
 \item \eqref{eq_-Ca} follows by upper-bounding
\begin{align}
  &~P_{Z|X}(z|x) 1\left\{ M \leq \exp\left(\jmath_X(x,d) -\gamma \right) \right\} \notag\\
  \leq&~\frac {\exp\left( -\gamma\right)} M \exp\left(\jmath_X(x,d) \right)
\end{align}
 for every $(x, z) \in A \times \left\{ 1, \ldots, M\right\}$,
 \item \eqref{eq_-Ca1} uses \eqref{eq_Pdball_Markov} particularized to $Y$ distributed according to $Q_Y$, and
 \item \eqref{eq_-Cb} is due to \eqref{eq_csiszar}.
\end{itemize}
\end{proof}
\begin{remark}
Theorem \ref{thm:Cg} gives a pleasing generalization of the almost-lossless data compression converse bound \cite{verdu2009notes},\cite[Lemma 1.3.2]{han2003information}. In fact, skipping \eqref{eq_-Ca1}, the above proof applies to the case $d = 0$ and $d(x, y) = 1 \left\{ x \neq y \right\}$ that corresponds to almost-lossless data compression.
\end{remark}
\begin{remark}
As explained in Appendix \ref{appx:csiszar}, condition \eqref{item:d} can be dropped from the assumptions of Theorem \ref{thm:Cg}. 
\end{remark}
Our next converse result, which is tighter than the one in Theorem \ref{thm:Cg} in some cases, is based on binary hypothesis testing. The optimal performance achievable among all randomized tests $P_{W|X}\colon A \rightarrow \left\{ 0, 1\right\}$ between probability distributions $P$ and $Q$ on $A$ is denoted by ($1$ indicates that the test chooses $P$):\footnote{Throughout, $P$, $Q$ denote distributions, whereas $\mathbb P$, $\mathbb Q$ are used for the corresponding probabilities of events on the underlying probability space.}
\begin{equation}
\label{eq_beta}
\beta_{\alpha}(P, Q) = \min_{\substack{P_{W|X}\colon \\ \Prob{W = 1} \geq \alpha}} \mathbb Q \left[ W = 1\right]
\end{equation}

\begin{thm}[Converse]
\label{thm:C}
Let $P_X$ be the source distribution defined on the alphabet $A$. Any $(M,d,\epsilon)$ code must satisfy
\begin{equation}
\label{eq_C}
M \geq \sup_{Q} \inf_{y \in B} \frac {\beta_{1 - \epsilon}(P_X, Q)}{\mathbb Q \left[ d(X, y) \leq d\right]}
\end{equation}
where the supremum is over all distributions on $A$.
\end{thm}
\begin{proof}
Let $(P_{Z|X}, P_{Y|Z})$ be an $(M, d, \epsilon)$ code. Fix a distribution $Q$ on $A$, and observe that  $W = 1 \left\{ d(X, Y) \leq d \right\}$ defines a (not necessarily optimal) hypothesis test between $P_X$ and $Q$ with $\mathbb P \left[ W = 1\right] \geq 1 - \epsilon$. Thus,
\begin{align}
&~\beta_{1 - \epsilon}(P_X, Q) \notag\\
\leq&~  \sum_{x \in A} Q_X(x) \sum_{m = 1}^M P_{Z|X}(m|x) \sum_{y \in B} P_{Y|Z} (y|m) 1\{ d(x, y) \leq d\} \notag\\
 \leq&~ \sum_{m = 1}^M   \sum_{y \in B} P_{Y|Z} (y|m) \sum_{x \in A} Q_X(x)  1\{ d(x, y) \leq d\}\\
 \leq&~ \sum_{m = 1}^M   \sum_{y \in B} P_{Y|Z} (y|m) \sup_{y \in B} \mathbb Q \left[d(X, y) \leq d\right]\\
=&~ M \sup_{y \in B} \mathbb Q \left[d(X, y) \leq d\right]
\end{align}
\end{proof}
 Suppose for a moment that $X$ takes values on a finite alphabet, and let us further lower bound \eqref{eq_C} by taking $Q$ to be the equiprobable distribution on $A$, $Q = U$. Consider the set $\Omega \subset A$ that has total probability $1 - \epsilon$ and contains the most probable source outcomes, i.e. for any source outcome $x \in \Omega$, there is no element outside $\Omega$ having probability greater than $P_{X}(x)$. For any $x \in \Omega$, the optimum binary hypothesis test (with error probability $\epsilon$) between $P_X$ and $Q$ must choose $P_X$. Thus the numerator of \eqref{eq_C} evaluated with $Q = U$ is proportional to the number of elements in $\Omega$, while the denominator is proportional to the number of elements in a distortion ball of radius $d$. Therefore \eqref{eq_C} evaluated with $Q = U$ yields a lower bound to the minimum number of $d$-balls required to cover $\Omega$.

\begin{remark}
 In general, the lower bound in Theorem \ref{thm:C} is not achievable due to overlaps between distortion $d-$balls that comprise the covering. One special case when it is in fact achievable is almost lossless data compression on a countable alphabet $A$. To encompass that case, it is convenient to relax the restriction in \eqref{eq_beta} that requires $Q$  to be a probability measure and allow it to be a $\sigma$-finite measure,  so that $\beta_\alpha(P_X, Q)$ is no longer bounded by 1.\footnote{The Neyman-Pearson lemma generalizes to $\sigma$-finite measures.}  Note that Theorem \ref{thm:C} would still hold. Letting $U$ to be the counting measure on $A$  (i.e. $U$ assigns unit weight to each letter), we have (Appendix \ref{appx:lossless})
 \begin{equation}
 \beta_{1 - \epsilon}(P_X, U) \leq M^\star(0, \epsilon) \leq  \beta_{1 - \epsilon}(P_X, U) + 1 \label{eq_Mhtlossless}
\end{equation}
The lower bound in \eqref{eq_Mhtlossless} is satisfied with equality whenever $ \beta_{1 - \epsilon}(P_X, U)$ is achieved by a non-randomized test. \end{remark}

\subsection{Achievability bounds}
The following result gives an exact analysis of the excess probability of random coding, which holds in full generality.
\begin{thm}[Exact performance of random coding]
\label{thm:ExactExcess}
Denote by $\epsilon_d\left( c_1, \ldots, c_M\right)$ the probability of exceeding distortion level $d$ achieved by the optimum encoder with codebook $\left(c_1, \ldots, c_M\right)$.  Let $Y_1, \ldots, Y_M$ be independent, distributed according to an arbitrary distribution on the reproduction alphabet $P_Y$. Then
\begin{equation}
\label{eq_ExactExcess}
\mathbb E \left[ \epsilon_d \left( Y_1, \ldots, Y_M\right) \right] = \E{ 1 - P_Y(B_d(X)) }^M
\end{equation}
\end{thm}
\begin{proof}
Upon observing the source output $x$, the optimum encoder chooses arbitrarily among the members of the set
\begin{equation*}
\arg\min_{i = 1, \ldots, M} d(x, c_i)
\end{equation*}
The indicator function of the event that the distortion exceeds $d$ is
\begin{equation}
1 \left\{ \min_{i = 1, \ldots, M} d(x, c_i) > d\right\} = \prod_{i = 1}^M 1 \left\{ d(x, c_i) > d\right\}
\end{equation}
 Averaging over both the input $X$ and the choice of codewords chosen independently of $X$, we get
 \begin{align}
&~\mathbb E \left[ \prod_{i = 1}^M 1 \left\{ d(X, Y_i) > d\right\}\right] \notag\\
=&~  \E{ \mathbb E \left[ \prod_{i = 1}^M 1 \left\{ d(X, Y_i) > d\right\}| X\right]} \\
=&~ \mathbb E \prod_{i = 1}^M \mathbb E \left[ 1 \left\{ d(X, Y_i) > d\right\}| X\right] \label{eq_condindep}\\
=&~ \mathbb E \left( \Prob{ d(X, Y) > d | X } \right)^M
 \end{align}
 where in \eqref{eq_condindep} we have used the fact that $Y_1, \ldots, Y_M$ are independent even when conditioned on $X$.
\end{proof}

Invoking Shannon's random coding argument, the following achievability result follows immediately from Theorem \ref{thm:ExactExcess}.
\begin{thm}[Achievability]
\label{thm:A}
There exists an $(M,d,\epsilon)$ code with
\begin{equation}
\label{eq_A}
\epsilon \leq \inf_{P_{Y}}  \E{ 1 - P_Y(B_d(X))}^M
\end{equation}
where the infimization is over all random variables defined on $B$, independent of $X$.
\end{thm}
While the right side of \eqref{eq_A} gives the exact performance of random coding, Shannon's random coding bound (Theorem \ref{thm:Ashannon}) was obtained by upper bounding the performance of random coding. As a consequence, the result in Theorem \ref{thm:A} is tighter than Shannon's random coding bound (Theorem \ref{thm:Ashannon}), but it is also harder to compute.

Applying $(1 - x)^M \leq e^{-Mx}$ to \eqref{eq_A}, one obtains the following more numerically stable bound.
\begin{cor}[Achievability]
\label{cor:A}
There exists an $(M,d,\epsilon)$ code with
\begin{equation}
\label{eq_Acor}
\epsilon \leq \inf_{P_{Y}}  \E{ e^{-M P_Y(B_d(X))}}
\end{equation}
where the infimization is over all random variables defined on $B$, independent of $X$.
\end{cor}

The last result in this section will come handy in the analysis of the bound in Theorem \ref{thm:A} (see Section \ref{ssec:dtiltedgen} for related notation).
\begin{lemma} For an arbitrary $P_Y$ on $B$,  
\begin{align}
P_{Y}(B_d(x)) \geq  \sup_{P_{\hat X}, \gamma > 0}\exp\left( -\Lambda_{Y}(x, \lambda^\star_{\hat X, Y}) - \lambda^\star_{\hat X, Y} \gamma \right) \notag\\
\cdot \Prob{ d - \gamma < d(x, \hat Z^\star) \leq d | \hat X = x}  \label{eq_aepAnonasymptotic}
\end{align}
where the supremization is over all $P_{\hat X}$ on $A$ such that $d_{\min \mid \hat X, Y} < d$, and $\hat Z^\star$ achieves $\mathbb R_{\hat X, Y}(d)$. 
\label{lemma:aepAnonasymptotic}
 \end{lemma}
\begin{proof}
 We streamline the treatment in \cite[(3.26)]{yang1999redundancy}.  Fix $\gamma > 0$ and distribution $P_{\hat X}$ on the input alphabet $A$. We have\begin{align}
&~ P_{Y}(B_d(x)) \notag\\
=&~ \sum_{y \in B_d(x)} P_{Y}(y)\\
 \geq&~ \sum_{y \in B_d(x) \setminus B_{d - \gamma}(x)} P_{Y}(y)\\
 \geq&~ \exp\left(- \lambda^\star_{\hat X, Y} \gamma \right)\notag \\
 \cdot&~\sum_{y \in B_d(x) \setminus B_{d - \gamma}(x)} P_{Y}(y) \exp\left( \lambda^\star_{\hat X, Y} d - \lambda^\star_{\hat X, Y} d(x, y) \right) \label{eq_-A1} \\
 =&~ \exp\left( -\Lambda_{Y}(x, \lambda^\star_{\hat X, Y}) - \lambda^\star_{\hat X, Y} \gamma \right)\notag\\
 \cdot&~ \sum_{y \in B_d(x) \setminus B_{d - \gamma}(x)} P_{\hat Z^\star | \hat X = x}(y) \label{eq_-A2}\\
 =&~  \exp\left( -\Lambda_{Y}(x, \lambda^\star_{\hat X, Y}) - \lambda^\star_{\hat X, Y} \gamma \right) \notag\\
 &~\Prob{ d - \gamma < d(x, \hat Z^\star) \leq d | \hat X = x} \label{eq_dballlb}
\end{align}
where \eqref{eq_-A1} holds because $y \notin B_{d - \gamma}(x)$ implies 
\begin{equation}
 \lambda d - \lambda d(x, y) - \lambda \gamma \leq 0
\end{equation}
for all $\lambda > 0$, and \eqref{eq_-A2} takes advantage of \eqref{eq_jlambdagendensity}. 
\end{proof}
\section{Gaussian approximation}
\label{sec:2order}
\subsection{Rate-dispersion function}
\label{sec:2orderA}
In the spirit of \cite{polyanskiy2010channel}, we introduce the following definition.
\begin{defn}
Fix $d \geq d_{\min}$. The rate-dispersion function (squared information units per source output) is defined as
\begin{align}
 V(d)&= \lim_{\epsilon \rightarrow 0} \limsup_{n \rightarrow \infty}
 n \left( \frac { R(n,d,\epsilon) - R(d)}{ \Qinv{\epsilon}}\right)^2\\
&= \lim_{\epsilon \rightarrow 0} \limsup_{n \rightarrow \infty}
 \frac { n \left(R(n,d,\epsilon) - R(d)\right)^2}{2 \log_e \frac 1
\epsilon} \label{eq_dispersiondef}
\end{align}
\end{defn}

Fix $d$, $0 < \epsilon < 1$, $\eta > 0$, and suppose the target is to sustain the probability of exceeding distortion $d$ bounded by $\epsilon$ at rate $R = (1 + \eta) R(d)$. As \eqref{eq_2orderIntro} implies, the required blocklength scales linearly with rate dispersion:
\begin{equation}
n(d, \eta, \epsilon) \approx \frac{V(d)}{R^2(d)} \left( \frac{\Qinv{\epsilon}}{\eta}\right)^2 \label{eq_nrequired}
\end{equation}
where note that only the first factor depends on the source, while the second depends only on the design specifications.

\subsection{Main result}
\label{ssec:2orderMain}
In addition to the basic conditions \eqref{item:a}-\eqref{item:d} of Section \ref{ssec:dtilted}, in the remainder of this section we impose the following restrictions on the source and on the distortion measure.
\begin{enumerate}[(i)]
\item The source $\{X_i\}$ is stationary and memoryless,  $P_{X^n}  = P_{\mathsf X} \times \ldots \times P_{\mathsf X}$. \label{item:first}
\item The distortion measure is separable, $d(x^n, y^n) = \frac 1 n \sum_{i = 1}^n d(x_i, y_i)$. \label{item:separable}
\item The distortion level satisfies $d_{\min} < d < d_{\max}$, where $d_{\min}$ is defined in \eqref{eq_dmin}, and $d_{\max} =\inf_{\mathsf y \in \mathcal B} \E{d(\mathsf X, \mathsf y)}$, where averaging is with respect to the unconditional distribution of $\mathsf X$. The excess-distortion probability satisfies $0 < \epsilon < 1$. \label{item:dminmax}
 \item $
 \E{d^{9}(\mathsf X, \mathsf Y^\star)} < \infty
$
where averaging is with respect to $P_{\mathsf X} \times P_{\mathsf Y^\star}$. \label{item:last} 
\end{enumerate}
The main result in this section is the following\footnote{Recently, using an approach based on typical sequences and error exponents, Ingber and Kochman \cite{ingber2011dispersion} independently found the dispersion of finite alphabet sources. The Gaussian i.i.d. source with mean-square error distortion was treated separately in \cite{ingber2011dispersion}. The result of Theorem \ref{thm:2order} is more general as it applies to sources with abstract alphabets.}.
\begin{thm}[Gaussian approximation]
\label{thm:2order}
Under restrictions \eqref{item:first}--\eqref{item:last},
\begin{align}
R(n,d,\epsilon) &= R(d) + \sqrt{\frac{V(d)}{n}}\Qinv{\epsilon} + \theta\left(\frac{\log n}{n}\right)
\label{eq_2order}\\
V(d) &= {\rm Var}\left[\jmath_{\mathsf X}(\mathsf X, d) \right] \label{eq_dispersion}
\end{align}
and the remainder term in \eqref{eq_2order} satisfies
\begin{align}
&-\frac 1 2\frac{\log n}{n} + \bigo{\frac 1 n} \leq \theta\left(\frac{\log n}{n}\right) \label{eq_Cremainder}\\
&\leq C\frac{\log n}{n} + \frac{\log\log n}{n} + \bigo{\frac 1 n} 	\label{eq_Aremainder}
\end{align}
where
 \begin{equation}
 C = \frac 1 2 + \frac{\Var{\Lambda^{\prime}_{\mathsf Y^\star}(\mathsf X, \lambda^\star) } }{ \E{\left| \Lambda^{\prime\prime}_{\mathsf Y^\star}(\mathsf X, \lambda^\star)\right|} \log e} \label{eq_remainderC}
\end{equation}
In \eqref{eq_remainderC}, $(\cdot)^\prime$ denotes differentiation with respect to $\lambda$, $\Lambda_{\mathsf Y^\star}(\mathsf x, \lambda)$ is defined in \eqref{eq_jlambdagen}, and $\lambda^\star = - R^\prime(d)$. 
\end{thm}

\begin{remark}
Since the rate-distortion function can be expressed as (see \eqref{eq_Ejd} in Section  \ref{sec:nonasymptotic})
\begin{equation}
\label{eq_R(d)alternative}
R(d) = \E{\jmath_{\mathsf X}(\mathsf X, d)}
\end{equation}
it is equal to the expectation of the random variable whose variance we take in \eqref{eq_dispersion}, thereby drawing a pleasing parallel with the channel coding results in \cite{polyanskiy2010channel}.
\end{remark}
\begin{remark}
For almost lossless data compression, Theorem \ref{thm:2order} still holds as long as the random variable $\imath_{\mathsf X}(\mathsf X)$ has finite third moment. Moreover, using \eqref{eq_Mhtlossless} the upper bound in \eqref{eq_Aremainder} can be strengthened (Appendix \ref{appx:2orderlossless}) to obtain for $\Var{ \imath_{\mathsf X}( \mathsf X)} > 0$
\begin{align}
R(n, 0, \epsilon) &= H(\mathsf X) + \sqrt { \frac{\Var{ \imath_{\mathsf X}( \mathsf X)}}{n} } \Qinv{\epsilon} \notag\\
&- \frac 1 2 \frac{\log n}{n} + \bigo{\frac 1 n} \label{eq_2orderLossless}
\end{align}
which is consistent with the second-order refinement for almost lossless data compression developed in \cite{strassen1962asymptotische}.
If $\Var{ \imath_{\mathsf X}( \mathsf X)} = 0$, then
\begin{equation}
R(n, 0, \epsilon) = H(\mathsf X) - \frac 1 n \log \frac 1 {1 - \epsilon} + o_n \label{eq_2orderLossless0}
\end{equation}
where 
\begin{equation}
 0 \leq o_n \leq \frac {\exp\left( -nH(\mathsf X)\right)} {(1 - \epsilon) n}
\end{equation}
As we will see in Section \ref{sec:BMS}, in contrast to the lossless case in \eqref{eq_2orderLossless}, the remainder term in the lossy case in \eqref{eq_2order} can be strictly larger than $- \frac 1 2 \frac {\log n} n$ appearing in \eqref{eq_2orderLossless} even when $V(d) > 0$.
\end{remark}
\begin{remark}
 As will become apparent in the proof of Theorem \ref{thm:2order}, if $V(d) = 0$, the lower bound in \eqref{eq_2order} can be strengthened non-asymptotically:
\begin{equation}
 R(n, d, \epsilon) \geq R(d) -\frac 1 n \log \frac 1 {1 - \epsilon} \label{eq_2orderLossy0}
\end{equation}
which aligns nicely with \eqref{eq_2orderLossless0}.
\end{remark}
\begin{remark}
Let us consider what happens if we drop restriction \eqref{item:d} of Section \ref{ssec:dtilted} that $R(d)$ is achieved by the unique conditional distribution $P_{\mathsf Y|\mathsf X}^\star$. If several $P_{\mathsf Y|\mathsf X}$  achieve $R(d)$,  writing $ \jmath_{\mathsf X; \mathsf Y}(\mathsf x, d)$ for the $d-$tilted information corresponding to $\mathsf Y$, Theorem \ref{thm:2order} still holds with
\begin{equation}
 V(d) =
\begin{cases}
 \max \Var{\jmath_{\mathsf X; \mathsf Y}(\mathsf X, d)} & 0 < \epsilon \leq \frac 1 2\\
 \min  \Var{\jmath_{\mathsf X; \mathsf Y}(\mathsf X, d)} & \frac 1 2 < \epsilon < 1
\end{cases}
\end{equation}
where the optimization is performed over all $P_{\mathsf Y|\mathsf X}$ that achieve the rate-distortion function. Moreover, as explained in Appendix \ref{appx:csiszar}, Theorem \ref{thm:Cg} and the converse part of Theorem \ref{thm:2order} do not even require existence of a minimizing $P_{\mathsf Y|\mathsf X}^\star$.
\end{remark}

Let us consider three special cases where $V(d)$ is constant as a function of $d$.


a) {\it Zero dispersion}.
For a particular value of $d$, $V(d) = 0$ if and only if $\jmath_{\mathsf X}(\mathsf X,d)$ is deterministic with probability 1.
In particular, for finite alphabet sources, $V(d) = 0$ if the source distribution $P_{\mathsf X}$ maximizes $\mathbb R_{\mathsf X}(d)$ over all source distributions defined on the same alphabet \cite{ingber2011dispersion}.  Moreover, Dembo and Kontoyiannis \cite{dembo2001critical} showed that under mild conditions, the rate-dispersion function can only vanish for at most finitely many distortion levels $d$ unless the source is equiprobable and the distortion matrix is symmetric with rows that are permutations of one another, in which case $V(d) = 0$ for all $d \in (d_{\min},d_{\max})$.

b) {\it Binary source with bit error rate distortion}.  Plugging $n = 1$ into \eqref{eq_jBMS}, we observe that the rate-dispersion function reduces to the varentropy \cite{verdu2009notes} of the source,
\beq
V(d) = V(0) = \Var{\imath_{\mathsf X}(\mathsf X)}
\eeq

c) {\it Gaussian source with mean-square error distortion}. Plugging $n = 1$ into \eqref{eq_jGMS}, we see that
\beq
V(d) = \frac 1 2 \log^2 e
\eeq
for all $0 < d < \sigma^2$.
 Similar to the BMS case, the rate dispersion is equal to the variance of $\log f_{\mathsf X}(\mathsf X)$, where $f_{\mathsf X}(\mathsf X)$ is the Gaussian probability density function.
 
\subsection{Proof of Theorem \ref{thm:2order}}
Before we proceed to proving Theorem \ref{thm:2order}, we state two auxiliary results. The first is an important tool in the Gaussian approximation analysis of $R(n, d, \epsilon)$.
\begin{thm}[{Berry-Esseen CLT, e.g. \cite[Ch. XVI.5 Theorem 2]{feller1971introduction}} ]
\label{thm:Berry-Esseen}
Fix a positive integer $n$. Let $Z_i$, $i = 1, \ldots, n$ be independent. Then, for any real $t$
\begin{equation}
\left| \mathbb P \left[ \sum_{i = 1}^n Z_i > n \left( \mu_n + t \sqrt {\frac{V_n}{ n}}\right) \right]  - Q(t) \right| \leq \frac {B_n}{\sqrt n},
\label{eq_BerryEsseen}
\end{equation}
where
\begin{align}
\mu_n &= \frac 1 n \sum_{i = 1}^n \E{ Z_i}\\
V_n &= \frac 1 n \sum_{i = 1}^n \Var{Z_i}\\
T_n &= \frac 1 n \sum_{i = 1}^n \E{ |Z_i - \mu_i|^3 }\\
B_n &= 6 \frac{T_n}{V_n^{3/2}} \label{eq_BerryEsseenBn}
\end{align}
\end{thm}

The second auxiliary result, proven in Appendix \ref{appx:aep}, is a nonasymptotic refinement of the lossy AEP (Theorem \ref{thm:AEP}) tailored to our purposes. 
\begin{lemma} Under restrictions \eqref{item:first}--\eqref{item:last}, there exist constants $n_0, c, K > 0$ such that for all $n \geq n_0$, 
\begin{align}
 &~ \Prob{ \log \frac 1 {P_{Y^{n\star}}(B_d(X^n))} \leq \sum_{i = 1}^n \jmath_{\mathsf X}(X_i, d) + C \log n + c } \notag\\
  \geq&~ 1 - \frac K {\sqrt n}
\end{align}
where $C$ is given by \eqref{eq_remainderC}.
\label{lemma:aepA}
\end{lemma}

We start with the converse part. Note that for the converse, restriction \eqref{item:last} can be replaced by the following weaker one: 
\begin{enumerate}
 \item[(iv$^\prime$)] The random variable $\jmath_{\mathsf X}(\mathsf X,d)$ has finite absolute third moment. \label{item:jmoment} 
\end{enumerate}
To verify that \eqref{item:last} implies (iv$^\prime$), observe that by the concavity of the logarithm,
\begin{equation}
0 \leq \jmath_{\mathsf X}(\mathsf x, d) + \lambda^\star d\leq   \lambda^\star \E{d(\mathsf x, \mathsf Y^\star)} 
\end{equation}
so
\begin{equation}
\E{ \left| \jmath_{\mathsf X}(\mathsf X, d) + \lambda^\star d \right|^3} \leq \lambda^{\star 3} \E{d^3(\mathsf X, \mathsf Y^\star)}
\end{equation}
\begin{proof}[Proof of the converse part of Theorem \ref{thm:2order}]
First, observe that due to \eqref{item:first} and \eqref{item:separable}, $P_{Y^n}^\star = P_{\mathsf Y}^\star \times \ldots \times P_{\mathsf Y}^\star$, and the $d-$tilted information single-letterizes, that is, for a.e. $x^n$,
\begin{equation}
\jmath_{X^n}(x^n,d) = \sum_{i = 1}^n \jmath_{\mathsf X}(x_i,d) \label{eq_id_iid}
\end{equation}
Consider the case $V(d) > 0$, so that $B_n$ in \eqref{eq_BerryEsseenBn} with $Z_i = \jmath_{\mathsf X}(X_i,d)$ is finite by restriction (iv$^\prime$). Let $\gamma = \frac 1 2 \log n$ in \eqref{eq_Cg}, and choose
\begin{align}
\log M &= n R(d) + \sqrt{nV(d)} \Qinv{\epsilon_n}  - \gamma \label{eq_-2orderCM}\\
\epsilon_n &= \epsilon + \exp(-\gamma) + \frac {B_n}{\sqrt n} \label{eq_-2orderCepsilon}
\end{align}
so that $R = \frac {\log M }{n}$ can be written as the right side of \eqref{eq_2order} with \eqref{eq_Cremainder} satisfied.  Substituting  \eqref{eq_id_iid} and \eqref{eq_-2orderCM} in \eqref{eq_Cg}, we conclude that for any $(M, d, \epsilon^\prime)$ code it must hold that
\begin{align}
\epsilon^\prime &\geq \Prob{ \sum_{i = 1}^n\jmath_{\mathsf X} (X_i,d) \geq nR(d) + \sqrt{nV(d)} \Qinv{\epsilon_n} }\notag\\ 
&- \exp( -\gamma)  \label{eq_-2orderCb}
\end{align}
The proof for $V(d) > 0$ is complete upon noting that the right side of \eqref{eq_-2orderCb} is lower bounded by $\epsilon$ by the Berry-Esseen inequality \eqref{eq_BerryEsseen} in view of \eqref{eq_-2orderCepsilon}.

If $V(d) = 0$, it follows that $\jmath_{\mathsf X}(\mathsf X, d) = R(d)$ almost surely. Choosing $\gamma = \log \frac 1 {1 - \epsilon}$ and $ \log M = nR(d) - \gamma$ in \eqref{eq_Cg} it is obvious that $\epsilon^\prime \geq \epsilon$.
\end{proof}

\begin{proof}[Proof of the achievability part of Theorem \ref{thm:2order}]
The proof consists of the asymptotic analysis of the bound in Corollary \ref{cor:A} using Lemma \ref{lemma:aepA}.  Denote
\begin{equation}
 G_n = \log M - \sum_{i = 1}^n \jmath_{\mathsf X}(x_i, d) - C \log n - c \label{eq_-2orderAGn}
\end{equation}
where constants $c$ and $C$ were defined in Lemma \ref{lemma:aepA}. Letting $X = X^{n}$ in \eqref{eq_Acor} and weakening the right side of \eqref{eq_Acor} by choosing $P_Y = P_{Y^n}^\star = P_{\mathsf Y}^\star \times \ldots \times P_{\mathsf Y}^\star$,  we conclude that there exists an $(n, M, d, \epsilon^\prime)$ code with
\begin{align}
\epsilon^\prime &\leq \E{ e^{- M  P_{Y^n}^\star(B_d(X^n))} }\\
&\leq \E{e^{-\exp\left( G_n\right)}} + \frac K {\sqrt n} \label{eq_-2orderAa}\\
&= \E{e^{-\exp(G_n)} 1 \left\{ G_n < \log \frac {\log_e n}{2} \right\}}  \notag\\
&+\E{ e^{-\exp(G_n)} 1 \left\{ G_n \geq \log \frac {\log_e n}{2}  \right\}} \notag\\
&+ \frac K {\sqrt n}\\
&\leq \Prob{ G_n < \log \frac {\log_e n}{2}} \notag\\
&+ \frac 1 {\sqrt n} \Prob{G_n \geq \log \frac {\log_e n}{2}} + \frac K {\sqrt n} \label{eq_-2orderAb}
\end{align}
where \eqref{eq_-2orderAa} holds for $n \geq n_0$ by Lemma \ref{lemma:aepA}, and \eqref{eq_-2orderAb} follows by upper bounding $e^{-\exp(G_n)} $ by $1$ and $\frac 1 {\sqrt n}$ respectively.  We need to show that \eqref{eq_-2orderAb} is upper bounded by $
\epsilon$ for some $R = \frac{\log M}{n}$ that can be written as \eqref{eq_2order} with the remainder satisfying \eqref{eq_Aremainder}.
Considering first the case $V(d) > 0$, let
\begin{align}
\log M &= n R(d) + \sqrt{n V(d)}\Qinv{\epsilon_n}\notag \\
&+ C \log n + \log \frac{ \log_e n}{2} + c \label{eq_-2orderAR}\\
\epsilon_n &= \epsilon - \frac{B_n + K + 1}{\sqrt{n}}\label{eq_-2orderAepsilon}
\end{align}
where $B_n$ is given by \eqref{eq_BerryEsseenBn} and is finite by restriction (iv$^\prime$). Substituting \eqref{eq_-2orderAR} into \eqref{eq_-2orderAb} and applying the Berry-Esseen inequality \eqref{eq_BerryEsseen} to the first term in \eqref{eq_-2orderAb}, we conclude 
that $\epsilon^\prime \leq \epsilon$ for all $n$ such that $\epsilon_n > 0$. 

It remains to tackle the case $V(d) = 0$, which implies $\jmath_{\mathsf X}(\mathsf X, d) = R(d)$ almost surely. Let
\begin{equation}
 \log M = n R(d) + C \log n +  c + \log \log_e \frac 1 {\epsilon - \frac K {\sqrt n}}
\end{equation}
Substituting $M$ into \eqref{eq_-2orderAa} we obtain immediately that $\epsilon^\prime \leq \epsilon$, as desired.
\end{proof}

\subsection{Distortion-dispersion function}

One can also consider the related problem of finding the minimum excess distortion $D(n, R, \epsilon)$ achievable at blocklength $n$, rate $R$ and excess-distortion probability $\epsilon$. We define the {\it distortion-dispersion function} at rate $R$ by
\begin{equation}
 \mathcal V(R)
= \lim_{\epsilon \rightarrow 0} \limsup_{n \rightarrow \infty}
 \frac { n \left(D(n,R,\epsilon) - D(R)\right)^2}{2 \log_e \frac 1
\epsilon} \label{eq_distortion-dispersiondef}
\end{equation}
For a  fixed $n$ and $\epsilon$, the functions $R(n, \cdot, \epsilon)$ and $D(n, \cdot, \epsilon)$ are functional inverses of each other. Consequently, the rate-dispersion and the distortion-dispersion functions also define each other. Under mild conditions, it is easy to find one from the other:
\begin{thm}(Distortion dispersion)
If $R(d)$ is twice differentiable, $R^\prime(d) \neq 0$ and $V(d)$ is differentiable in some interval $(\underline d, \bar d]\subseteq (d_{\min}, d_{\max}]$
then for any rate $R$ such that $R = R(d)$ for some $d \in (\underline d, \bar d)$ the distortion-dispersion function is given by
\beq
\mathcal V (R) = (D^\prime(R))^2 V(D(R))
\eeq
and
\beq
D(n,R,\epsilon) = D(R) + \sqrt{\frac {\mathcal V(R)} n} \Qinv{\epsilon} - D^\prime(R)\theta\left(\frac{\log n}{n}\right)  \label{eq_D2order}
\eeq
where $\theta(\cdot)$ satisfies \eqref{eq_Cremainder}, \eqref{eq_Aremainder}. 
\label{thm:D2order}
\end{thm}
\begin{proof}
Appendix \ref{appx:D2order}.
\end{proof}
In parallel to \eqref{eq_nrequired}, suppose that the goal is to compress at rate $R$ while exceeding distortion $d = (1 + \eta) D(R)$ with probability not higher than $\epsilon$. As \eqref{eq_D2order} implies, the required blocklength scales linearly with the distortion-dispersion function:
\begin{equation}
n(R, \eta, \epsilon) \approx \frac{\mathcal V(R)}{D^2(R)} \left( \frac{\Qinv{\epsilon}}{\eta}\right)^2 \label{eq_nrequiredD}
\end{equation}

The distortion-dispersion function assumes a particularly simple form for the Gaussian memoryless source with mean-square error distortion, in which case for any $0 < d < \sigma^2$
\begin{align}
D(R) &= \sigma^2 \exp(-2R) \\
\frac{\mathcal V(R)}{D^2(R)} &= 2\\
n(R, \eta, \epsilon) &\approx 2 \left( \frac{\Qinv{\epsilon}}{\eta}\right)^2
\end{align}
so in the Gaussian case, the required blocklength is essentially independent of the target distortion.

\section{Binary memoryless source}
\label{sec:BMS}

This section particularizes the nonasymptotic bounds in Section \ref{sec:new} and the asymptotic analysis in Section \ref{sec:2order} to the stationary binary memoryless source with bit error rate distortion measure,  i.e. $d(x^n, y^n) = \frac 1 n \sum_{i= 1}^n 1\left\{x_i \neq y_i\right\}$.
For convenience, we denote
\begin{equation}
\binosum{n}{k} = \sum_{j = 0}^{k} {n \choose j} \label{eq_binosum}
\end{equation}
with the convention $\binosum{n}{k} = 0$ if $k < 0$ and $\binosum{n}{k} = \binosum{n}{n}$ if $k > n$.
\subsection{Equiprobable BMS (EBMS)}
The following results pertain to the i.i.d. binary equiprobable source and hold for $0 \leq d < \frac 1 2$, $0 < \epsilon < 1$.

Particularizing \eqref{eq_jBMS} to the equiprobable case, one observes that for all binary $n-$strings $x^n$
\begin{equation}
 \jmath_{X^n}(x^n,d) = n\log 2 - nh(d) = nR(d)
\end{equation}
Then, Theorem \ref{thm:Cg} reduces  to \eqref{eq_2orderLossy0}. Theorem \ref{thm:C} leads to the following stronger converse result. 
\begin{thm}[Converse, EBMS]
\label{thm:CEBMS}
Any $(n, M, d, \epsilon)$ code must satisfy:
\begin{equation}
\label{eq_CEBMS}
\epsilon \geq
1 - M 2^{-n} \binosum{n}{\lfloor nd \rfloor}
\end{equation}
\end{thm}
\begin{proof}
Invoking Theorem \ref{thm:C} with the $n-$dimensional source distribution playing the role of $P_X$ therein, we have
\begin{align}
M &\geq \sup_Q \inf_{y^n \in \{0,1\}^n} \frac{\beta_{1 - \epsilon}(P_{X^n}, Q)}{\mathbb Q \left[ d(X^n, y^n) \leq d\right]}\\
&\geq \inf_{y^n \in \{0,1\}^n} \frac{\beta_{1 - \epsilon}(P_{X^n}, P_{X^n})}{\mathbb P \left[ d(X^n, y^n) \leq d\right]} \label{eq_P=Q BMS}\\
&= \frac{1 - \epsilon}{\mathbb P \left[ d(X^n, \mathbf 0) \leq d\right] }\\
&= \frac{1 - \epsilon}{2^{-n} \binosum{n}{\lfloor nd \rfloor}}
\end{align}
where \eqref{eq_P=Q BMS} is obtained by substitution $Q = P_{X}$.
\end{proof}

\begin{thm}[Exact performance of random coding, EBMS]
\label{thm:ExactExcessEBMS}
The minimal averaged probability that bit error rate exceeds $d$ achieved by random coding with $M$ codewords is
\begin{equation}
\label{eq_ExactExcessEBMS}
\min_{P_Y}\mathbb E \left[ \epsilon_d \left( Y_1, \ldots, Y_M\right) \right] = \left( 1 - 2^{-n}\binosum{n}{\lfloor nd \rfloor}\right)^M
\end{equation}
attained by $P_{Y}$ equiprobable on $\{0, 1\}^n$.  
\end{thm}
\begin{proof}
For all  $M \geq 1$, $(1 - z)^M$ is a convex function of $z$ on $0 \leq z < 1$, so the right side of \eqref{eq_ExactExcess} is lower bounded by Jensen's inequality:
\begin{equation}
 \E{1 - P_{Y^n}(B_d(X^n))}^M \geq \left(  1 - \E{P_{Y^n}(B_d(X^n))}\right)^M \label{eq_-ExactExcessEBMS}
\end{equation}
Equality in \eqref{eq_-ExactExcessEBMS} is attained by  $Y^n$ equiprobable on $\{0, 1\}^n$, because then
\begin{equation}
 P_{Y^n}(B_d(X^n)) = 2^{-n} \binosum{n}{\lfloor nd \rfloor} \text{ a.s.}
\end{equation}
\end{proof}
Theorem \ref{thm:ExactExcessEBMS} leads to an achievability bound since there must exist an $\left(M, d, \mathbb E \left[ \epsilon_d \left( Y_1, \ldots, Y_M\right) \right]\right)$ code.

\begin{figure}[htp]
\begin{center}
    \epsfig{file=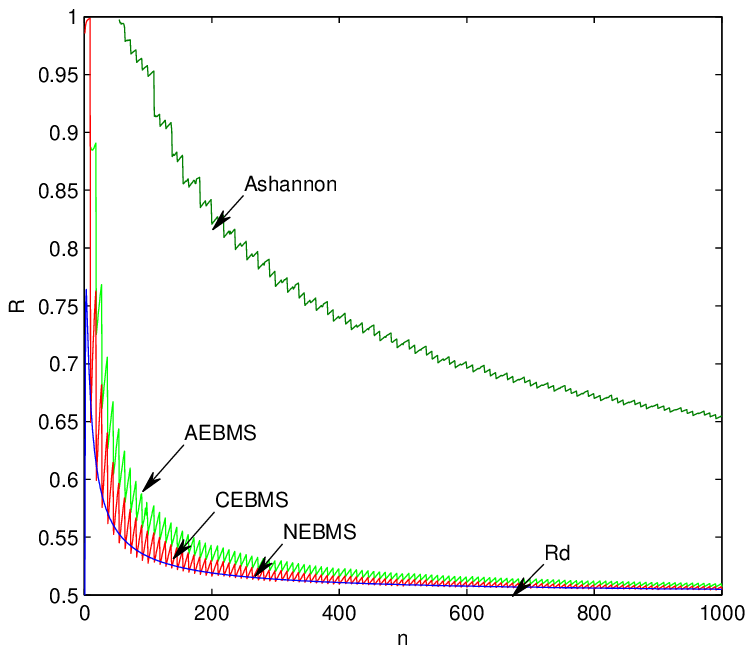,width=1\linewidth}
\end{center}
 \caption[]{Bounds to $R(n,d,\epsilon)$ and Gaussian approximation for EBMS, $d = 0.11$, $\epsilon = 10^{-2}$.} \label{fig:EBMS1e-2}
\end{figure}

\begin{cor}[Achievability, EBMS]
\label{thm:AEBMS}
There exists an $(n, M, d, \epsilon)$ code such that
\begin{align}
\label{eq_AEBMS}
\epsilon \leq  \left( 1 - 2^{-n}\binosum{n}{\lfloor nd \rfloor} \right)^M\end{align}
\end{cor}
As mentioned in Section \ref{sec:2order} after Theorem \ref{thm:2order}, the EBMS with bit error rate distortion has zero rate-dispersion function for all $d$. The asymptotic analysis of the bounds in \eqref{eq_AEBMS} and \eqref{eq_CEBMS} allows for the following more accurate characterization of $R(n,d,\epsilon)$.
\begin{thm}[Gaussian approximation, EBMS]
\label{thm:2orderEBMS}
The minimum achievable rate at blocklength $n$ satisfies
\begin{equation}
\label{eq_2orderEBMS}
R(n,d,\epsilon) = \log 2 - h(d) + \frac 1 2 \frac{\log n}{n} + O\left( \frac 1 {n}\right)
\end{equation} 
if $0 < d < \frac 1 2$, and 
\begin{equation}
 R(n, 0, \epsilon) = \log 2 - \frac 1 n \log \frac{1}{1 - \epsilon}+ o_n
\end{equation}
where $0 \leq o_n \leq \frac {2^{-n}} {(1 - \epsilon)n}$.
\end{thm}
\begin{proof}
Appendix \ref{appx:2orderEBMS}.
\end{proof}
A numerical comparison of the achievability bound \eqref{eq_Ashannon} evaluated with stationary memoryless $P_{Y^n|X^n}$, the new bounds in \eqref{eq_AEBMS} and \eqref{eq_CEBMS} as well as the approximation in \eqref{eq_2orderEBMS} neglecting the $\bigo{\frac 1 n}$ term is presented in Fig. \ref{fig:EBMS1e-2}. Note that Marton's converse (Theorem \ref{thm:Cmarton}) is not applicable to the EBMS because the region in \eqref{eq_martonRegion} is empty. The achievability bound in \eqref{eq_Ashannon}, while asymptotically optimal, is quite loose in the displayed region of blocklengths. The converse bound in \eqref{eq_CEBMS} and the achievability bound in \eqref{eq_AEBMS} tightly sandwich the finite blocklength fundamental limit. Furthermore, the approximation in \eqref{eq_2orderEBMS} is quite accurate, although somewhat optimistic, for all but very small blocklengths.

\subsection{Non-equiprobable BMS}
The results in this subsection focus on the i.i.d. binary memoryless source  with $\mathbb P \left[ \mathsf X = 1\right] = p < \frac 1 2$ and apply for $0 \leq d < p$, $0 < \epsilon < 1$. The following converse result is a simple calculation of the bound in Theorem \ref{thm:Cg} using \eqref{eq_jBMS}.
\begin{thm}[Converse, BMS]
\label{thm:CgBMS}
For any $(n,M, d, \epsilon)$ code, it holds that
\begin{align}
\epsilon &\geq \sup_{\gamma \geq 0} \left\{\Prob{  g_n(Z) \geq \log M +  \gamma  } - \exp\left(-\gamma \right) \right\}\label{eq_CgBMS}\\
g_n(Z) &= Z \log \frac{1}{p} + (n - Z) \log \frac1 {1 - p} - n h(d)
\end{align}
where $Z$ is binomial with success probability $p$ and $n$ degrees of freedom.
\end{thm}

An application of Theorem \ref{thm:C} to the specific case of non-equiprobable BMS yields the following converse bound:
\begin{thm}[Converse, BMS]
\label{thm:CBMS}
Any $(n, M,d,\epsilon)$ code must satisfy
\begin{align}
\label{eq_CBMS}
M \geq \frac{\binosum{n}{r^\star} +\alpha {n \choose r^\star + 1} }{\binosum{n}{\lfloor nd \rfloor}}
\end{align}
where we have denoted the integer
\begin{align}
\label{eq_rstarBMS}
r^\star &= {\max }\left\{ r: \ \sum_{k = 0}^r {n \choose k} p^k (1 - p)^{n-k} \leq 1 - \epsilon \right\}
\end{align}
 and $\alpha \in [0, 1)$ is the solution to
\begin{align}
&~\sum_{k = 0}^{r^\star} {n \choose k} p^k (1 - p)^{n-k} + \alpha  p^{r^\star + 1} (1 - p)^{n - r^\star - 1} {n \choose r^\star + 1} \notag\\
=&~ 1 - \epsilon
\end{align}
\end{thm}
\begin{proof}
In Theorem \ref{thm:C}, the $n-$dimensional source distribution $P_{X^n}$ plays the role of $P_X$, and we make the possibly suboptimal choice $Q = U$, the equiprobable distribution on $A = \{0, 1\}^n$. The optimal randomized test to decide between $P_{X^n}$ and $U$ is given by
\begin{equation}
P_{W|X^n}(1| x^n) = \begin{cases} 0, &| x^n | > r^\star + 1\\
						1, &| x^n | \leq r^\star\\
						\alpha, &| x^n | = r^\star + 1
						\end{cases}
\end{equation}
where $| x^n |$ denotes the Hamming weight of $x^n$, and $\alpha$ is such that $\sum_{ x^n \in A}P( x^n )P_{W|X}(1| x^n) = 1- \epsilon$, so
\begin{align}
&~\beta_{1 - \epsilon}(P_X, U) \notag\\
=&~ \min_{\substack{P_{W|X}: \\ \sum_{x^n \in A} P(x^n) P_{W|X}(1| x^n) \geq 1 - \epsilon}}  2^{-n} \sum_{x^n \in A} P_{W|X}(1| x^n) \notag\\
=&~2^{-n} \left[ \binosum{n}{r^\star} + \alpha  {n \choose r^\star + 1}\right]
\end{align}
The result is now immediate from \eqref{eq_C}.
\end{proof}

An application of Theorem \ref{thm:A} to the non-equiprobable BMS yields the following achievability bound:
\begin{thm}[Achievability, BMS]
\label{thm:ABMS}
There exists an $(n, M,d,\epsilon)$ code with
\begin{equation}
\label{eq_ABMS}
\epsilon \leq \sum_{k = 0}^n {  n \choose k } p^k (1 - p)^{n - k} \left[ 1 - \sum_{t = 0}^n  L_n(k, t) q^t (1 - q)^{n - t}\right]^M
\end{equation}
where 
\begin{equation}
q = \frac {p - d}{1 - 2d} \label{eq_q}
\end{equation}
and
\begin{equation}
 L_n(k, t) = 
{k \choose t_0} { n - k \choose t - t_0} \label{eq_ABMSL}
\end{equation}
with $t_0 = \left \lceil\frac{t+k-nd}{2}\right\rceil^+$ if $ t - nd \leq k \leq t + nd $, and $ L_n(k, t) = 0$ otherwise.
\end{thm}
\begin{proof}
We compute an upper bound to \eqref{eq_A} for the specific case of the BMS. Let $P_{Y^n} = P_{\mathsf Y} \times \ldots \times P_{\mathsf Y}$, where $P_{\mathsf Y} (1) = q$. Note that $P_{\mathsf Y}$ is the marginal of the joint distribution that achieves the rate-distortion function (e.g. \cite{berger1971rate}). The number of binary strings of Hamming weight $t$ that lie within Hamming distance $nd$ from a given string of Hamming weight $k$ is
\begin{equation}
 \sum_{i = t_0}^{k} {k \choose i} { n - k \choose t - i} \geq {k \choose t_0} { n - k \choose t - t_0}
\end{equation}
as long as $t - nd \leq k \leq t + nd$ and is $0$ otherwise. It follows that if $x^n$ has Hamming weight $k$,
\begin{equation}
 P_{Y^n}\left( B_d(x^n)\right) \geq \sum_{t = 0}^n L_n(k, t) q^t (1 - q)^{n - t}  \label{eq_-ABMS}
\end{equation}
Relaxing \eqref{eq_A} using \eqref{eq_-ABMS}, \eqref{eq_ABMS} follows.
\end{proof}

The following bound shows that good constant composition codes exist.

\begin{thm}[Achievability, BMS]
\label{thm:ABMS'}
There exists an $(n, M,d,\epsilon)$ constant composition code with
\begin{equation}
\label{eq_ABMS'}
\epsilon \leq \sum_{k = 0}^n {  n \choose k } p^k (1 - p)^{n - k} \left[ 1 -   {n \choose \lceil nq \rceil}^{-1}L_n(k, \lceil nq \rceil) \right]^M
\end{equation}
where $q$ and $L_n(\cdot, \cdot)$ are defined in \eqref{eq_q} and \eqref{eq_ABMSL} respectively.
\end{thm}
\begin{proof}
The proof is along the lines of the proof of Theorem \ref{thm:ABMS}, except that now we let $P_{Y^n}$ be equiprobable on the set of binary strings of Hamming weight $\lceil q n \rceil$.
\end{proof}
The following asymptotic analysis of $R(n,d,\epsilon)$ strengthens Theorem \ref{thm:2order}. 
\begin{thm}[Gaussian approximation, BMS]
\label{thm:2orderBMS}
The minimum achievable rate at blocklength $n$ satisfies \eqref{eq_2order}
where
\begin{align}
R(d) &= h(p) - h(d)\\
V(d) &= \Var{\imath_{\mathsf X}(\mathsf X)} = p (1-p) \log^2 \frac {1-p}{p} \label{eq_dispersionBMS}
\end{align}
and the remainder term in \eqref{eq_2order} satisfies
\begin{align}
\bigo{\frac 1 n } &\leq \theta \left( \frac {\log n}{n} \right) \label{eq_remCBMS}\\
 &\leq \frac 1 2 \frac {\log n}{n} + \frac {\log \log n} n + \bigo{\frac 1 n} \label{eq_remABMS}
\end{align}
if $0 < d < p$, and
\begin{equation}
\theta \left( \frac {\log n}{n} \right) = - \frac 1 2 \frac {\log n}{n} + \bigo{\frac 1 n }
\end{equation}
if $d = 0$.
\end{thm}

\begin{proof} The case $d = 0$ follows immediately from \eqref{eq_2orderLossless}. For $0 < d< p$,  the dispersion \eqref{eq_dispersionBMS} is easily obtained plugging $n = 1$ into \eqref{eq_jBMS}. The tightened upper bound for the remainder \eqref{eq_remABMS} follows via the asymptotic analysis of Theorem \ref{thm:ABMS'} shown in Appendix \ref{appx:2orderABMS}. We proceed to show the converse part, which yields a better $\frac {\log n} n$ term than Theorem  \ref{thm:2order}.

According to the definition of $r^\star$ in \eqref{eq_rstarBMS},
\begin{equation}
\Prob{\sum_{i = 1}^n X_i > r} \geq \epsilon \label{eq_-CBMS}
\end{equation}
for any $r \leq r^\star$, where $\{X_i\}$ are binary i.i.d. with $P_{X_i}(1) = p$. In particular, due to \eqref{eq_BerryEsseen}, \eqref{eq_-CBMS} holds for
\begin{align}
r &= np + \sqrt{n p (1-p)} \Qinv{\epsilon + \frac{B_n}{\sqrt{n}}}\\
&= np + \sqrt{n p (1-p)} \Qinv{\epsilon} + \bigo{1} \label{eq_-2orderCBMS}
\end{align}
where \eqref{eq_-2orderCBMS} follows because in the present case $B_n = 6 \frac{1 - 2p + 2p^2}{\sqrt{p(1-p)}}$, which does not depend on $n$. Using \eqref{eq_CBMS}, we have
\begin{equation}
M \geq \frac{\binosum{n}{\lfloor r \rfloor}}{\binosum{n}{\lfloor nd \rfloor}} \label{eq_-2orderCBMSa}
\end{equation}
Taking logarithms of both sides of \eqref{eq_-2orderCBMSa}, we have
\begin{align}
&~\log M \notag\\
\geq&~ \log {\binosum{n}{\lfloor r \rfloor}} - \log {\binosum{n}{\lfloor nd \rfloor}}\\
=&~ nh\left(p + \frac 1 {\sqrt n} \sqrt{p (1-p)} \Qinv{\epsilon}\right) - n h(d) + \bigo{1} \label{eq_-2orderCBMS-a}\\
=&~ nh(p) - n h(d) + \sqrt n \sqrt{p (1-p)} h^\prime(p) \Qinv{\epsilon}+ \bigo{1} \notag
\end{align}
where \eqref{eq_-2orderCBMS-a} is due to \eqref{eq_BinosumAsymptotics} in Appendix \ref{appx:2orderEBMS}. The desired bound \eqref{eq_remABMS} follows since $h^\prime(p) = \log \frac {1-p}{p}$.
\end{proof}

\begin{figure}[htp]
\begin{center}
    \epsfig{file=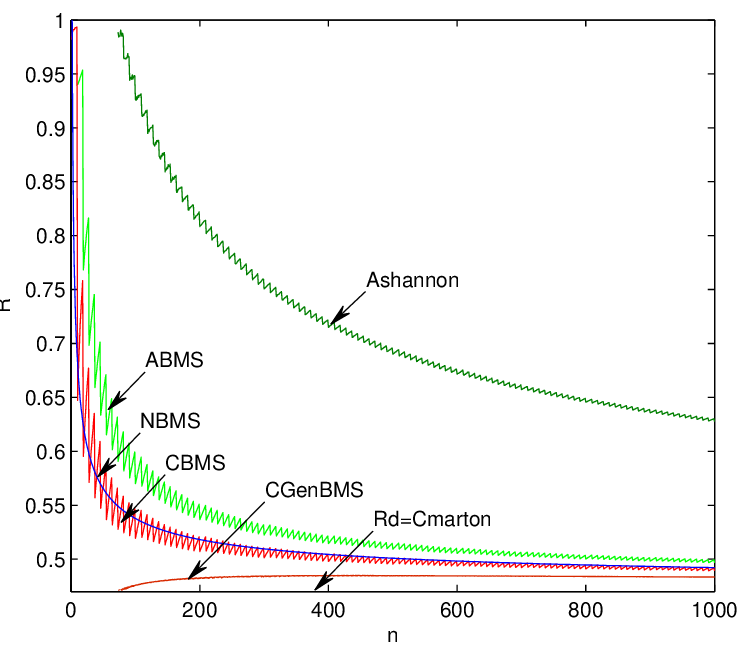,width=1\linewidth}
\end{center}
 \caption[]{Bounds to $R(n,d,\epsilon)$ and Gaussian approximation for BMS with $p = 2/5$, $d = 0.11$ , $\epsilon = 10^{-2}$.} \label{fig:BMS1e-2}
\end{figure}

\begin{figure}[htp]
\begin{center}
    \epsfig{file=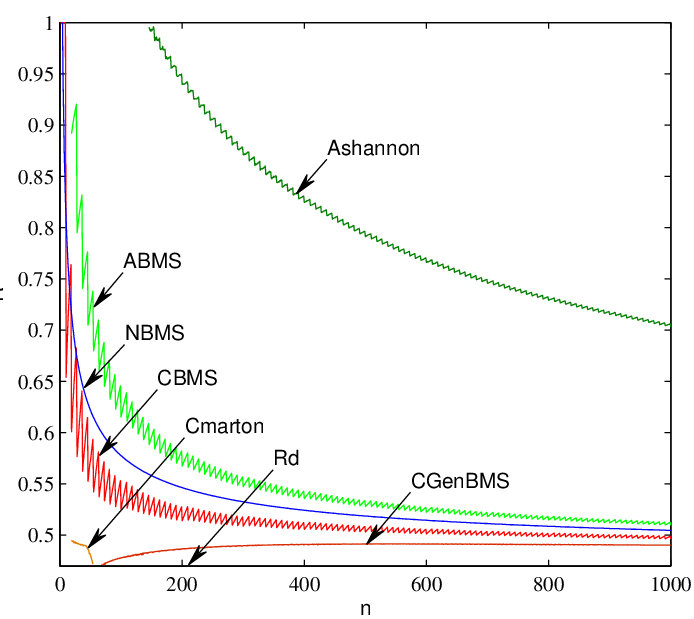,width=1\linewidth}
\end{center}
 \caption[]{Bounds to $R(n,d,\epsilon)$ and Gaussian approximation for BMS with $p = 2/5$, $d = 0.11$ , $\epsilon = 10^{-4}$.} \label{fig:BMS1e-4}
\end{figure}

Figures \ref{fig:BMS1e-2} and \ref{fig:BMS1e-4} present a numerical comparison of Shannon's achievability bound \eqref{eq_Ashannon}, the new bounds in \eqref{eq_ABMS}, \eqref{eq_CBMS} and \eqref{eq_CgBMS} as well as the Gaussian approximation in \eqref{eq_2order} in which we have neglected $\theta\left( \frac {\log n}{n}\right)$. The achievability bound \eqref{eq_Ashannon} is very loose and so is Marton's converse which is essentially indistinguishable from $R(d)$. The new finite blocklength bounds \eqref{eq_ABMS} and \eqref{eq_CBMS} are fairly tight unless the blocklength is very small. In Fig. \ref{fig:BMS1e-4} obtained with a more stringent $\epsilon$, the approximation of Theorem \ref{thm:2orderBMS} is essentially halfway between the converse and achievability bounds.

\section{Discrete memoryless source}
\label{sec:DMS}
This section particularizes the bounds in Section \ref{sec:new} to stationary memoryless sources with alphabet $\mathcal A$ and symbol error rate distortion measure, i.e. $d(x^n, y^n) = \frac 1 n \sum_{i= 1}^n 1\left\{x_i \neq y_i\right\}$. For convenience, we denote the number of strings within Hamming distance $k$ from a given string by
\begin{equation}
\label{eq_HammingSum}
\hammingsum{n}{k} = \sum_{j = 0}^{k} {n \choose j} (|\mathcal A|-1)^{j}
\end{equation}

\subsection{Equiprobable DMS (EDMS)}
In this subsection we fix $0 \leq d < 1 - \frac {1} {|\mathcal A|}$, $0 < \epsilon < 1$ and assume that all source letters are equiprobable, in which case the rate-distortion function is given by  \cite{erokhin1958epsilon}
\begin{equation}
  R(d) = \log |\mathcal A| - h(d) - d \log (|\mathcal A| - 1) \label{eq_RdEDMS}
\end{equation}

As in the equiprobable binary case, Theorem \ref{thm:Cg}  reduces to \eqref{eq_2orderLossy0}. A stronger converse bound is obtained using Theorem \ref{thm:C} in a manner analogous to that of Theorem \ref{thm:CEBMS}. 
\begin{thm}[Converse, EDMS]
\label{thm:CEDMS}
Any $(n, M, d, \epsilon)$ code must satisfy:
\begin{equation}
\label{eq_CEDMS}
\epsilon \geq
1 - M |\mathcal A|^{-n} \hammingsum{n}{\lfloor nd \rfloor}
\end{equation}
\end{thm}

The following result is a straightforward generalization of Theorem \ref{thm:ExactExcessEBMS} to the non-binary case. 
\begin{thm}[Exact performance of random coding, EDMS]
\label{thm:ExactExcessEDMS}
The minimal averaged probability that symbol error rate exceeds $d$ achieved by random coding with $M$ codewords is
\begin{equation}
\label{eq_ExactExcessEDMS}
\min_{P_{Y}}\mathbb E \left[ \epsilon_d \left( Y_1, \ldots, Y_M\right) \right] = \left( 1 - |\mathcal A|^{-n}\hammingsum{n}{\lfloor nd \rfloor}\right)^M
\end{equation}
attained by $P_{Y}$ equiprobable on $\mathcal A^n$.  
\end{thm}

Theorem \ref{thm:ExactExcessEDMS} leads to the following achievability bound. 
\begin{thm}[Achievability, EDMS]
\label{thm:AEDMS}
There exists an $(n, M, d, \epsilon)$ code such that
\begin{align}
\label{eq_AEDMS}
\epsilon \leq  \left( 1 - \hammingsum{n}{\lfloor nd \rfloor}  |\mathcal A|^{-n}\right)^M\end{align}
\end{thm}
The asymptotic analysis of the bounds in \eqref{eq_AEDMS} and \eqref{eq_CEDMS} yields the following tight approximation. 
\begin{thm}[Gaussian approximation, EDMS]
\label{thm:2orderEDMS}
The minimum achievable rate at blocklength $n$ satisfies
\begin{equation}
 R(n,d,\epsilon) =R(d) + \frac 1 2 \frac{\log n}{n} + O\left( \frac 1 {n}\right)
\end{equation}
if $0 < d < 1 - \frac {1} {|\mathcal A|}$, and
\begin{align}
R(n, 0, \epsilon) &= \log |\mathcal A| - \frac 1 n \log \frac{1}{1 - \epsilon}+ o_n
\end{align}
where $0 \leq o_n \leq \frac {|\mathcal A|^{-n}} {(1 - \epsilon)n}$.
\end{thm}
\begin{proof}
Appendix \ref{appx:2orderEDMS}.
\end{proof}

\subsection{Nonequiprobable DMS}
In this subsection we assume that the source is stationary memoryless on an alphabet of $m = |\mathcal A|$ letters labeled by $\mathcal A = \{1, \ldots, m\}$. We assume
\beq
P_{\mathsf X}(1) \geq P_{\mathsf X}(2) \geq \ldots \geq P_{\mathsf X}(m)
\eeq
and $0 \leq d < 1 - P_{\mathsf X}(1)$, $0 < \epsilon < 1$.

Recall that the rate-distortion function is achieved by \cite{erokhin1958epsilon}
\begin{align}
P_{\mathsf Y^\star}(b) &= \begin{cases}
\frac{P_{\mathsf X}(b) - \eta}{1 - d - \eta} & b \leq m_\eta\\
0 & \text{otherwise}
\end{cases} \label{eq_DMSPYstar}\\
P_{\mathsf X| \mathsf Y}^\star( a|b ) &= \begin{cases}
1 - d & a=b, ~ a \leq m_\eta\\
\eta & a \neq b, ~ a \leq m_\eta\\
P_{\mathsf X}(a) & a > m_\eta
\end{cases} \label{eq_DMSPX|Ystar}
\end{align}
where $0 \leq \eta \leq 1$ is the solution to
\begin{align}
d &= \sum_{a = m_\eta + 1}^{m} P_{\mathsf X}(a) + (m_\eta - 1)\eta \label{eq_DMSeta}\\
m_\eta &= \max\{a: \ P_{\mathsf X}(a) > \eta\} \label{eq_DMSmeta}
\end{align}
The rate-distortion function can be expressed as \cite{erokhin1958epsilon}
\begin{equation}
R(d) = \sum_{a = 1}^{m_\eta} P_{\mathsf X}(a) \imath_{\mathsf X}(a) + (1 - d)\log (1 - d) + (m_\eta - 1) \eta \log \eta \label{eq_RdDMS}\\
\end{equation}
Note that if $0 \leq d < (m-1) P_{\mathsf X}(m)$, then $m_\eta = m$, $\eta = \frac d {m -1}$, and \eqref{eq_DMSPYstar}, \eqref{eq_DMSPX|Ystar} and \eqref{eq_RdDMS} can be simplified. In particular, the rate-distortion function on that region is given by
\begin{equation}
  R(d) = H(\mathsf X) - h(d) - d \log (m - 1) \label{eq_RdDMSsimplified}
\end{equation}
The first result of this section is a particularization of the bound in Theorem \ref{thm:Cg} to the DMS case. 
\begin{thm}[Converse, DMS]
\label{thm:CgDMS}
For any $(n, M, d, \epsilon)$ code, it holds that
\begin{align}
\epsilon \geq \sup_{\gamma \geq 0}\left\{ \Prob{ \sum_{i = 1}^n \jmath_{\mathsf X}(X_i, d) \geq \log M +  \gamma  } - \exp\left\{-\gamma \right\} \right\} \label{eq_CgDMS}
\end{align}
where 
\begin{align}
  \jmath_{\mathsf X}(a, d) &=  (1 - d) \log (1 - d) + d \log \eta \notag\\
  &+ \min\left\{ \imath_{\mathsf X}(a), \log \frac 1 \eta \right\} \label{eq_jDMS}
\end{align}
and $\eta$ is defined in \eqref{eq_DMSeta}.
\end{thm}
\begin{proof}
Case $d = 0$ is obvious. For $0< d < 1- P_{\mathsf X}(1)$, differentiating \eqref{eq_RdDMS} with respect to $d$ yields
\beq
\lambda^\star = \log \frac {1-d}{\eta}
\eeq
Plugging \eqref{eq_DMSPX|Ystar} and $\lambda^\star$ into \eqref{eq_iddensity}, one obtains \eqref{eq_jDMS}.
\end{proof}

We adopt the notation of \cite{szpankowski2009minimum}:
\begin{itemize}
\item type of the string: $\mathbf k = (k_1, \ldots, k_{m}), \ k_1 + \ldots + k_{m} = n$
\item probability of a given string of type $\mathbf k$: $p^{\mathbf k} = P_{\mathsf X}(1)^{k_1} \ldots P_{\mathsf X}(m)^{k_{m}}$
\item type ordering: $\mathbf j \preceq \mathbf k$ if and only if $p^{\mathbf j} \geq p^{\mathbf k}$
\item type $\mathbf 1$ denotes $[n, 0, \ldots, 0]$
\item previous and next types: $\mathbf j - 1$ and $\mathbf j + 1$, respectively
\item multinomial coefficient: $\displaystyle{{n \choose \mathbf k} = \frac{n!}{k_1! \ldots k_m!}}$
\end{itemize}

The next converse result is a particularization of Theorem \ref{thm:C}. 
\begin{thm}[Converse, DMS]
\label{thm:CDMS}
Any $(n, M,d,\epsilon)$ code must satisfy
\begin{align}
\label{eq_CDMS}
M \geq \frac{\displaystyle{\sum_{\mathbf i =\mathbf 1}^{\mathbf k^\star} {n \choose \mathbf i} +\alpha {n \choose \mathbf k^\star + 1} }}{ \hammingsum{n}{\lfloor nd \rfloor}}
\end{align}
where
\begin{align}
\mathbf k^\star &= {\max }\left\{ \mathbf k: \ \sum_{\mathbf i = \mathbf 1}^ {\mathbf k} {n \choose \mathbf i} p^{\mathbf i} \leq 1 - \epsilon \right\}
\label{eq_kstarDMS}
\end{align}
 and $\alpha \in [0, 1)$ is the solution to
\begin{align}
\sum_{\mathbf i = \mathbf 1}^{\mathbf k^\star} {n \choose \mathbf i} p^{\mathbf i} + \alpha  {n \choose \mathbf k^\star + 1} p^{\mathbf k^\star + 1} = 1 - \epsilon
\end{align}
\end{thm}

\begin{proof}
Consider a binary hypothesis test between the $n-$dimensional source distribution $P_{X^n}$ and $U$, the equiprobable distribution on $\mathcal A^n$. From Theorem \ref{thm:C},
\begin{equation}
\label{eq_htDMS}
M \geq |\mathcal A|^n \frac{\beta_{1 - \epsilon}(P_{X^n}, U)}{\hammingsum{n}{\lfloor nd \rfloor}}
\end{equation}
 The calculation of $\beta_{1 - \epsilon}(P_{X^n} U)$ is analogous to the BMS case.
\end{proof}

 The following result guarantees existence of a good code with all codewords of type $\mathbf t^\star = ( [nP_{\mathsf Y}^\star(1)], \ldots, [nP_{\mathsf Y}^\star(m_\eta)], 0, \ldots, 0 )$ where $[\cdot]$ denotes rounding off to a neighboring integer so that $\sum_{b = 1}^{m_\eta} [nP_{\mathsf Y}^\star(b) ] = n$ holds. 

\begin{thm}[Achievability, DMS]
\label{thm:ADMS}
There exists an $(n, M,d,\epsilon)$ fixed composition code with codewords of type $\mathbf t^\star$ and
\begin{align}
\epsilon &\leq \sum_{\mathbf k} {  n \choose \mathbf k } p^{\mathbf k} \left( 1 - {n \choose \mathbf t^\star}^{-1} L_n(\mathbf k, \mathbf t^\star)\right)^M \label{eq_ADMS}\\
 L_n(\mathbf k, \mathbf t^\star) &= \prod_{a = 1}^{m} {k_a \choose \mathbf t_{a}} \label{eq_rhoDMS}
\end{align}
 where $\mathbf k = [k_1, \ldots, k_m]$ ranges over all $n$-types, and $k_a$-types $\mathbf t_{a} = (t_{a, 1}, \ldots, t_{a, m_\eta})$ are given by
 \begin{equation}
t_{a, b} =  \left[ P_{\mathsf X|\mathsf Y}^\star(a | b) t_b^\star+ \delta(a, b)n\right]
  \label{eq_tik} 
 \end{equation}
 where
\begin{align}
\delta(a,b) &= \frac{\Delta_a}{m_\eta} +
\begin{cases}
 \frac {1}{m_\eta^2} \sum_{i = m_\eta + 1}^m \Delta_i & a = b, a \leq m_\eta\\
  \frac {-1}{m_\eta^2(m_\eta - 1)} \sum_{i = m_\eta + 1}^m \Delta_i & a \neq b, a \leq m_\eta\\
  0 & a >m_\eta
 \end{cases}\\
n \Delta_a &= k_a - n P_{\mathsf X}(a), ~ a = 1, \ldots, m
\end{align}

 In \eqref{eq_tik}, $a = 1, \ldots, m$, $b = 1, \ldots, m_\eta$ and $[\cdot]$ denotes rounding off to a neighboring nonnegative integer so that
\begin{align}
\sum_{b = 1}^{m_\eta}t_{b, b} &\geq n(1 - d) \label{eq_ADMStik}\\
\sum_{b = 1}^{m_\eta} t_{a, b}  &= k_a \label{eq_ADMStik-a}\\
\sum_{a = 1}^m t_{a, b}  &= t_b^\star \label{eq_ADMStik-b}
\end{align}
and among all possible choices the one that results in the largest value for \eqref{eq_rhoDMS} is adopted. If no such choice exists, $L_n(\mathbf k, \mathbf t^\star) = 0$. 
\end{thm}

\begin{proof}
We compute an upper bound to \eqref{eq_A} for the specific case of the DMS. Let $P_{Y^n}$ be equiprobable on the set of $m-$ary strings of type  $\mathbf t^\star$. To compute the number of strings of type $\mathbf t^\star$ that are within distortion $d$ from a given string $x^n$ of type $\mathbf k$, observe that by fixing $x^n$ we have divided an $n$-string into $m$ bins, the $a$-th bin corresponding to the letter $a$ and having size $k_a$. If $t_{a, b}$ is the number of the letters $b$ in a sequence $y^n$ of type $\mathbf t^\star$ that fall into $a$-th bin, the strings $x^n$ and $y^n$ are within Hamming distance $nd$ from each other as long as \eqref{eq_ADMStik} is satisfied. Therefore, the number of strings of type $\mathbf t^\star$ that are within Hamming distance $nd$ from a given string of type $\mathbf k$ is bounded by
\begin{equation}
\sum \prod_{a = 1}^{m} {k_a \choose \mathbf t_{a}} \label{eq_-ADMSa}\\
 \geq L_n(\mathbf k, \mathbf t^\star)
\end{equation}
where the summation in the left side is over all collections of $k_a$-types $\mathbf t_{a} = (t_{a, 1}, \ldots, t_{a, m_\eta})$, $a = 1, \ldots m$ that satisfy \eqref{eq_ADMStik}-\eqref{eq_ADMStik-b}, and inequality \eqref{eq_-ADMSa} is obtained by lower bounding the sum by the term with $t_{a, b}$ given by \eqref{eq_tik}. It follows that if $x^n$ has type $\mathbf k$,
\begin{equation}
 P_{Y^n}\left( B_d(x^n)\right) \geq {n \choose \mathbf t^\star}^{-1} L_n(\mathbf k, \mathbf t^\star)  \label{eq_-ADMSb}
\end{equation}
Relaxing \eqref{eq_A} using \eqref{eq_-ADMSb}, \eqref{eq_ADMS} follows.
\end{proof}

\begin{remark}
As $n$ increases, the bound in \eqref{eq_-ADMSa} becomes increasingly tight. This is best understood by checking that all strings with $k_{a, b}$ given by \eqref{eq_tik} lie at a Hamming distance of approximately $nd$ from some fixed string of type $\mathbf k$, and recalling \cite{zhang1997redundancy} that most of the volume of an $n-$dimensional ball is concentrated near its surface (a  similar phenomenon occurs in Euclidean spaces as well), so that the largest contribution to the sum on the left side of  \eqref{eq_-ADMSa} comes from the strings satisfying  \eqref{eq_tik}. 
\end{remark}

\begin{figure}
\includegraphics[width=.5\textwidth]{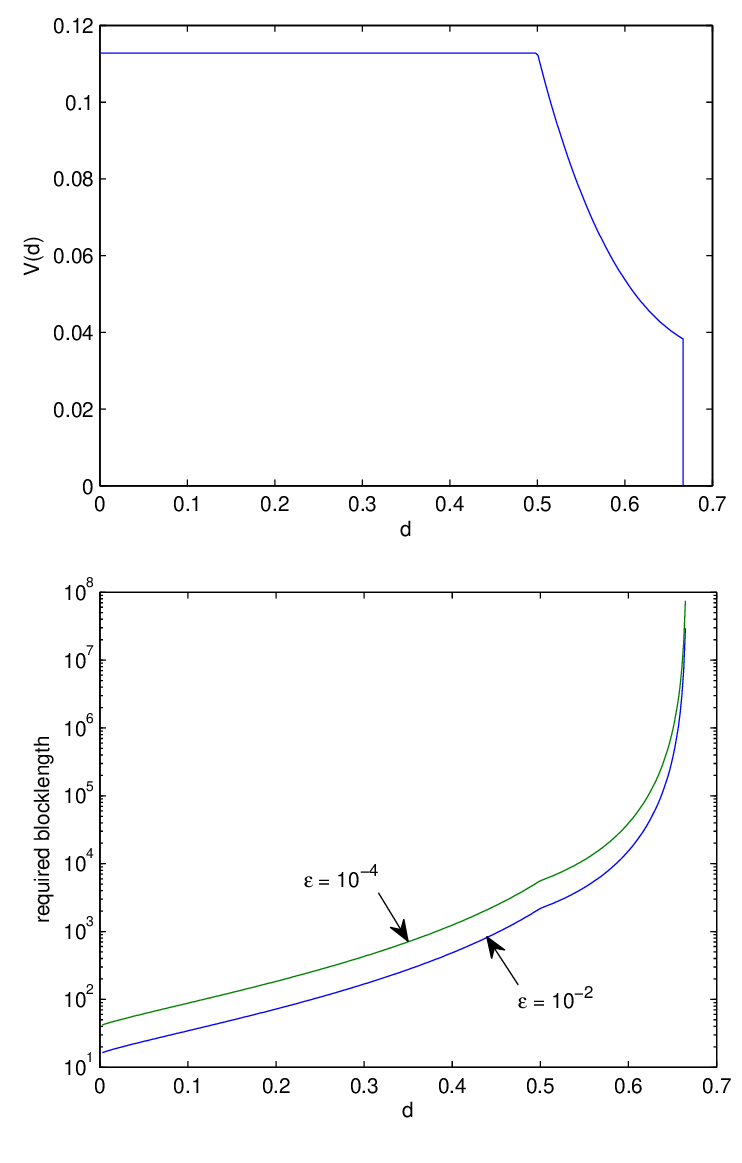}
\caption{Rate-dispersion function (bits) and the blocklength \eqref{eq_nrequired} required to sustain $R = 1.1 R(d)$ provided that excess-distortion probability is bounded by $\epsilon$ for DMS with $P_{\mathsf X} = \left[\frac 1 3, \frac 1 4, \frac 1 4, \frac 1 6\right]$ }
\label{fig:dispersionDMS}
\end{figure}

The following second-order analysis makes use of Theorem \ref{thm:2order} and, to strengthen the bounds for the remainder term, of Theorems \ref{thm:CDMS} and \ref{thm:ADMS}. 
\begin{thm}[Gaussian approximation, DMS]
\label{thm:2orderDMS}
The minimum achievable rate at blocklength $n$, $R(n, d, \epsilon)$, satisfies \eqref{eq_2order} where $R(d)$ is given by \eqref{eq_RdDMS}, and $V(d)$ can be characterized parametrically:
\begin{equation}
V(d) = \Var{ \min\left\{ \imath_{\mathsf X}(\mathsf X), \log \frac 1 \eta \right\} } \label{eq_dispersionDMS}
\end{equation}
where $\eta$ depends on $d$ through \eqref{eq_DMSeta}, \eqref{eq_DMSmeta}. Moreover, \eqref{eq_Aremainder} can be replaced by:
\begin{equation}
 \theta \left(\frac {\log n}{n} \right) \leq \frac {(m - 1)(m_\eta - 1)} 2 \frac{\log n}{n} + \frac{\log \log n}{n} + \bigo{\frac 1 n} \label{eq_ADMSremainder}
\end{equation}
If $0 \leq d < (m-1) P_{\mathsf X}(m)$, \eqref{eq_dispersionDMS} reduces to
\begin{equation}
V(d) = \Var{\imath_{\mathsf X}(\mathsf X)} \label{eq_dispersionDMSsimplified}
\end{equation}
and if $d > 0$, \eqref{eq_Cremainder} can be strengthened to 
\begin{equation}
O\left(\frac {1} n\right) \leq \theta \left(\frac {\log n}{n} \right)  \label{eq_CDMSremainder}
\end{equation}
while if $d = 0$,
\begin{equation}
 \theta \left(\frac {\log n}{n} \right)  = - \frac 1 2 \frac{\log n}{n} + \bigo{\frac{1}{n}}
\end{equation}
\end{thm}
\begin{proof} 
Using the expression for $d-$tilted information \eqref{eq_jDMS}, we observe that $\Var{ \jmath_{\mathsf X}(\mathsf X, d)} = \Var{ \min\left\{ \imath_{\mathsf X}(\mathsf X), \log \frac 1 \eta \right\} }$, and \eqref{eq_dispersionDMS} follows. The case $d = 0$ is verified using \eqref{eq_2orderLossless}.  Theorem \ref{thm:ADMS} leads to \eqref{eq_ADMSremainder}, as we show in Appendix \ref{appx:2orderADMS}. 

When $0 < d < (m-1) P_{\mathsf X}(m)$, not only \eqref{eq_RdDMS} and \eqref{eq_dispersionDMS} reduce to \eqref{eq_RdDMSsimplified} and \eqref{eq_dispersionDMSsimplified} respectively, but a tighter converse for the $\frac{\log n}{n}$ term \eqref{eq_CDMSremainder} can be shown. Recall the asymptotics of $\hammingsum{n}{\lfloor nd \rfloor}$ in \eqref{eq_HammingSumAsymptotics} (Appendix \ref{appx:2orderEDMS}). Furthermore, it can be shown \cite{szpankowski2009minimum} that
\begin{equation}
\sum_{\mathbf i = 1} ^{\mathbf k}{n \choose \mathbf i} = \frac{C}{\sqrt n}\exp\left\{nH\left(\frac{\mathbf k}{n} \right)\right\}\label{eq_multisumszpankowski}
\end{equation}
 for some constant $C$. Armed with \eqref{eq_multisumszpankowski} and \eqref{eq_HammingSumAsymptotics}, we are ready to proceed to the second-order analysis of \eqref{eq_CDMS}. From the definition of $\mathbf k^\star$ in \eqref{eq_kstarDMS},
\begin{equation}
\Prob{\frac 1 n\sum_{i = 1}^n \imath_{\mathsf X}(X_i) > H(\mathsf X) + \sum_{a = 1}^m \Delta_a \imath_{\mathsf X}(a) } \geq \epsilon
\label{eq_-2orderCDMS}
\end{equation}
for any $\mathbf \Delta$ with $\sum_{a = 1}^m \Delta_a = 0$ satisfying $n (\mathbf p + \mathbf \Delta) \preceq \mathbf k^\star$, where $\mathbf p = [P_{\mathsf X}(1), \ldots, P_{\mathsf X}(m)]$ (we slightly abused notation here as $n (\mathbf p + \mathbf \Delta) $ is not always precisely an $n$-type; naturally, the definition of the type ordering $\preceq$ extends to such cases).  Noting that $\E{ \imath_{\mathsf X}(X_i)} = H(\mathsf X)$ and $\Var{ \imath_{\mathsf X}(X_i) }= \Var{\imath_{\mathsf X}(\mathsf X)}$, we conclude from the Berry-Esseen CLT \eqref{eq_BerryEsseen} that \eqref{eq_-2orderCDMS} holds for
\begin{equation}
\sum_{a = 1}^m \Delta_a \imath_{\mathsf X}(a) = \sqrt{\frac {\Var{\imath_{\mathsf X}(\mathsf X)}} n} \Qinv{\epsilon - \frac {B_n}{\sqrt{n}}} \label{eq_-2orderCDMSa}
\end{equation}
where $B_n$ is given by \eqref{eq_BerryEsseenBn}.
Taking logarithms of both sides of \eqref{eq_CDMS}, we have
\begin{align}
&~\log M \notag\\
\geq&~ \log \left[ \sum_{\mathbf i = \mathbf 1}^{\mathbf k^\star}{n \choose \mathbf i} + \alpha {n \choose \mathbf k^\star}\right] - \log \hammingsum{n}{\lfloor nd \rfloor}\\
\geq&~ \log \sum_{\mathbf i = \mathsf 1}^{\mathbf k^\star} {n \choose \mathbf i} - \log \hammingsum{n}{\lfloor nd \rfloor}\\
\geq&~ nH(\mathbf p + \mathbf \Delta ) - nh(d) - nd \log (m-1) + O(1) \label{eq_-2orderCDMS-a}\\
=&~ nH(\mathbf p ) + n\sum_{a = 1}^m \Delta_a \imath_{\mathsf X}(a) - nh(d) - nd \log (m-1) + O(1)
\label{eq_-2orderCDMS-b}
\end{align}
where we used \eqref{eq_HammingSumAsymptotics} and \eqref{eq_multisumszpankowski} to obtain \eqref{eq_-2orderCDMS-a}, and \eqref{eq_-2orderCDMS-b} is obtained by applying a Taylor series expansion to $H(\mathbf p + \mathbf \Delta)$. The desired result in \eqref{eq_CDMSremainder} follows by substituting \eqref{eq_-2orderCDMSa} in \eqref{eq_-2orderCDMS-b}, applying a Taylor series expansion to $\Qinv{\epsilon - \frac {B_n}{\sqrt{n}}}$ in the vicinity of $\epsilon$ and noting that $B_n$ is a finite constant.
\end{proof}
The rate-dispersion function and the blocklength \eqref{eq_nrequired} required to sustain $R = 1.1R(d)$ are plotted in Fig. \ref{fig:dispersionDMS} for a quaternary source with distribution $[\frac 1 3, \frac 1 4, \frac 1 4, \frac 1 6]$. Note that according to \eqref{eq_nrequired}, the blocklength required to approach $1.1R(d)$ with a given probability of excess distortion grows rapidly as $d \to d_{\max}$.

\section{Erased binary memoryless source}
\label{sec:BES}
Let $S^n \in \left\{0, 1\right\}^n$ be the output of the binary equiprobable source, $X^n$ be the output of the binary erasure channel with erasure rate $\delta$ driven by $S^n$. The compressor only observes $X^n$, and the goal is to minimize the bit error rate with respect to $S^n$. For $d = \frac \delta 2$, codes with rate approaching the rate-distortion function were constructed in \cite{martinian2004iterative}. For $\frac \delta 2 \leq d \leq \frac 1 2$, the rate-distortion function is given by
\begin{equation}
R(d) = (1-\delta)\left(\log 2 - h\left( \frac{d - \frac \delta 2}{1 - \delta}\right)\right)
\end{equation}
Throughout the section, we assume $\frac \delta 2 < d < 1 - \frac \delta 2$ and $0 < \epsilon < 1$. 

\begin{thm}[Converse, BES]
\label{thm:CBES}
Any $(n, M,d,\epsilon)$ code must satisfy
\begin{align}
\epsilon \geq& \sum_{k = 0}^{n} {n \choose k} \delta^{k}(1-\delta)^{n-k}\notag\\ 
\cdot& \sum_{j = 0}^{k} 2^{-k}{k \choose j} \left[ 1- M 2^{-(n-k)}\binosum{n-k}{\lfloor n d - j\rfloor }\right]^+
\label{eq_CBES}
\end{align}
\end{thm}
\begin{proof}
 Fix an $(n, M, d, \epsilon)$ code $(P_{Z^n|X^n}, P_{Y^n|Z^n})$. Even if the decompressor knows erasure locations, the probability that $k$ erased bits are at Hamming distance $\ell$ from their representation is
 \begin{equation}
\mathbb P \left[k ~ d(S^k, Y^k) = \ell \mid X^k = (?\ldots ?)\right] =  2^{-k} {k \choose\ell } \label{eq_erased}
\end{equation}
because given $X^k = (?\ldots ?)$, $S_i$'s are i.i.d. binary independent of $Y^k$. 

The probability that $n-k$ nonerased bits lie within Hamming distance $\ell$ from their representation can be upper bounded using Theorem \ref{thm:CEBMS}:
\begin{align}
&~\Prob { (n-k)d(S^{n-k}, Y^{n-k}) \leq \ell \mid X^{n-k} = S^{n-k}}\notag\\
\leq&~ M2^{-n+k} \binosum{n-k}{\ell}
\end{align}
Since the errors in the erased symbols are independent of the errors in the nonerased ones,
\begin{align}
&~\mathbb P \left[ d(S^n,Y^n )\leq d \right] \notag\\
=&~
\sum_{k = 0}^{n} \mathbb P[ k \text{ erasures in } S^n] \notag \\
\cdot&~ \sum_{j = 0}^{k} \mathbb P \left[ k \ d(S^{k}, Y^{k}) = j | X^k = ?\ldots ?\right] \notag \\
\cdot&~ \mathbb P \left[ (n-k) d(S^{n-k}, Y^{n-k}) \leq nd - j | X^{n-k} = S^{n-k} \right] \notag\\
\leq &~ \sum_{k = 0}^n {n \choose k} \delta^{k}(1-\delta)^{n-k} \notag \\
\cdot&~\sum_{j = 0}^{k}2^{-k} {k \choose j} \min\left\{ 1, \ M 2^{-(n-k)} \binosum{n-k}{\lfloor nd - j\rfloor }\right\} 
\end{align}
\end{proof}

\begin{figure}
\includegraphics[width=.5\textwidth]{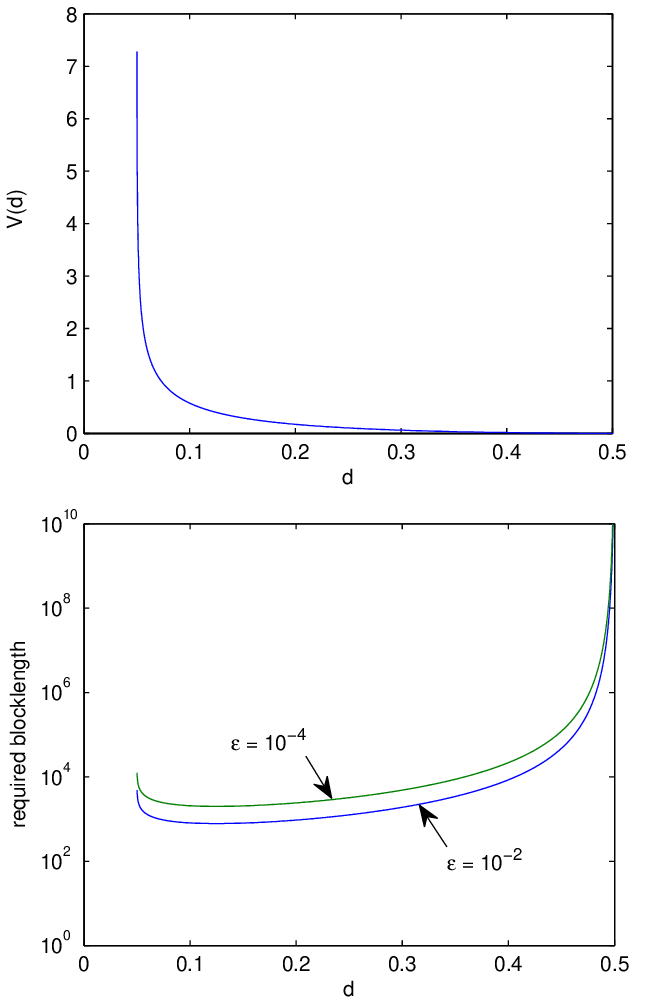}
\caption{Rate-dispersion function (bits) and the blocklength \eqref{eq_nrequired} required to sustain $R = 1.1 R(d)$ provided that excess-distortion probability is bounded by $\epsilon$ for BES with erasure rate $\delta = 0.1$. }
\label{fig:dispersionBES}
\end{figure}

\begin{thm}[Achievability, BES]
\label{thm:ABES}
There exists an $(n, M,d,\epsilon)$ code such that
\begin{align}
\epsilon \leq& \sum_{k = 0}^n {n \choose k} \delta^{k}(1-\delta)^{n-k} \notag\\
\cdot& \sum_{j = 0}^{k}2^{-k} {k \choose j}\left( 1 - 2^{-(n-k)}\binosum{n-k}{\lfloor nd - j \rfloor} \right)^M
\label{eq_ABES}
\end{align}
\end{thm}
\begin{proof}
Consider the ensemble of codes with $M$ codewords drawn i.i.d. from the equiprobable distribution on $\{0,1\}^n$. As discussed in the proof of Theorem \ref{thm:CBES}, the distortion in the erased symbols does not depend on the codebook and is given by \eqref{eq_erased}. The probability that the Hamming distance between the nonerased symbols and their representation exceeds $
\ell$, averaged over the code ensemble is found as in Theorem \ref{thm:AEBMS}:
\begin{align}
&~\mathbb P \left[ (n-k)d(S^{n-k}, \mathsf C(\mathsf f(X^{n-k}))) > \ell | S^{n-k} = X^{n-k}\right] \notag\\
=&~ \left( 1 - 2^{-(n-k)}\binosum{n-k}{\ell}\right)^M
\end{align}
where $\mathsf C(m)$, $m = 1, \ldots, M$ are i.i.d on $\{0,1\}^{n-k}$.
Averaging over the erasure channel, we have
\begin{align}
&~\mathbb P \left[ d(S^n, \mathsf C(\mathsf f(X^n)))) > d \right]\notag\\
=&~\sum_{k = 0}^n \mathbb P[ k \text{ erasures in } S^n] \notag\\
\cdot &~\sum_{j = 0}^{k} \mathbb P \left[ k\ d(S^{k}, \mathsf C(\mathsf f(X^{k}))) = j| X^{k} = ?\ldots ?\right] \notag \\
\cdot &~\mathbb P \left[ (n-k) d(S^{n-k}, \mathsf C(\mathsf f(X^{n-k}))) > nd - j | X^{n-k} = S^{n-k} \right]  \notag\\
= &~ \sum_{k = 0}^n {n \choose k} \delta^{k}(1-\delta)^{n-k} \notag \\
\cdot &~\sum_{j = 0}^{k}2^{-k} {k \choose j} \left( 1 - 2^{-(n-k)}\binosum{n-k}{\lfloor nd - j \rfloor}  \right)^M
\end{align}
Since there must exist at least one code whose excess-distortion probability is no larger than the average over the ensemble, there exists a code satisfying \eqref{eq_ABES}.
\end{proof}
\begin{thm}[Gaussian approximation, BES]
\label{thm:2orderBES}
The minimum achievable rate at blocklength $n$ satisfies \eqref{eq_2order} where
\begin{align}
\label{eq_dispersionBES}
V(d) &= \delta(1-\delta) \log^2 \cosh \left( Ê\frac{\lambda^\star}{2 \log e} \right)  + \frac \delta 4 \lambda^{\star 2} \\
\lambda^\star &= - R^\prime(d) = \log \frac{1 - \frac \delta 2 - d}{d - \frac \delta 2 }
\end{align}
and the remainder term in \eqref{eq_2order} satisfies
\begin{equation}
\bigo{\frac 1 n} \leq \theta\left(\frac {\log n} n\right) \leq \frac 1 2 \frac{\log n}{n} + \frac{\log \log n}{n} + \bigo{\frac 1 n} \label{eq_2orderBESremainder}
\end{equation}
\end{thm}
\begin{proof}
Appendix \ref{appx:2orderBES}.
\end{proof}
\begin{remark}
It is satisfying to observe that even though Theorem \ref{thm:2order} is not directly applicable, still $V(d) = \Var{\jmath_{\mathsf S,\mathsf X}( \mathsf S, \mathsf X, d)}$, where $\jmath_{\mathsf S, \mathsf X}(\mathsf s, \mathsf x, d)$ is spelled out in \eqref{eq_jdensityBES} below. Indeed,
since the rate-distortion function is achieved by $P_{\mathsf Y}^\star(0) = P_{\mathsf Y}^\star(1) = \frac 1 2$ and 
\begin{equation}
 P_{\mathsf X|\mathsf Y}^\star(a|b) = 
\begin{cases}
  1 - d - \frac \delta 2 & b = a\\
  d - \frac \delta 2 & b \neq a \neq ?\\
  \delta	& a = ?
\end{cases}
\end{equation}
where $a \in \{0, 1, ?\}$ and $b \in \{0, 1\}$, we may adapt \eqref{eq_iddensity} to obtain
\begin{align}
 &~\jmath_{\mathsf S,\mathsf X}( \mathsf S, \mathsf X, d) \notag\\
 =&~ \imath_{\mathsf X; \mathsf Y^\star}(\mathsf X; 0) + \lambda^\star d(\mathsf S, 0) - \lambda^\star d \label{eq_jdensityBES}\\
 =&~ - \lambda^\star d + 
\begin{cases}
 \log \frac{2}{1 + \exp( -\lambda^\star)} & \text{w.p. } 1 - \delta \\
\lambda^\star & \text{w.p. } \frac \delta 2\\
0  & \text{w.p. } \frac \delta 2 \label{eq_jBES}
\end{cases}
\end{align}
The variance of \eqref{eq_jBES} is \eqref{eq_dispersionBES}. 
\end{remark}
The rate-dispersion function and blocklength required to sustain a given excess distortion are plotted in Fig. \ref{fig:dispersionBES}. Note that as $d$ approaches $\frac \delta 2$, the rate-dispersion function grows without limit. This should be expected, because for $d = \frac \delta 2$, a code that reconstructs a sequence with vanishingly small excess-distortion probability does not exist, as about half of the erased bits will always be reconstructed incorrectly, regardless of the blocklength.

\begin{figure}[htp]
\begin{center}
    \epsfig{file=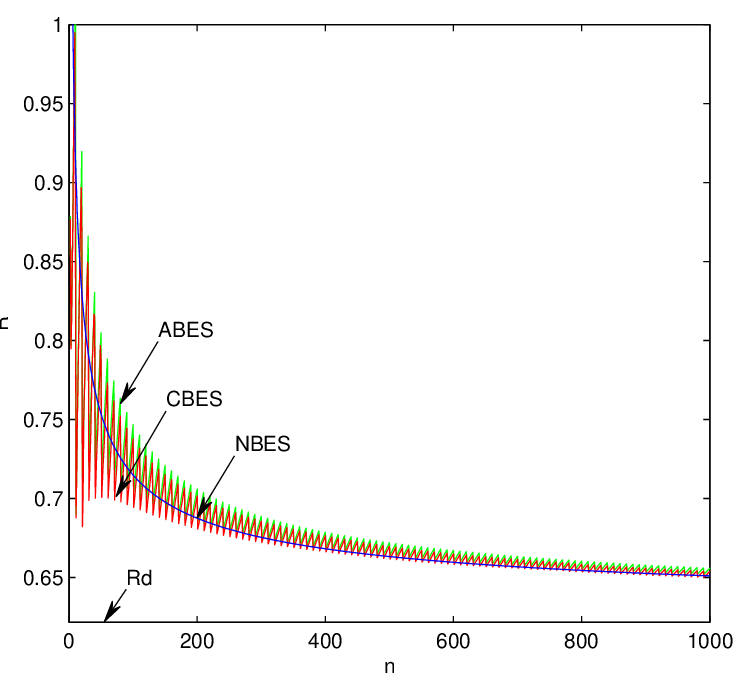,width=1\linewidth}
\end{center}
 \caption[]{Bounds to $R(n,d,\epsilon)$ and Gaussian approximation for BES with $\delta = 0.1$, $d = 0.1$, $\epsilon = 0.1$} \label{fig:BES}
\end{figure}

The bounds in Theorems \ref{thm:CBES} and \ref{thm:ABES} as well as the approximation in Theorem \ref{thm:2orderBES} are plotted in Fig. \ref{fig:BES}. The achievability and converse bounds are extremely tight. At blocklength 1000, the penalty over the rate-distortion function is $9 \%$.

\section{Gaussian memoryless source}
\label{sec:GMS}
\begin{table*}[!b]
\newcounter{mytempeqncnt}
\normalsize
\setcounter{mytempeqncnt}{\value{equation}}
\setcounter{equation}{231}
\vspace*{4pt}
\hrulefill
\begin{equation}
\mathbb M (r)=\\
\begin{cases}
e \left( n \log_e n + n \log_e \log_e n + 5n\right) r^n &  \ r \geq n\\
n \left( n \log_e n + n \log_e \log_e n + 5n\right) r^n &  \ \frac n {\log_e n} \leq r < n\\
\frac{7^{4 \log_e 7/7}}{4}\sqrt{2\pi}\frac{n\sqrt{n} \left[ (n-1)\log_e rn + (n-1) \log_e \log_e n + \frac 1 2 \log_e n + \log_e \frac{\pi \sqrt{2n}}{\sqrt{\pi n} - 2}\right]}{r \left(1 - \frac 2 {\log_e n}\right)\left(1 - \frac 2 {\sqrt{\pi n}}\right) \log_e^2 n}r^n & 2 < r < \frac{n}{\log_e n}\\
\sqrt{2\pi}\frac{\sqrt{n} \left[ (n-1)\log_e rn + (n-1) \log_e \log_e n + \frac 1 2 \log_e n + \log_e \frac{\pi \sqrt{2n}}{\sqrt{\pi n} - 2}\right]}{r \left(1 - \frac 2 {\log_e n}\right)\left(1 - \frac 2 {\sqrt{\pi n}}\right) }r^n & 1 < r \leq 2
\end{cases}
\label{eq_rogers}
\end{equation}
\setcounter{equation}{\value{mytempeqncnt}}
\end{table*}

This section applies Theorems \ref{thm:Cg}, \ref{thm:C} and \ref{thm:A} to the i.i.d. Gaussian source with mean-square error distortion, $d(x^n, y^n) = \frac 1 n \sum_{i = 1}^n (x_i- y_i)^2$, and refines the second-order analysis in Theorem \ref{thm:2order}. Throughout the section, it is assumed that $X_i \sim \mathcal N(0, \sigma^2)$, $0 < d < \sigma^2$ and $0 < \epsilon < 1$.

The particularization of Theorem \ref{thm:Cg} to the GMS using \eqref{eq_jGMS} yields the following result.
\begin{thm}[Converse, GMS]
\label{thm:CgGMS}
Any $(n, M,d,\epsilon)$ code must satisfy
\begin{align}
\epsilon &\geq \sup_{\gamma \geq  0}\left\{  \Prob{ g_n(Z) \geq \log M + \gamma } - \exp( -\gamma)\right\} \label{eq_CgGMS}\\
g_n(Z) &= \frac n 2 \log \frac{ \sigma^2}{d} + \frac{Z - n} 2 \log e
\end{align}
where $Z \sim \chi_2^n$ (i.e. chi square distributed with $n$ degrees of freedom).
\end{thm}
The following result can be obtained by an application of Theorem \ref{thm:C} to the GMS. 
\begin{thm}[Converse, GMS]
\label{thm:CGMS}
Any $(n, M,d,\epsilon)$ code must satisfy
\begin{align}
\label{eq_CGMS}
M \geq \left( \frac{\sigma}{\sqrt d} r_n(\epsilon)\right)^{n}
\end{align}
where $r_n(\epsilon)$ is the solution to
\begin{equation}
\label{eq_rGMS}
\Prob{ Z < n\ r^2_n(\epsilon) }  = 1 - \epsilon,
\end{equation}
and $Z \sim \chi^2_n$.
\end{thm}
\begin{proof}
Inequality \eqref{eq_CGMS} simply states that the minimum number of $n$-dimensional balls of radius $\sqrt {n d}$ required to cover an $n$-dimensional ball of radius $\sqrt n \sigma r_n(\epsilon)$ cannot be smaller than the ratio of their volumes. Since 
\begin{equation}
 Z = \frac 1 {\sigma^2}\sum_{i = 1}^n X_i^2
\end{equation}
is $\chi^2_n$-distributed, the left side of \eqref{eq_rGMS} is the probability that the source produces a sequence that falls inside $\mathbb B$, the $n$-dimensional ball of radius $\sqrt n \sigma r_n(\epsilon)$ with center at $\mathbf 0$. But as follows from the spherical symmetry of the Gaussian distribution, $\mathbb B$ has the smallest volume among all sets in $\mathbb R^n$ having probability $1 - \epsilon$.  Since any $(n, M,d,\epsilon)$-code is a covering of a set that has total probability of at least $1 - \epsilon$, the result follows.
\end{proof}
Note that the proof of Theorem \ref{thm:CGMS} can be formulated in the hypothesis testing language of Theorem \ref{thm:C} by choosing $Q$ to be the Lebesgue measure on $\mathbb R^n$. 

The following achievability result can be regarded as the rate-distortion counterpart to Shannon's geometric analysis of optimal coding for the Gaussian channel \cite{shannon1959probability}. 
\begin{thm}[Achievability, GMS]
\label{thm:AGMS}
There exists an $(n, M,d,\epsilon)$ code with
\begin{equation}
\epsilon \leq
n \int_0^\infty \left[ 1 - \rho(n, z)\right]^M f_{\chi^2_n}\left(n z\right) dz \label{eq_AGMS}
\end{equation}
where $f_{\chi^2_n}(\cdot)$ is the $\chi^2_n$ probability density function, and
\begin{equation}
 \rho(n, z) = 
\frac{\Gamma \left( \frac n 2 + 1\right)}{\sqrt \pi n \Gamma \left( \frac{n-1}{2} + 1\right) }\left( 1 - \frac{\left(1 + z - 2\frac d {\sigma^2}\right)^2}{4 \left(1 - \frac d {\sigma^2}\right) z} \right) ^{\frac{n-1} 2}
\end{equation}
if $a^2 \leq z \leq b^2$, where
\begin{align}
a &= \sqrt{1 - \frac{d}{\sigma^2}} - \sqrt{\frac{d}{\sigma^2}}\\
b &= \sqrt{1 - \frac{d}{\sigma^2}} + \sqrt{\frac{d}{\sigma^2}}
\end{align}
and $\rho(n, z) = 0$ otherwise.
\end{thm}
\begin{proof}
We compute an upper bound to \eqref{eq_A} for the specific case of the GMS. Let $P_{Y^n} $ be the uniform distribution on the surface of the $n$-dimensional sphere with center at $\mathbf 0$ and radius 
\begin{equation}
 r_0 = \sqrt n \sigma \sqrt{1 - \frac{d}{\sigma^2}}
\end{equation}
This choice corresponds to a positioning of representation points that is optimal in the limit of large $n$, see Fig. \ref{fig:geometry}(a), \cite{wyner1968communication,sakrison1968geometric}. Indeed, for large $n$, most source sequences will be concentrated within a thin shell near the surface of the sphere of radius $\sqrt n \sigma$. The center of the sphere of radius $\sqrt{nd}$ must be at distance $r_0$ from the origin in order to cover the largest area of the surface of the sphere of radius $\sqrt n \sigma$. 

We proceed to lower-bound $P_{Y^n}(B_d(x^n))$, $x^n \in \mathbb R^n$. Observe that $P_{Y^n}(B_d(x^n)) = 0$ if $x^n$ is either too close or too far from the origin, that is, if $|x^n| < \sqrt n  \sigma a$ or $|x^n| >  \sqrt n \sigma b$, where $|\cdot |$ denotes the Euclidean norm. To treat the more interesting case $\sqrt n  \sigma a \leq |x^n | \leq \sqrt n  \sigma b$, it is convenient to introduce the following notation.
\begin{itemize}
\item $S_n(r) = \frac {n\pi ^{\frac n 2}} {\Gamma \left( \frac n 2 + 1\right)} r^{n-1}$: surface area of an $n$-dimensional sphere of radius $r$;
\item $S_n(r, \theta)$: surface area of an $n$-dimensional polar cap of radius $r$ and polar angle $\theta$.
\end{itemize}
Similar to \cite{wyner1968communication,sakrison1968geometric}, from Fig. \ref{fig:geometry}(b),
\begin{equation}
S_n(r, \theta) \geq \frac{\pi^{\frac{n-1}2}}{\Gamma\left( \frac {n-1}2 + 1\right)} (r \sin \theta)^{n-1} \label{eq_GMSdisc}
\end{equation}
where the right side of \eqref{eq_GMSdisc} is the area of an $(n-1)$-dimensional disc of radius $r\sin \theta$. 
So if $\sqrt n  \sigma a \leq |x^n | = r \leq \sqrt n  \sigma b$,
\begin{align}
 P_{Y^n}\left( B_d(x^n) \right) &= \frac{S_n(|x^n|, \theta)}{S_n(|x^n|)}\\
 &\geq \frac{\Gamma \left( \frac n 2 + 1\right)}{\sqrt \pi n \Gamma \left( \frac{n-1}{2} + 1\right) }\left( \sin \theta\right) ^{n-1} \label{eq_GMSdball}
\end{align}
where $\theta$ is the angle in Fig.  \ref{fig:geometry}(b); by the law of cosines 
\begin{equation}
 \cos \theta = \frac{r^2 + r_0^2 - nd}{2 r r_0}
\end{equation}
 Finally, by Theorem \ref{thm:A}, there exists an $(n, M, d, \epsilon)$ code with
\begin{align}
 \epsilon &\leq \E{1 - P_{Y^n}(B_d(X^n))}^M\\
 &=  \E{ \left[1 - P_{Y^n}(B_d(X^n))\right] ^M \mid \sqrt n  \sigma a \leq |X^n | \leq \sqrt n  \sigma b } \notag\\
  &+ \Prob{|X^n| < \sqrt n  \sigma a} + \Prob{|X^n| > \sqrt n  \sigma a} \label{eq_-AGMS}
\end{align}
Since $\frac{|X^n|^2}{\sigma^2}$ is $\chi_n^2$-distributed, one obtains \eqref{eq_AGMS} by plugging $\sin^2 \theta = 1 - \cos^2 \theta$ into \eqref{eq_GMSdball} and substituting the latter in \eqref{eq_-AGMS}.
\begin{figure}[htp]
\begin{center}
    \subfigure[]{
    \epsfig{file=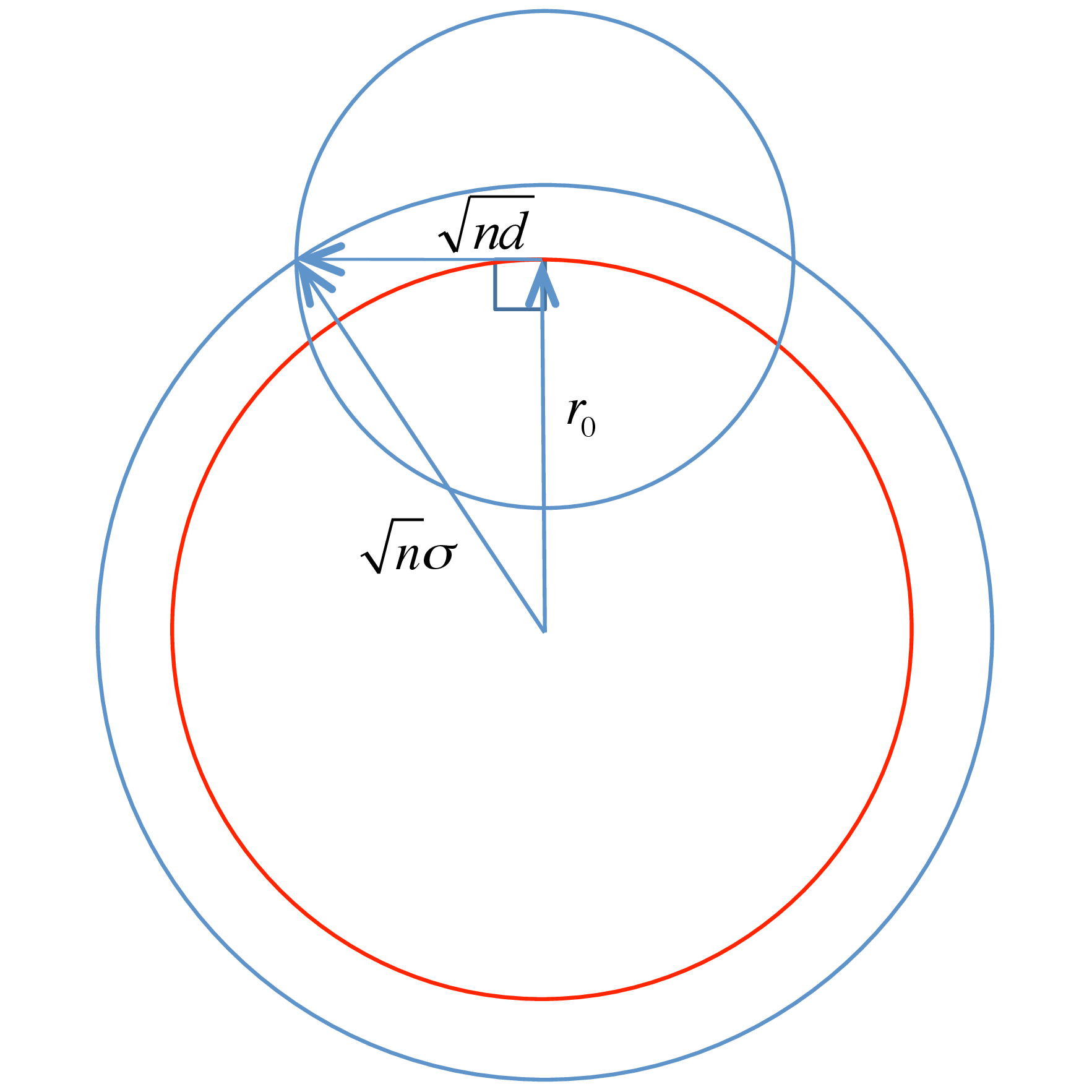,width=.7\linewidth}
    }
    \subfigure[]{
    \epsfig{file=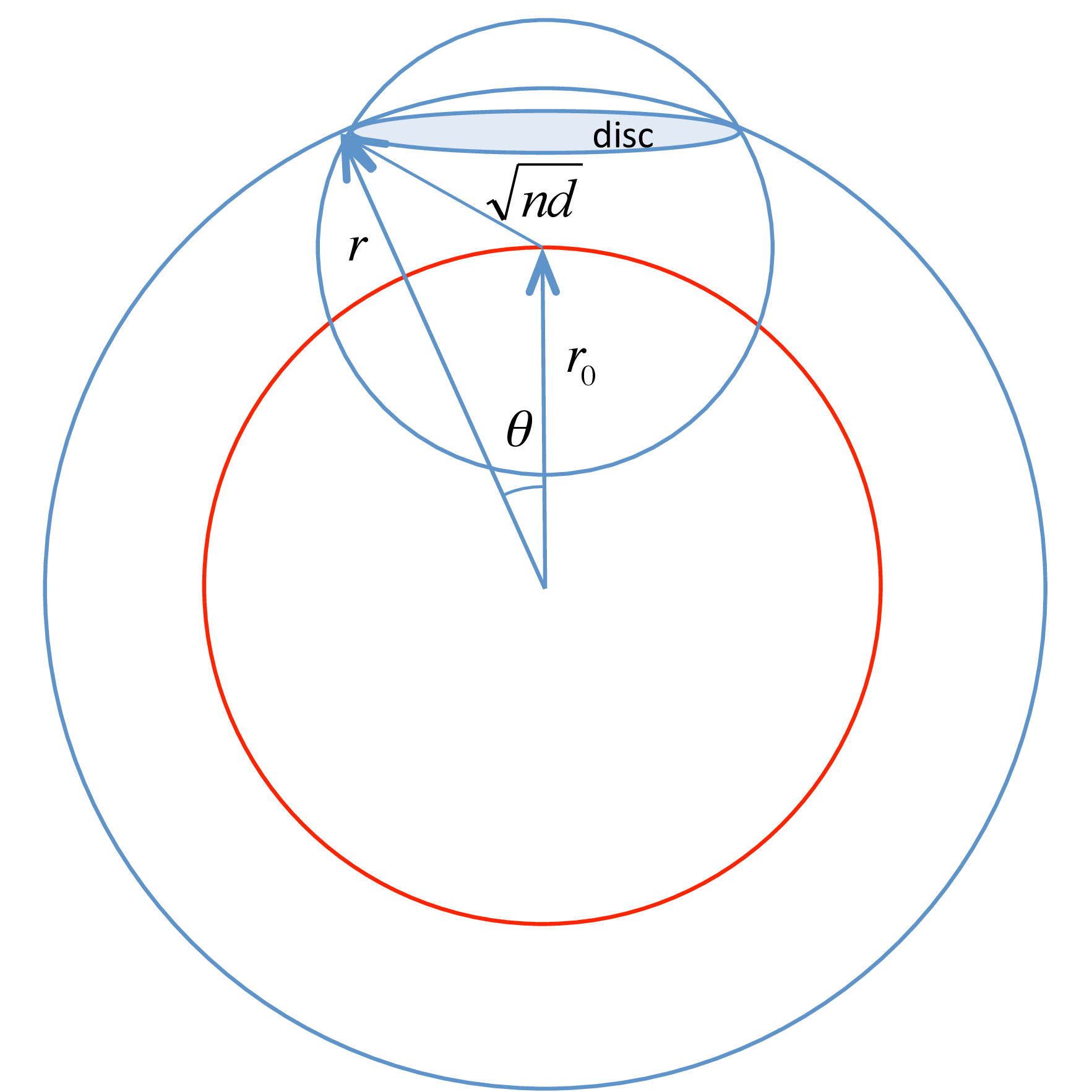,width=.7\linewidth }
    }
\end{center}
 \caption[]{Optimum positioning of the representation sphere (a) and the geometry of the excess-distortion probability calculation (b).
 } \label{fig:geometry}
\end{figure}
\end{proof}
Essentially Theorem \ref{thm:AGMS} evaluates the performance of Shannon's random code with all codewords lying on the surface of a sphere contained inside the sphere of radius $\sqrt n \sigma$. The following result allows us to bound the performance of a code whose codewords lie inside a ball of radius slightly larger than $\sqrt n \sigma$.

\begin{thm}[Rogers \cite{rogers1963covering} - Verger-Gaugry \cite{verger2005covering}]
\label{thm:rogers}
If $r > 1$ and $n \geq 2$, an $n-$dimensional sphere of radius $r$ can be covered by $\lfloor \mathbb M(r)\rfloor$ spheres of radius $1$, where $\mathbb M(r)$ is defined in \eqref{eq_rogers}.
\addtocounter{equation}{1}
\end{thm}
The first two cases in \eqref{eq_rogers} (at the bottom of the page) are encompassed by the classical result of Rogers \cite{rogers1963covering} that appears not to have been improved since 1963, while the last two are due to the recent improvement by Verger-Gaugry \cite{verger2005covering}. An immediate corollary to Theorem \ref{thm:rogers} is the following:
\begin{thm}[Achievability, GMS]
\label{thm:AGMSrogers}
For $n \geq 2$, there exists an $(n, M,d,\epsilon)$ code such that
\begin{equation}
M \leq \mathbb M \left(\frac{\sigma}{\sqrt d} r_n(\epsilon)\right)
\label{eq_AGMSrogers}
\end{equation}
where $r_n(\epsilon)$ is the solution to \eqref{eq_rGMS}.
\end{thm}
\begin{proof}
Theorem \ref{thm:rogers} implies that there exists a code with no more than $\mathbb M \left(\frac{\sigma}{\sqrt d} r_n(\epsilon)\right)$ codewords such that all  source sequences that fall inside $\mathbb B$, the $n$-dimensional ball of radius $\sqrt n \sigma r_n(\epsilon)$ with center at $\mathbf 0$, are reproduced within distortion $d$.  The excess-distortion probability is therefore given by the probability that the source produces a  sequence that falls outside $\mathbb B$.
\end{proof}
Note that Theorem \ref{thm:AGMSrogers} studies the number of balls of radius $\sqrt {n d}$ to cover $\mathbb B$ that is provably achievable, while the converse in Theorem \ref{thm:CGMS} lower bounds the minimum number of balls of radius $\sqrt {n d}$ required to cover $\mathbb B$ by the ratio of their volumes.
\begin{thm}[Gaussian approximation, GMS]
\label{thm:2orderGMS}
The minimum achievable rate at blocklength $n$ satisfies
\begin{equation}
\label{eq_2orderGMS}
R(n,d,\epsilon)  = \frac 1 {2} \log \frac{\sigma^2 }{d}+ \sqrt{\frac{1}{2n}}\Qinv{\epsilon} \log e + \theta\left( \frac {\log n}{n}\right)
\end{equation}
where the remainder term satisfies
\begin{align}
\bigo{\frac 1 n} &\leq \theta\left( \frac {\log n}{n}\right) \label{eq_2orderCGMS}\\
&\leq \frac 1 2 \frac {\log n}{n} + \frac {\log \log n}{n} + \bigo{\frac 1 n} \label{eq_2orderAGMS}
\end{align}
\end{thm}
\begin{proof}
We start with the converse part, i.e. \eqref{eq_2orderCGMS}.

Since in Theorem \ref{thm:CGMS} $Z = \frac 1 {\sigma^2}\sum_{i = 1}^n X_i^2$, $X_i \sim \mathcal N(0, \sigma^2)$,
we apply the Berry-Esseen CLT (Theorem \ref{thm:Berry-Esseen}) to $\frac{1}{\sigma^2} X_i^2$. Each $\frac{1}{\sigma^2} X_i^2$ has mean, second and third central moments equal to $ 1$, $2$ and $8$, respectively.
Let
\begin{align}
\label{eq_rlbGMS}
r^2 &=  1 +\sqrt{\frac 2 n} Q^{-1}\left( \epsilon + \frac {12 \sqrt 2}{\sqrt n}\right)\\
&= 1 + \sqrt{\frac 2 n} Q^{-1}\left( \epsilon \right) + O\left(\frac 1 {n}\right)
\end{align}
 Then by the Berry-Esseen inequality \eqref{eq_BerryEsseen}
\begin{equation}
\mathbb P \left[ Z > n \bar r^2\right] \geq \epsilon
\end{equation}
and therefore $r_n(\epsilon)$ that achieves the equality in \eqref{eq_rGMS} must satisfy $r_n(\epsilon) \geq r$. Weakening \eqref{eq_CGMS} by plugging $r$ instead of $r_n(\epsilon)$ and taking logarithms of both sides therein, one obtains:
\begin{align}
\log M &\geq \frac n 2 \log \frac{\sigma^2 r^2} {d} \label{eq_CGMSa}\\
&=  \frac n 2 \log \frac{\sigma^2 }{d}+ \sqrt{\frac{n}{2}}Q^{-1}\left( \epsilon \right)\log e + O\left(1\right) \label{eq_CGMSb}
\end{align}
where \eqref{eq_CGMSb} is a Taylor approximation of the right side of \eqref{eq_CGMSa}. 

The achievability part \eqref{eq_2orderAGMS} is proven in Appendix \ref{appx:2orderAGMS} using Theorem \ref{thm:AGMS}. Theorem \ref{thm:AGMSrogers} leads to the correct rate-dispersion term but a weaker remainder term.
\end{proof}

\begin{figure}[htp]
\begin{center}
    \epsfig{file=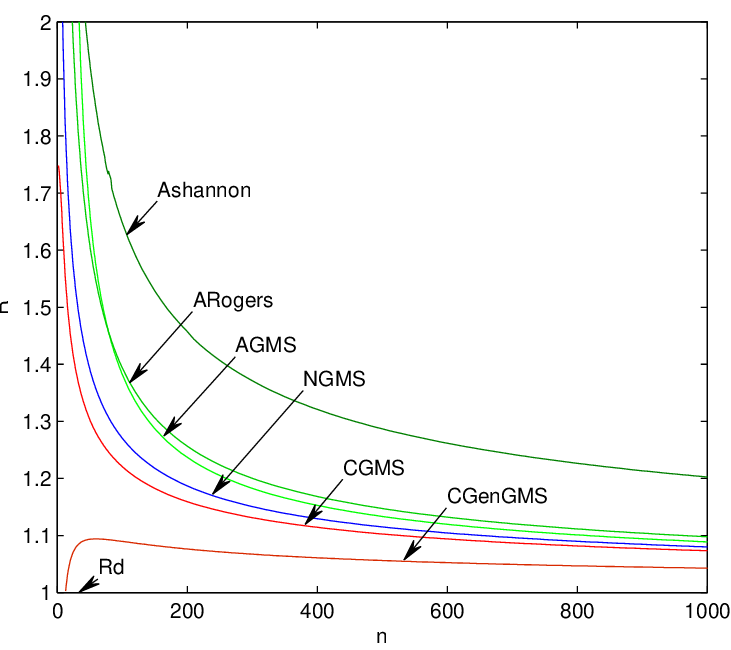,width=1\linewidth}
\end{center}
 \caption[]{Bounds to $R(n,d,\epsilon)$ and Gaussian approximation for GMS with $\sigma = 1$, $d = \frac 1 4$ , $\epsilon = 10^{-2}$.} \label{fig:GMS1e-2}
\end{figure}

\begin{figure}[htp]
\begin{center}
    \epsfig{file=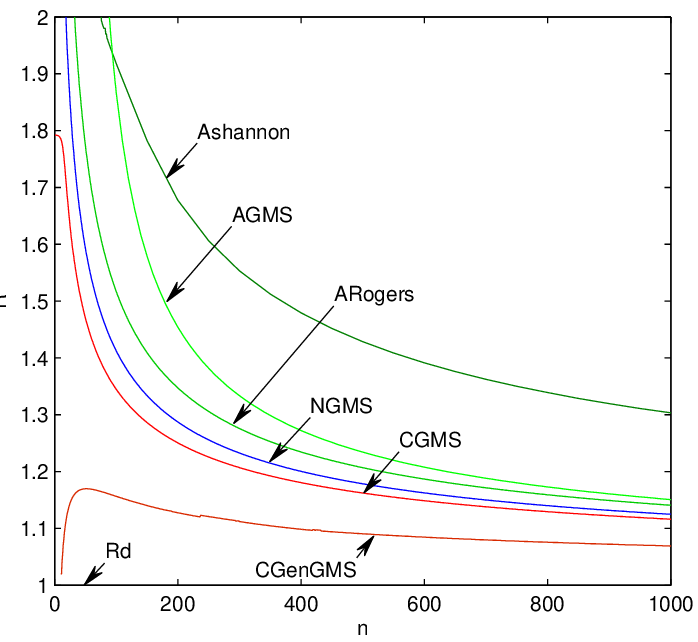,width=1\linewidth}
\end{center}
 \caption[]{Bounds to $R(n,d,\epsilon)$ and Gaussian approximation for GMS with $\sigma = 1$, $d = \frac 1 4$ , $\epsilon = 10^{-4}$.} \label{fig:GMS1e-4}
\end{figure}

Figures \ref{fig:GMS1e-2} and \ref{fig:GMS1e-4} present a numerical comparison of Shannon's achievability bound \eqref{eq_Ashannon} and the new bounds in \eqref{eq_AGMS}, \eqref{eq_AGMSrogers}, \eqref{eq_CGMS} and \eqref{eq_CgGMS} as well as the Gaussian approximation in \eqref{eq_2orderGMS} in which we took $\theta\left( \frac {\log n}{n}\right) = \frac 1 2 \frac {\log n}{n}$.  The achievability bound in \eqref{eq_AGMSrogers} is tighter than the one in \eqref{eq_AGMS} at shorter blocklengths. Unsurprisingly, the converse bound in \eqref{eq_CGMS} is quite a bit tighter than the one in \eqref{eq_CgGMS}.

\section{Conclusion}
\label{sec:conclusion}
To estimate the minimum rate required to sustain a given fidelity at a given blocklength, we have shown new achievability and converse bounds, which apply in full generality and which are tighter than existing bounds. The tightness of these bounds for stationary memoryless sources allowed us to obtain a compact closed-form expression that approximates the excess rate over the rate-distortion function incurred in the nonasymptotic regime (Theorem \ref{thm:2order}). For those sources and unless the blocklength is small, the rate dispersion (along with the rate-distortion function) serves to give tight approximations to the fundamental fidelity-rate tradeoff.

\section*{Acknowledgement}
Useful discussions with Dr. Yury Polyanskiy are gratefully acknowledged. In particular, Theorem \ref{thm:C} was suggested by him.

\appendices
\section{Hypothesis testing \\and almost lossless data compression}
\label{appx:lossless}
To show \eqref{eq_Mhtlossless}, without loss of generality, assume that the letters of the alphabet $A$ are labeled $1, 2, \ldots $ in order of decreasing probabilities:
\begin{equation}
 P_X(1) \geq P_X(2) \geq \ldots
\end{equation}

Observe that
\begin{equation}
 M^\star(0, \epsilon) = \min \left\{ m \geq 1: \Prob{X \leq m} \geq 1 - \epsilon \right\},
\end{equation}
and the optimal randomized test to decide between $P_X$ and $U$ is given by
\begin{equation}
 P_{W|X}(1|a) = 
\begin{cases}
 1, & a \leq M^\star(0, \epsilon) - 1\\
 \alpha, & a = M^\star(0, \epsilon)\\
 0, & a \geq M^\star(0, \epsilon) + 1
\end{cases}
\end{equation}
It follows that
\begin{equation}
  \beta_{1 - \epsilon}(P_X, U) = M^\star(0, \epsilon) - 1 + \alpha
\end{equation}
where $\alpha \in (0, 1]$ is the solution to
\begin{equation}
 \Prob{X \leq M^\star(0, \epsilon) - 1} + \alpha P_X(M^\star(0, \epsilon)) = 1 - \epsilon,
\end{equation}
hence \eqref{eq_Mhtlossless}.

\section{Gaussian approximation analysis \\of almost lossless data compression}
\label{appx:2orderlossless}
In this appendix we strenghten the remainder term in Theorem \ref{thm:2order} for $d = 0$ (cf. \eqref{eq_2orderLossless}). 
Taking the logarithm of \eqref{eq_Mhtlossless}, we have
\begin{align}
&~\log \beta_{1 - \epsilon}(P_X, U) \notag\\
\leq&~ \log M^\star(0, \epsilon) \label{eq_-2orderlossless}\\
\leq&~ \log  \left(\beta_{1 - \epsilon}(P_X, U) + 1 \right) \\
=&~ \log \beta_{1 - \epsilon}(P_X, U) + \log\left( 1 + \frac 1 {\beta_{1 - \epsilon} (P_X, U) } \right)\\
\leq&~ \log \beta_{1 - \epsilon}(P_X, U) + \frac 1 {\beta_{1 - \epsilon}(P_X, U)} \log e\label{eq_-2orderlosslessa}
\end{align}
where in \eqref{eq_-2orderlosslessa} we used $\log(1+x) \leq x \log e$, $x > -1$.

Let $P_{X^n} = P_{\mathsf X} \times \ldots \times P_{\mathsf X}$ be the source distribution, and let $U^n$ to be the counting measure on $\mathcal A^n$. Examining the proof of Lemma 58 of \cite{polyanskiy2010channel} on the asymptotic behavior of $\beta_{1 - \epsilon}(P,Q)$ it is not hard to see that it extends naturally to $\sigma$-finite $Q$'s; thus if $\Var{\imath_{\mathsf X}(\mathsf X)} > 0$,
\begin{align}
\log \beta_{1 - \epsilon} (P_{X^n}, U^n) &= n H(\mathsf X) + \sqrt {n \Var{ \imath_{\mathsf X}( \mathsf X)}} \Qinv{\epsilon} \notag\\
&- \frac 1 2 \log n + \bigo{1} \label{eq_beta2order}
\end{align}
  and if $\Var{\imath_{\mathsf X}(\mathsf X)} = 0$,
\begin{equation}
\log \beta_{1 - \epsilon} (P_{X^n}, U^n) = n H(\mathsf X) - \log \frac 1 {1 - \epsilon} \label{eq_beta2order0}
\end{equation}
Letting $P_{X^n}$ and $U^n$ play the roles of $P_X$ and $U$ in \eqref{eq_-2orderlossless} and \eqref{eq_-2orderlosslessa} and invoking \eqref{eq_beta2order} and \eqref{eq_beta2order0}, we obtain \eqref{eq_2orderLossless} and \eqref{eq_2orderLossless0}, respectively.

\section{Generalization of Theorems \ref{thm:Cg} and \ref{thm:2order}}
\label{appx:csiszar}
We show that even if the rate-distortion function is not achieved by any output distribution, the definition of $d-$tilted information can be extended appropriately, so that Theorem \ref{thm:Cg} and the converse part of Theorem \ref{thm:2order} still hold.

We use the following general representation of the rate-distortion function due to Csisz{\'a}r \cite{csiszar1974extremum}. 

\begin{thm} [Alternative representation of $\mathbb R(d)$  \cite{csiszar1974extremum}]
 \label{thm:csiszar}
Under the basic restrictions \eqref{item:a}-\eqref{item:c} of Section \ref{ssec:dtilted}, for each $d > d_{\min}$, it holds that
\begin{equation}
 \mathbb R_X(d) = \max_{\alpha(x), ~ \lambda}\left\{ \E{ \alpha(X)} - \lambda d\right\} \label{eq_RR(d)csiszar}
\end{equation}
where the maximization is over $\alpha(x)\geq 0$ and $\lambda\geq 0$ satisfying the constraint
\begin{equation}
\E{ \exp\left\{ \alpha(X) - \lambda d(X, y)\right\}} \leq 1 ~ \forall y \in B \label{eq_csiszarg}
\end{equation}
\end{thm}

Let $\left( \alpha^\star(x), ~\lambda^\star\right)$ achieve the maximum in \eqref{eq_RR(d)csiszar} for some $d > d_{\min}$, and define the $d-$tilted information in $x$ by
 \begin{equation}
\jmath_{X}(x, d) = \alpha^\star(x)  - \lambda^\star d \label{eq_idg}
\end{equation}
Note that \eqref{eq_csiszar}, the only property of $d-$tilted information we used in the proof of Theorem \ref{thm:Cg}, still holds due to \eqref{eq_csiszarg}, thus Theorem \ref{thm:Cg} remains true.

The proof of the converse part of Theorem \ref{thm:2order} generalizes immediately upon making the following two observations. First, \eqref{eq_R(d)alternative} is still valid due to \eqref{eq_RR(d)csiszar}. Second, $d$-tilted information in \eqref{eq_idg} still single-letterizes  for memoryless sources:

\begin{lemma} \label{lemma:id_iid} Under restrictions \eqref{item:first} and \eqref{item:separable} in Section \ref{ssec:2orderMain}, \eqref{eq_id_iid} holds. 
\end{lemma}
\begin{proof}
Let $\left( \alpha^\star(\mathsf x), ~ \lambda^\star \right) $ attain the maximum in \eqref{eq_RR(d)csiszar} for the single-letter distribution $P_{\mathsf X}$. It suffices to check that
$
 \left( \sum_{i = 1}^n  \alpha^\star( x_i), n \lambda^\star \right)
$ attains the maximum in \eqref{eq_RR(d)csiszar} for $P_{X^n} = P_{\mathsf X} \times \ldots \times P_{\mathsf X}$.

As desired,
\begin{equation}
 \E{\sum_{i = 1}^n \alpha^\star(X_i)} - n\lambda^\star d = n \mathbb R_{\mathsf X}(d) = \mathbb R_{X^n}(d)
\end{equation}
and  we just need to verify the constraints in \eqref{eq_csiszarg} are satisfied:
\begin{align}
&~\E{ \exp\left\{\sum_{i = 1}^n \alpha^\star(X_i) - \lambda^\star \sum_{i = 1}^n d(X_i, y) \right\}  } \notag \\
=&~ \prod_{i = 1}^n \E{ \exp\left\{\alpha^\star(X_i) - \lambda^\star d(X_i, y) \right\} } \\
\leq&~ 1 ~ \forall y^n \in \mathcal B^n \label{eq_-2orderC}
\end{align}

\end{proof}

\section{Proof of Lemma \ref{lemma:aepA}}
\label{appx:aep}
Before we prove Lemma \ref{lemma:aepA}, let us present some background results we will use. 
For $k = 1, 2, \ldots$, denote 
\begin{equation}
 \bar d_{Y, k}(x, \lambda) = \frac{\E{d^k(x, Y) \exp\left( -\lambda d(x, Y)\right)}} {\E{\exp\left( -\lambda d(x, Y)\right)}}\label{eq_dk}
\end{equation} 
Observe that
\begin{equation}
 \bar d_{Y, k}(x, 0) = \E{d^k(x, Y)}\label{eq_dk0}
\end{equation}
(the expectations in \eqref{eq_dk} and \eqref{eq_dk0} are with respect to the unconditional distribution of $Y$). Denoting by $(\cdot)^{\prime}$ differentiation with respect to $\lambda > 0$, we state the following properties whose proofs can be found in \cite{yang1999redundancy}.

\begin{enumerate}[A.]
 \item $\left(\E{ \Lambda_{ Y}( X, \lambda^\star_{X, Y})}\right)^{\prime} = 0$ where $\lambda^\star_{X, Y} = - \mathbb R^{\prime}_{X, Y}(d)$.\label{propertyoptlambda}
   \item $\E{\Lambda_Y^{\prime\prime}(X, \lambda)} < 0$ for all $\lambda > 0$ if $\E{ \bar d_{Y,2}(X, 0) } < \infty$. \label{propertyELambda''}
 \item 
 $ \Lambda^\prime_{Y}( x, \lambda) = -d + \bar d_{Y, 1}(x, \lambda)$.
 \label{propertyLambda'}
 \item  $\Lambda_Y^{\prime\prime}(x, \lambda) = \left[\bar d_{Y,1}^2(x, \lambda) - \bar d_{Y,2}(x, \lambda)\right]\left(\log e\right)^{-1} \leq 0$\\ if $ \bar d_{Y,1}(x, 0) < \infty$. \label{propertyLambda''}
  \item  $\bar d_{Y, k}^\prime(x, \lambda) \leq 0$ if $\bar d_{Y, k}(x, 0) < \infty$. \label{propertydk}
  \item $d_{\min \mid X, Y} = \E{ \alpha_Y(X)}$, where $\alpha_Y(x) = \mathrm{ess} \inf d(x, Y)$. \label{propertydmin}
 \end{enumerate}
\begin{remark}
By Properties \ref{propertyoptlambda} and \ref{propertyELambda''},
\begin{equation}
 \E{ \Lambda_{ Y}( X, \lambda^\star_{X, Y})} = \sup_{\lambda > 0}  \E{ \Lambda_{ Y}( X, \lambda)} \label{eq_Lambdadual}
\end{equation}
\end{remark}

\begin{remark}
 Properties \ref{propertyLambda'} and \ref{propertyLambda''} imply that
\begin{equation}
- d \leq  \Lambda^\prime_{Y}(x, \lambda) \leq - d + \bar d_{Y, 1}(x, 0) \label{eq_Lambda'dominated}
\end{equation}
Therefore, as long as $\E{ \bar d_{Y,1}(X, 0) } < \infty$, the differentiation in Property \ref{propertyoptlambda} can be brought inside the expectation invoking the dominated convergence theorem. Keeping this in mind while averaging the equation in Property \ref{propertyLambda'} with $\lambda = \lambda^\star_{X, Y}$ with respect to $P_X$, we observe that
\begin{equation}
\E{ \bar d_{Y, 1}(X, \lambda^\star_{X, Y})} = d \label{eq_optlambda}
\end{equation}
\end{remark}
\begin{remark}
 Properties \eqref{eq_iddensity} and \eqref{eq_Ejd} of $d-$tilted information imply that the equality in \eqref{eq_optlambda} holds if $\lambda^\star_{X, Y}$ is replaced by $\lambda^\star = -\mathbb R_X^\prime(d)$, and $Y$ is replaced by $Y^\star$ - the $\mathbb R_X(d)$-achieving random variable. It follows that
\begin{equation}
 \lambda^\star = \lambda^\star_{X, Y^\star} \label{eq_slopes}
\end{equation}
\end{remark}
\begin{remark}
 By virtue of Properties \ref{propertyLambda''} and \ref{propertydk} we have
\begin{align}
- \bar d_{Y, 2}( x, 0) \leq \Lambda^{\prime\prime}_{Y}(x, \lambda) \log e \leq 0  \label{eq_Lambda''dominated}
\end{align}
\end{remark}
\begin{remark}
 Using \eqref{eq_optlambda}, derivatives of $\mathbb R_{X, Y}(d)$ are conveniently expressed via  $\E{ \bar d_{Y, k}(x, \lambda^\star_{X, Y})}$; in particular, at any 
\begin{equation}
 d_{\min \mid X, Y} < d \leq d_{\max \mid X, Y} = \E{\bar d_{Y, 1}(X, 0))}
\end{equation}
we have
\begin{align}
  \mathbb R_{X, Y}^{\prime\prime}(d) &= - \frac{1}{\left(\E{ \bar d_{Y, 1}(X, \lambda^\star_{X, Y})}\right)^{\prime}}\\
  &= \frac {\log e} {\E{\bar d_{Y,2}(X, \lambda^\star_{X, Y})} - \E{\bar d_{Y,1}^2(X, \lambda^\star_{X, Y})} }  \label{eq_R''}\\
  &> 0 \label{eq_R''>0}
\end{align}
where \eqref{eq_R''} holds by Property \ref{propertyLambda''} and the dominated convergence theorem due to  \eqref{eq_Lambda''dominated} as long as $\E{ \bar d_{Y,2}(X, 0) } < \infty$, and \eqref{eq_R''>0} is by Property \ref{propertyELambda''}. 
\label{remark:R''}
\end{remark}

The proof of Lemma \ref{lemma:aepA} consists of Gaussian approximation analysis of the bound in Lemma \ref{lemma:aepAnonasymptotic}. First, we weaken the bound in Lemma \ref{lemma:aepAnonasymptotic} by choosing $P_{\hat X}$ and $\gamma$ in \eqref{eq_aepAnonasymptotic} in the following manner. Fix $\tau > 0$, and let $\gamma = \frac \tau n$, $P_{Y} = P_{Y^{n \star}} = P_{\mathsf Y^\star} \times \ldots \times P_{\mathsf Y^\star}$, where $\mathsf Y^\star$ achieves $\mathbb R_{\mathsf X}(d)$, and choose $P_{\hat X} = P_{\hat X^n} = P_{\hat {\mathsf X}} \times \ldots P_{\hat {\mathsf X}}$, where $P_{\hat {\mathsf X}}$ is the measure on $\mathcal A$ generated by the empirical distribution of $x^n \in \mathcal A^n$: 
\begin{equation}
P_{\hat {\mathsf X}}(a) = \frac 1 n \sum_{i = 1}^n 1\{x_i = a\} \label{eq_Pempirical}
\end{equation}
Since the distortion measure is separable, for any $\lambda > 0$ we have
\begin{equation}
 \Lambda_{Y^{n \star} }(x^n, \lambda n) = \sum_{i = 1}^n \Lambda_{\mathsf Y^\star}(x_i, \lambda)
\end{equation}
so by Lemma \ref{lemma:aepAnonasymptotic}, for all 
\begin{equation}
 d > d_{\min \mid \hat X^n, Y^{n \star}} \label{eq_Ad>dmin}
\end{equation}
it holds that
\begin{align}
P_{Y^{n \star}}(B_d(x^n)) &\geq \exp\left( - \sum_{i = 1}^n \Lambda_{\mathsf Y^\star}(x_i, \lambda(x^n)) - \lambda(x^n) \tau \right) \notag\\
&\cdot \Prob{ nd - \tau < \sum_{i = 1}^n d(x_i, \hat Z^\star_i) \leq nd | \hat X^n = x^n}  \label{eq_aepAnonasymptotic1}
\end{align}
where we denoted 
\begin{equation}
 \lambda(x^n) = -\mathbb R^\prime_{\hat {\mathsf X}; \mathsf Y^\star}(d)
\end{equation}
 ($\lambda(x^n)$ depends on $x^n$ through the distribution of 
$\hat {\mathsf X}$
 in \eqref{eq_Pempirical}),
and 
$P_{\hat Z^{ n \star} }= P_{\hat{\mathsf Z}^\star} \times \ldots \times P_{\hat{\mathsf Z}^\star}$
, where 
$P_{\hat{\mathsf Z}^\star| \hat {\mathsf X}}$
 achieves 
 $\mathbb R_{\hat {\mathsf X}; \mathsf Y^\star}(d)$. 
 The probability appearing in \eqref{eq_aepAnonasymptotic1} can be lower bounded by the following lemma.

\begin{lemma}
Assume that restrictions \eqref{item:first}-\eqref{item:last} in Section \ref{ssec:2orderMain} hold. Then, there exist  $\delta_0, n_0 > 0$ such that for all $\delta \leq \delta_0$, $n \geq n_0$, there exist a set $F_n \subseteq \mathcal A^n$ and constants $\tau, C_1, K_1 > 0$ such that 
\begin{equation}
 \Prob{X^n \notin F_n} \leq \frac {K_1} {\sqrt n} \label{eq_-APTyp}
\end{equation}
and for all $x_n \in F_n$, 
\begin{align}
  \Prob{ nd - \tau < \sum_{i = 1}^n d(x_i, \hat Z^\star_i) \leq nd | \hat X^n = x^n} &\geq  \frac {C_1} {\sqrt n} \label{eq_aepA1}\\
  \left| \lambda(x^n) - \lambda^\star\right| &< \delta \label{eq_-ALambdaTyp}
\end{align}
where $\lambda^\star = - \mathbb R_{\mathsf X}^\prime(d)$. 
\label{lemma:aepA1}
\end{lemma}

\begin{proof}
The reasoning is similar to the proof of \cite[(4.6)]{yang1999redundancy}. Fix 
\begin{equation}
0 < \Delta < \frac 1 3 \min\left\{ d - d_{\min \mid \mathsf X, \mathsf Y^\star}, ~ d_{\max \mid \mathsf X, \mathsf Y^\star} - d \right\} \label{eq_-ADeltaub}
\end{equation}
 (the right side of \eqref{eq_-ADeltaub} is positive by restriction \eqref{item:dminmax} in Section \ref{ssec:2orderMain}) and denote
\begin{align}
\underline \lambda &= -\mathbb R^\prime_{\mathsf X, \mathsf Y^\star} \left( d + \frac {3 \Delta}{2} \right) \label{eq_Alowerlambda}\\
\bar \lambda &=  -\mathbb R^\prime_{\mathsf X, \mathsf Y^\star} \left( d - \frac {3 \Delta}{2} \right) \label{eq_Aupperlambda}\\
\mu^{\prime\prime} &=  \E{ \left| \Lambda_{\mathsf Y^\star}^{\prime\prime}( \mathsf X, \lambda^\star) \right| } \label{eq_mu''}\\
\delta &=  \frac {3\Delta} 2 \sup_{|\theta| < \frac {3 \Delta}{2} }\mathbb R^{\prime\prime}_{\mathsf X, \mathsf Y^\star}(d + \theta) \label{eq_ATypdelta}\\
 \overline V(x^n) &= \frac 1 n \sum_{i = 1}^n \sup_{|\theta| < \delta} \left| \Lambda^{\prime\prime} (x_i, \lambda^\star + \theta) \right| \log e \label{eq_Vupper}\\
 \underline V(x^n) &= \frac 1 n \sum_{i = 1}^n \inf_{|\theta| < \delta} \left| \Lambda^{\prime\prime} (x_i, \lambda^\star + \theta) \right| \log e \label{eq_Vlower}
\end{align}
We say that $x^n \in F_n$ if it meets the following conditions:
\begin{align}
\frac 1 n \sum_{i = 1}^n \alpha_{\mathsf Y^\star}(x_i) &< d_{\min \mid \mathsf X, \mathsf Y^\star} + \Delta  \label{eq_ATypdmin} \\
\frac 1 n \sum_{i = 1}^n \bar d_{\mathsf Y^\star\!, 1} (x_i, 0) &> d_{\max \mid \mathsf X, \mathsf Y^\star} - \Delta  \label{eq_ATypdmax}\\
\frac 1 n \sum_{i = 1}^n \bar d_{\mathsf Y^{\star}\!, 1}(x_i, \underline \lambda) &> d + \Delta \label{eq_ATypd1a}\\
\frac 1 n \sum_{i = 1}^n \bar d_{\mathsf Y^\star\!, 1}(x_i, \bar \lambda) &< d - \Delta \label{eq_ATypd1b}\\
\frac 1 n \sum_{i = 1}^n \bar d_{\mathsf Y^\star\!, 3}(x_i, 0)  &\leq \E{\bar d_{\mathsf Y^\star\!, 3}(\mathsf X, 0)} + \Delta  \label{eq_ATypd3}\\
 \overline V(x^n) &\geq \frac {\mu^{\prime\prime}}{2} \log e  \label{eq_ATypVupper}\\
 \underline V(x^n) &\leq \frac {3 \mu^{\prime\prime}} 2 \log e \label{eq_ATypVlower}
\end{align}
Let us first show that \eqref{eq_-ALambdaTyp} holds with $\delta$ given by \eqref{eq_ATypdelta} for all $x^n$ satisfying the conditions \eqref{eq_ATypdmin}--\eqref{eq_ATypd1b}. From \eqref{eq_ATypd1a} and \eqref{eq_ATypd1b}, 
\begin{equation}
\frac 1 n \sum_{i = 1}^n \bar d_{\mathsf Y^\star\!, 1}(x_i, \bar \lambda) < d < \frac 1 n \sum_{i = 1}^n \bar d_{\mathsf Y^\star\!, 1}(x_i, \underline \lambda)
\end{equation}
On the other hand, from \eqref{eq_optlambda} we have 
\begin{equation}
  d =  \frac 1 n \sum_{i = 1}^n \bar d_{\mathsf Y^\star, 1}(x_i, \lambda(x^n)) \label{eq_Adempirical}
\end{equation}
Therefore, since the right side of \eqref{eq_Adempirical} is decreasing (Property \ref{propertyELambda''}), 
\begin{equation}
\underline \lambda < \lambda(x^n) < \bar \lambda \label{eq_-Alambdabounds}
\end{equation}
Finally, an application Taylor's theorem to \eqref{eq_Alowerlambda} and \eqref{eq_Aupperlambda} using \eqref{eq_slopes} expands \eqref{eq_-Alambdabounds} as
\begin{equation}
 - \frac {3\Delta} 2 \mathbb R^{\prime\prime}_{\mathsf X, \mathsf Y^\star}( \bar d) + \lambda^\star < \lambda(x^n) < \lambda^{\star} + \frac {3\Delta} 2 \mathbb R^{\prime\prime}_{\mathsf X, \mathsf Y^\star}( \underline d) \label{eq_slopeempirical}
\end{equation}
for some $\bar d \in [d, d + \frac {3 \Delta} 2]$, $\underline d \in [d, d - \frac {3 \Delta} 2]$. Note that \eqref{eq_-ADeltaub}, \eqref{eq_ATypdmin} and \eqref{eq_ATypdmax} ensure that
\begin{equation}
 d_{\min \mid \mathsf X, \mathsf Y^\star} + 2 \Delta < d < d_{\max \mid \mathsf X, \mathsf Y^\star} - 2\Delta \label{eq_-Adbounds}
\end{equation}
so the derivatives in \eqref{eq_slopeempirical} exist and are positive by Remark \ref{remark:R''}. Therefore \eqref{eq_-ALambdaTyp} holds with $\delta$ given by \eqref{eq_ATypdelta}.

We are now ready to show that as long as $\Delta$ (and, therefore, $\delta$) is small enough, there exists a $K_1 \geq 0$ such that \eqref{eq_-APTyp} holds. H\"{o}lder's inequality and assumption \eqref{item:last} in Section \ref{ssec:2orderMain} imply that the third moments of the random variables involved in conditions \eqref{eq_ATypd1a}--\eqref{eq_ATypd3} are finite. By the Berry-Esseen inequality, the probability of violating these conditions is $\bigo{\frac 1 {\sqrt n}}$. To bound the probability of violating conditions \eqref{eq_ATypVupper} and \eqref{eq_ATypVlower}, 
observe that since $\Lambda_{\mathsf Y^\star}^{\prime\prime}(\mathsf X, \lambda)$ is dominated by integrable functions due to \eqref{eq_Lambda''dominated},  we have by Fatou's lemma and continuity of $\Lambda^{\prime\prime}_{\mathsf Y^\star}(\mathsf x, \cdot)$
\begin{align}
\mu^{\prime\prime} &\leq
 \liminf_{\delta \downarrow 0}\E{ \inf_{|\theta^\prime| \leq \delta} \left|\Lambda_{\mathsf Y^\star}^{\prime\prime}( \mathsf X, \lambda^\star + \theta^\prime)\right|  }\\
&\leq
 \limsup_{\delta \downarrow 0}\E{ \sup_{|\theta^\prime| \leq \delta} \left|\Lambda_{\mathsf Y^\star}^{\prime\prime}( \mathsf X, \lambda^\star + \theta^\prime)\right|  }\\
 &\leq \mu^{\prime\prime}\label{eq_-ALambdaContinuity}
\end{align}
Therefore, if $\delta$ is small enough,
\begin{equation}
\frac { 3 \mu^{\prime\prime}} 4 \log e  \leq \E{\underline V(X^n)} \leq \E{\overline V(X^n)} \leq \frac {5  \mu^{\prime\prime}} 4 \log e \label{eq_-AVbounds}
\end{equation}
The third absolute moments of $\overline V(X^n)$ and $\underline V(X^n)$ are finite by H\"{o}lder's inequality, \eqref{eq_Lambda''dominated} and assumption \eqref{item:last} in Section \ref{ssec:2orderMain}. Thus, the probability of violating conditions \eqref{eq_ATypVupper} and \eqref{eq_ATypVlower} is also $\bigo{\frac 1 {\sqrt n}}$. Now, \eqref{eq_-APTyp} follows via the union bound.

To complete the proof of Lemma \ref{lemma:aepA1}, it remains to show \eqref{eq_aepA1}. Toward this end, observe, recalling Properties \ref{propertyLambda''} and \ref{propertydk} that the corresponding moments in the Berry-Esseen theorem are given by
\begin{align}
 \mu(x^n) &=  \frac 1 n \sum_{i = 1}^n \E{ d(x_i, \hat{\mathsf Z}^\star ) | \hat { \mathsf X} = x_i }\\
 &=  \frac 1 n \sum_{i = 1}^n \bar d_{\mathsf Y^\star, 1}(x_i, \lambda(x^n))\\
 &= d\\
 V(x^n) 
&= \frac 1 n \sum_{i = 1}^n \left[\bar d_{\mathsf Y^\star, 2}(x_i, \lambda(x^n)) - \bar d_{\mathsf Y^\star, 1}^2(x_i, \lambda(x^n)) \right]\\
&= - \frac 1 n \sum_{i = 1}^n \Lambda^{\prime\prime} (x_i, \lambda(x^n)) \log e \\
 T(x^n) 
 &= \\
 \frac 1 n \sum_{i = 1}^n &~\E{ \left| d(x_i, \hat{\mathsf Z}^\star ) - \E{ d(x_i, \hat{\mathsf Z}^\star ) \mid \hat {\mathsf X} = x_i }\right|^3 \mid \hat {\mathsf X} = x_i } \notag\\
 &\leq \frac 8 n \sum_{i = 1}^n \E{ \left| d(x_i, \hat{\mathsf Z}^\star )\right|^3 \mid \hat {\mathsf  X} = x_i }\\
  &= \frac 8 n \sum \bar d_{\mathsf Y^\star\!, 3}(x_i, \lambda(x^n))\\
  &\leq \frac 8 n \sum \bar d_{\mathsf Y^\star\!, 3}(x_i, 0)
  \end{align}
Due to \eqref{eq_-ALambdaTyp}, \eqref{eq_ATypVupper} and \eqref{eq_ATypVlower}, 
$\frac{\mu^{\prime\prime} }{2} \log e \leq V(x^n) \leq \frac{3 \mu^{\prime\prime}}{2} \log e$
 as long as $x^n \in F_n$. Furthermore, 
\begin{equation}
T(x^n) \leq 8 \E{\bar d_{\mathsf Y^\star\!, 3}(\mathsf X, 0)} + 8 \Delta
\end{equation}
for such $x^n$ due to \eqref{eq_ATypd3}. Therefore, by the Berry-Esseen inequality we have for all $x^n \in F_n$:

\begin{align}
&~ \Prob{ nd - \tau < \sum_{i = 1}^n d(x_i, \hat Z^\star_i) \leq nd | \hat X^n = x^n}\\
  \geq&~ \frac 1 {\sqrt{2 \pi}} \int_{0}^{\frac \tau {\sqrt{n V(x^n)}}} e^{-\frac {u^2} 2 }du - \frac {12T(x^n)}{V^{\frac 3 2}(x^n)} \frac 1{\sqrt n}\\
 \geq&~ \left( \frac {\tau}{\sqrt{2 \pi V(x^n)}}e^{-\frac{\tau^2}{2 n V(x_n)}} - \frac {12T(x^n)}{V^{\frac 3 2}(x^n)} \right) \frac 1 {\sqrt n}\\
 \geq&~  \left( \frac {\tau}{\sqrt{3 \pi \mu^{\prime\prime} \log e}}e^{-\frac{\tau^2}{ n \mu^{\prime\prime} \log e}} -2 \bar B\right) \frac 1 {\sqrt n} \label{eq_AProblb}
\end{align}
where $\bar B = 96\sqrt 2   \frac {\E{\bar d_{\mathsf Y^\star\!, 3}(\mathsf {X}, 0)} + \Delta }{ \left(\mu^{\prime\prime} \log e\right)^{\frac 3 2}} $. 
The proof is complete upon observing that as long as $n$ is large enough, we can always choose $\tau > 0$ so that \eqref{eq_AProblb} is positive. 
\end{proof}

To upper-bound $ \sum_{i = 1}^n \Lambda_{\mathsf Y^\star}(x_i, \lambda(x^n))$ appearing in \eqref{eq_aepAnonasymptotic1}, we invoke the following result. 
\begin{lemma}
Assume that restrictions \eqref{item:first}-\eqref{item:last} in Section \ref{ssec:2orderMain} hold. There exist constants $n_0, K_2 > 0$ such that for $n \geq n_0$, 
\begin{align}
&~\Prob{\sum_{i = 1}^n \Lambda_{\mathsf Y^\star}(X_i, \lambda (X^n) ) \leq \sum_{i = 1}^n \Lambda_{\mathsf Y^\star}(X_i, \lambda^\star) + C_2 \log n} \notag\\
 >&~ 1 - \frac { K_2} {\sqrt n}  \label{eq_lemmaB}
\end{align}
where 
\begin{equation}
 C_2 = \frac{\Var{\Lambda^{\prime}_{\mathsf Y^\star}(\mathsf X, \lambda^\star) } }{ \E{\left| \Lambda^{\prime\prime}_{\mathsf Y^\star}(\mathsf X, \lambda^\star)\right|} \log e}
\end{equation}
\label{lemma:aepA2}
\end{lemma}

\begin{proof}
Using \eqref{eq_-ALambdaTyp}, we have for all $x_n \in F_n$,
\begin{align}
 &~\sum_{i = 1}^n \left[ \Lambda_{\mathsf Y^\star}(x_i, \lambda (x^n) ) - \Lambda_{\mathsf Y^\star}(x_i, \lambda^\star) \right] \notag\\
 =&~ \sup_{|\theta| < \delta}  \sum_{i = 1}^n \left[ \Lambda_{\mathsf Y^\star}(x_i, \lambda^\star + \theta) - \Lambda_{\mathsf Y^\star}(x_i, \lambda^\star) \right] \label{eq_-Asup}\\
 =&~ \sup_{|\theta| < \delta} \theta \sum_{i = 1}^n \Lambda_{\mathsf Y^\star}^\prime(x_i, \lambda^\star) + \frac {\theta^2}{2} \sum_{i = 1}^n \Lambda_{\mathsf Y^\star}^{\prime\prime}(x_i, \lambda^\star + \xi_n) \label{eq_-ATaylor}\\
 \leq&~ \sup_{|\theta| < \delta} \theta S^\prime(x^n) - \frac {\theta^2}{2} S^{\prime \prime}(x^n)  \label{eq_-A3}\\
 \leq&~ \frac{\left(S^\prime(x^n) \right)^2}{2S^{ \prime \prime} (x^n) }\label{eq_-A4}
\end{align}
where
\begin{itemize}
\item \eqref{eq_-Asup} is due to \eqref{eq_Lambdadual};
 \item \eqref{eq_-ATaylor} holds for some $|\xi_n| \leq \delta$ by Taylor's theorem;
 \item   in \eqref{eq_-A3} we denoted
\begin{align}
 S^\prime(x^n) &= \sum_{i = 1}^n \Lambda_{\mathsf Y^\star}^\prime(x_i, \lambda^\star)\\
 S^{\prime \prime}(x^n) &= - \sum_{i = 1}^n \inf_{|\theta^\prime| < \delta} \left|\Lambda_{\mathsf Y^\star}^{\prime\prime}(x_i, \lambda^\star + \theta^\prime)\right|
\end{align}
and used Property \ref{propertyLambda''}; 
\item in \eqref{eq_-A4} we maximized the quadratic equation in \eqref{eq_-A3} with respect to $\theta$.
\end{itemize}
Note that the reasoning leading to \eqref{eq_-A4} is due to \cite[proof of Theorem 3]{dembo1999asymptotics}.  We now proceed to upper-bound the ratio in the right side of \eqref{eq_-A4}.  Since $\E{ \bar d_{\mathsf Y^\star,1}(\mathsf X, 0) } < \infty$ by assumption \eqref{item:last} in Section \ref{ssec:2orderMain}, the differentiation in Property \ref{propertyoptlambda} can be brought inside the expectation by \eqref{eq_Lambda'dominated} and the dominated convergence theorem, so
\begin{equation}
\E{ \frac 1 n S^\prime(X^n)} =\E{\Lambda^\prime_{\mathsf Y^\star}(\mathsf X, \lambda^\star)} = 0
\end{equation}
Denote
\begin{align}
 V^\prime &= \Var{\Lambda^\prime_{\mathsf Y^\star}(\mathsf X, \lambda^\star)}\\
 T^{\prime} &= \E{ \left|\Lambda^\prime_{\mathsf Y^\star}(\mathsf X, \lambda^\star) - \E{\Lambda^\prime_{\mathsf Y^\star}(\mathsf X, \lambda^\star)} \right|^3}
\end{align}

If $V^\prime = 0$, there is nothing to prove as that means $S^{\prime}(X^n) = 0$ a.s. Otherwise, since \eqref{eq_Lambda'dominated} with H\"{o}lder's inequality  and assumption \eqref{item:last} in Section \ref{ssec:2orderMain} guarantee that $T^\prime$ is finite, the Berry-Esseen inequality \eqref{eq_BerryEsseen} implies
\begin{align}
 &~\Prob{ \left(S^{\prime}(X^n) \right)^2 > V^\prime  n \log_e n } \notag\\
  \leq&~ \frac {12 T^\prime}{V^{\prime \frac 3 2}\sqrt n} + 2Q\left(\sqrt{\log_e n}\right) \\
<&~ \left(\frac {12 T^\prime}{V^{\prime \frac 3 2}} + \sqrt{\frac {2}{\pi}}\frac 1 {\sqrt{\log_e n} }\right) \frac {1}{\sqrt n} \label{eq_-APSprime}\\
\leq&~\left( \frac {12 T^\prime}{V^{\prime \frac 3 2}} + \sqrt{\frac 2 {\pi\log_e 2}} \right)  \frac {1}{\sqrt n} \label{eq_-AK2prime1}\\
=& \frac {K_2^\prime}{\sqrt n}\label{eq_-AK2prime}
\end{align}
In \eqref{eq_-APSprime}, we used
\begin{equation}
 Q(t) < \frac {1}{\sqrt{2 \pi } t} e^{- \frac{t^2}{2}} \label{eq_Qub}
 \end{equation}
and \eqref{eq_-AK2prime1} obviously holds for $n \geq 2$. To treat $S^{\prime\prime}(X^n)$, observe that $S^{\prime\prime}(x^n) = n\underline V(x^n) \left( \log e\right)^{-1}$ (see \eqref{eq_Vlower}), so as before, the variance $V^{\prime\prime}$ and the third absolute moment $T^{\prime\prime}$ of 
$Z_i = \inf_{|\theta^\prime| \leq \delta}\left| \Lambda_{\mathsf Y^\star}^{\prime\prime}( X_i, \lambda^\star + \theta^\prime)\right|$
are finite, and $\E{Z_i} \geq \frac{3\mu^{\prime\prime}} 4 $ by \eqref{eq_-AVbounds},  where $\mu^{\prime\prime} > 0$ is defined in \eqref{eq_mu''}. 
If $V^{\prime\prime} = 0$, 
we have $Z_i > \frac{\mu^{\prime\prime}} 2$ almost surely. Otherwise, by the Berry-Esseen inequality \eqref{eq_BerryEsseen}, 
\begin{align}
 \Prob{ S^{\prime \prime}(X^n) < n \frac{\mu^{\prime\prime}}{2} } &\leq \left(\frac {6 T^{\prime\prime}}{V^{\prime\prime \frac 3 2 }} +\sqrt \frac {8V^{\prime\prime}}{\pi \mu^{\prime\prime 2}} e^{ - \frac{n \mu^{\prime\prime 2}}{32 V^{\prime\prime}}}\right)\frac 1 {\sqrt n} \label{eq_-APS2prime}\\
 &< \left(\frac {6 T^{\prime\prime}}{V^{\prime\prime \frac 3 2 }}  + \sqrt{\frac {8V^{\prime\prime}}{\pi \mu^{\prime\prime 2}}}\right)\frac 1 {\sqrt n}\\
 &= \frac{K_2^{\prime\prime}}{\sqrt n}\label{eq_-AK2primeprime}
\end{align}
where in \eqref{eq_-APS2prime} we used \eqref{eq_Qub}.
Finally, denoting
\begin{equation}
 g(x^n) = \sum_{i = 1}^n \Lambda_{\mathsf Y^\star}(x_i, \lambda (x^n) ) - \sum_{i = 1}^n \Lambda_{\mathsf Y^\star}(x_i, \lambda^\star) 
\end{equation}
and letting $G_n$ be the set of $x^n \in \mathcal A^n$ satisfying both
\begin{align}
 \left(S^{\prime}(x^n) \right)^2 &\leq V^\prime  n \log_e n\\
 S^{\prime \prime}(x^n) &\geq n \frac{\mu^{\prime\prime}}{2}
\end{align}
we see from \eqref{eq_-APTyp}, \eqref{eq_-AK2prime}, \eqref{eq_-AK2primeprime} applying elementary probability rules that
\begin{align}
&~\Prob{g(X^n) > C_2 \log n} \notag\\
=&~ \Prob{g(X^n) > C_2 \log n, \  g(X^n) \leq \frac{\left(S^\prime(X^n) \right)^2}{2S^{ \prime \prime} (X^n) }} \notag\\
+&~ \Prob{g(X^n) > C_2 \log n, \  g(X^n) > \frac{\left(S^\prime(X^n) \right)^2}{2S^{ \prime \prime} (X^n) }} \\
\leq&~ \Prob{ \frac{\left(S^\prime(X^n) \right)^2}{2S^{ \prime \prime} (X^n) } > C_2 \log n} 
+ \frac{K_1}{\sqrt n}\\
=&~\Prob{ \frac{\left(S^\prime(X^n) \right)^2}{2S^{ \prime \prime} (X^n) } > C_2 \log n, \ X^n \in G_n } \notag\\
+&~ \Prob{ \frac{\left(S^\prime(X^n) \right)^2}{2S^{ \prime \prime} (X^n) } > C_2 \log n, \ X^n \notin G_n } 
+ \frac{K_1}{\sqrt n}\\
<&~ 0 
+ \frac{K_2^\prime}{\sqrt n} + \frac{K_2^{\prime\prime}}{\sqrt n}
+ \frac{K_1}{\sqrt n} 
\end{align}
We conclude that \eqref{eq_lemmaB} holds for $n \geq n_0$ with $K_2 = K_1 + K_2^\prime + K_2^{\prime\prime}$.
\end{proof}
To apply Lemmas  \ref{lemma:aepA1} and \ref{lemma:aepA2} to \eqref{eq_aepAnonasymptotic1}, note that \eqref{eq_Ad>dmin} (and hence \eqref{eq_aepAnonasymptotic1}) holds for $x^n \in F_n$ due to \eqref{eq_-Adbounds}. Weakening \eqref{eq_aepAnonasymptotic1} using Lemmas  \ref{lemma:aepA1} and \ref{lemma:aepA2} and the union bound we conclude that Lemma \ref{lemma:aepA} holds with
\begin{align}
 C &= \frac 1 2 + C_2\\
 K &= K_1 + K_2\\
 c &= (\lambda^\star + \delta)\tau - \log C_1
\end{align}
\section{Proof of Theorem \ref{thm:D2order}}
\label{appx:D2order}
\begin{figure}
 \epsfig{file=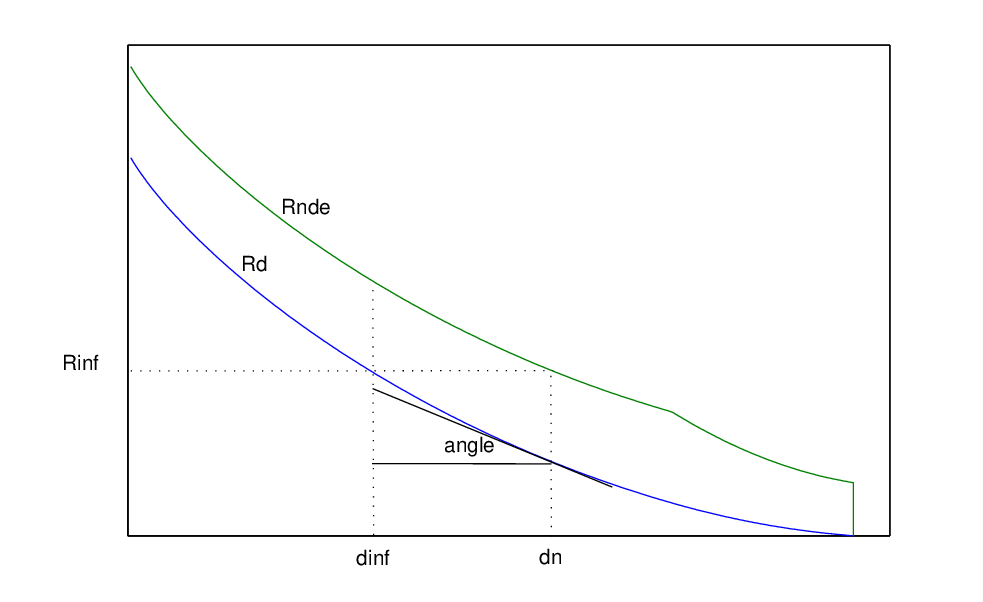,width=1\linewidth}
\caption{Estimating $d_n - d_\infty$ from $R(n, d, \epsilon) - R(d)$. }
\label{fig:illustration}
\end{figure}

In this appendix, we show that \eqref{eq_D2order} follows from \eqref{eq_2order}. Fix a point $(d_\infty, R_\infty) $ on the rate-distortion curve such that $d_\infty \in (\underline d, \bar d)$. Let $d_n = D(n,R_\infty,\epsilon)$, and let $\alpha$ be the acute angle between the tangent to the $R(d)$ curve at $d = d_n$ and the $d$ axis (see Fig. \ref{fig:illustration}). We are interested in the difference $d_n - d_\infty$. Since \cite{kieffer1991strong}
\beq
\lim_{n \to \infty} D(n, R, \epsilon) = D(R),
\eeq
there exists a $\delta > 0$ such that for large enough $n$,
\begin{equation}
 d_n \in \mathbb B_\delta(d_\infty) = [ d_\infty - \delta, d_\infty + \delta]  \subset (\underline d, \bar d)
\end{equation}
For such $d_n$, 
\begin{align}
|d_n - d_\infty| &\leq  \left| \frac {R(d_n) - R_\infty}{\tan \alpha_n} \right|\ \label{eq_-D2orderb1}\\
&\leq  \left| \frac {R(n,d_n,\epsilon) - R(d_n)}{ \min_{d \in \mathbb B_\delta(d_\infty)} R^\prime(d)} \right| \label{eq_-D2orderb2}\\
&= \bigo{\frac 1 {\sqrt n}} \label{eq_-D2orderb3}
\end{align}
where
\begin{itemize}
 \item \eqref{eq_-D2orderb1} is by convexity of $R(d)$;
 \item \eqref{eq_-D2orderb2} follows by substituting $R(n, d_n, \epsilon) = R_{\infty}$
and $\tan \alpha_n = |R^\prime(d_n)|$; 
\item \eqref{eq_-D2orderb3} follows by Theorem \ref{thm:2order}. Note that we are allowed to plug $d_n$ into \eqref{eq_2order}  because the remainder in \eqref{eq_2order} can be uniformly bounded over all $d$ from the compact set $\mathbb B_\delta(d_\infty)$ (just swap $B_n$ in \eqref{eq_-2orderCepsilon} for the maximum of $B_n$'s over $\mathbb B_\delta(d_\infty)$, and similarly swap $c, K, B_n$ in \eqref{eq_-2orderAR} and \eqref{eq_-2orderAepsilon} for the corresponding maxima); thus \eqref{eq_2order} holds not only for a fixed $d$ but also for any sequence $d_n \in \mathbb B_\delta(d_\infty)$. 
\end{itemize}
It remains to refine \eqref{eq_-D2orderb3} to show \eqref{eq_D2order}. Write
\begin{align}
V(d_n) &= V(d_\infty) + \bigo{\frac 1 {\sqrt n}} \label{eq_-D2orderc}\\
R(d_n) &= R(d_\infty) + R^\prime(d_\infty)(d_n - d_\infty) + \bigo{\frac 1 n}  \label{eq_-D2orderc1}\\
&= R(d_n) + \sqrt{\frac {V(d_n)} n} \Qinv{\epsilon}\notag\\
&+ R^\prime(d_\infty)(d_n - d_\infty) + \theta\left(\frac{\log n}{n}\right) \label{eq_-D2orderd} \\
&= R(d_n) + \sqrt{\frac {V(d_\infty)} n} \Qinv{\epsilon} \notag\\
&+ R^\prime(d_\infty)(d_n - d_\infty) + \theta\left(\frac{\log n}{n}\right)\label{eq_-D2ordere}
\end{align}
where 
\begin{itemize}
 \item \eqref{eq_-D2orderc} and \eqref{eq_-D2orderc1} follow by Taylor's theorem and \eqref{eq_-D2orderb3} using finiteness of $V^\prime(d)$ and $R^{\prime\prime}(d)$ for all $d \in \mathbb B_\delta(d_\infty) $;
 \item \eqref{eq_-D2orderd} expands $R_{\infty} = R(n, d_n, \epsilon)$ using \eqref{eq_2order};
 \item \eqref{eq_-D2ordere} invokes \eqref{eq_-D2orderc}.
\end{itemize}
Rearranging \eqref{eq_-D2ordere}, we obtain the desired approximation \eqref{eq_D2order} for the difference $d_n - d_\infty$. 

 \section{Proof of Theorem \ref{thm:2orderEBMS}}
\label{appx:2orderEBMS}
From the Stirling approximation, it follows that (e.g. \cite{gallager1968information})
\begin{align}
 &~\sqrt{\frac{n}{8k(n-k)}} \exp\left\{nh\left(\frac{k}{n}\right)\right\} \notag\\
 \leq &~{n \choose k} \label{eq_StirlingLower}\\
 \leq &~ \sqrt{\frac{n}{2 \pi k(n-k)}} \exp\left\{nh\left(\frac{k}{n}\right)\right\} \label{eq_StirlingUpper}
\end{align}
In view of the inequality
\begin{equation}
{n \choose k-j} \leq {n \choose k} \left( \frac{k}{n-k} \right)^j
\label{eq_binoineq}
\end{equation}
 we can write
 \begin{align}
{n \choose k } &\leq \binosum{n}{k} \label{eq_binosumLower}\\
&\leq {n \choose k } \sum_{j = 0}^\infty \left( \frac{k }{n - k}\right)^j\\
 &= {n \choose k} \frac{n - k}{n - 2k} \label{eq_binosumUpper}
 \end{align}
 where \eqref{eq_binosumUpper} holds as long as the series converges, i.e. as long as $2k < n$. Furthermore, combining \eqref{eq_binosumLower} and \eqref{eq_binosumUpper} with Stirling's approximation \eqref{eq_StirlingLower} and \eqref{eq_StirlingUpper}, we conclude that for any $0 < \alpha < \frac 1 2$,
 \begin{equation}
 \label{eq_BinosumAsymptotics}
 \log \binosum{n}{\lfloor n \alpha\rfloor} = nh\left(\alpha\right) - \frac 1 2 \log n + \bigo{1}
 \end{equation}
 Taking logarithms in \eqref{eq_CEBMS} and letting $\log M = nR$ for any $R \geq R(n, d, \epsilon)$, we obtain
\begin{align}
\log (1 - \epsilon) &\leq n(R - \log 2) + \log \binosum{n}{\lfloor nd \rfloor}\\
&\leq n(R - \log 2 + h(d)) - \frac 1 2 \log n + O\left( 1\right) \label{eq_-2orderCEBMS}
\end{align}
Since \eqref{eq_-2orderCEBMS} holds for any $R \geq R(n, d, \epsilon)$, we conclude that
\begin{equation}
R(n, d, \epsilon) \geq R(d) + \frac 1 2 \frac{\log n}{n} + O\left( \frac 1 n\right)
\end{equation}
Similarly, Corollary \ref{thm:AEBMS} implies that there exists an $(\exp(nR), d , \epsilon)$ code with
\begin{align}
\log \epsilon &\leq \exp\left(nR\right) \log \left(1 - \frac{\binosum{n}{\lfloor nd \rfloor}}{2^n}\right) \\
&\leq - \exp\left(nR\right) \frac{\binosum{n}{\lfloor nd \rfloor}}{2^n } \log e\label{eq_-2orderAEBMS}
\end{align}
where we used $\log (1 + x) \leq x \log e$, $x > -1$. Taking the logarithm of the negative of both sides in \eqref{eq_-2orderAEBMS}, we have
\begin{align}
\log \log \frac{1}{\epsilon} &\geq n(R - \log 2) + \log \binosum{n}{\lfloor nd \rfloor} + \log \log e\\
&= n(R - \log 2 + h(d)) - \frac 1 2 \log n  +  O\left( 1 \right)  \label{eq_-2orderAEBMS-a},
\end{align}
where \eqref{eq_-2orderAEBMS-a} follows from \eqref{eq_BinosumAsymptotics}. Therefore,
\begin{equation}
R(n, d, \epsilon) \leq R(d) + \frac 1 2 \frac{\log n}{n} + O\left( \frac 1 n\right)
\end{equation}
The case $d = 0$ follows directly from \eqref{eq_2orderLossless0}.  Alternatively, it can be easily checked by substituting $\binosum{n}{0} = 1$ in the analysis above.

\section{Gaussian approximation\\ of the bound in Theorem \ref{thm:ABMS'}}
\label{appx:2orderABMS}
By analyzing the asymptotic behavior of \eqref{eq_ABMS'}, we prove that
\begin{align}
R(n, d, \epsilon ) &\leq h(p) - h(d) + \sqrt{\frac{V(d)}{n}} \Qinv{\epsilon} \notag\\
&+ \frac 12 \frac {\log n}{n} + \frac {\log \log n}{n} + \bigo{\frac 1 n} \label{eq_2orderABMS}
\end{align}
where $V(d)$ is as in \eqref{eq_dispersionBMS}, thereby showing that a constant composition code that attains the rate-dispersion function exists. Letting $M = \exp\left(nR\right)$ and using $(1-x)^M \leq e^{-Mx}$ in \eqref{eq_ABMS'}, we can guarantee existence of an $(n, M, d, \epsilon^\prime)$ code with
\begin{equation}
 \epsilon^\prime \leq \sum_{k = 0}^n{  n \choose k } p^{k} (1 - p)^{n - k} e^{- {n \choose \lceil nq \rceil }^{-1}L_n(k, \lceil nq \rceil )\exp\left(nR\right)} \label{eq_ABMS''}
\end{equation}
In what follows we will show that one can choose an $R$ satisfying the right side of \eqref{eq_2orderABMS} so that the right side of \eqref{eq_ABMS''}  is upper bounded by $\epsilon$ when $n$ is large enough.
Letting $k  = np + n\Delta$, $t = \lceil nq \rceil $, $t_0 = \lceil \frac {\lceil nq \rceil + k - nd}{2} \rceil^+$ and using Stirling's formula \eqref{eq_StirlingLower}, it is an algebraic exercise to show that there exist positive constants $\delta$ and $C$ such that for all $\Delta \in [- \delta, \delta]$, 
 \begin{align}
{n \choose t }^{-1} {k \choose t_0} {n - k \choose t - t_0} 
&=  {n \choose k }^{-1} {t \choose t_0} {n - t \choose k - t_0} \\
&\geq \frac {C} {\sqrt{n}} \exp\left\{n g(\Delta)\right\} \label{eq_-ABMSrho}
\end{align}
where
\begin{equation}
g(\Delta) = h(p + \Delta) - qh\left( d - \frac \Delta {2q} \right) - (1 - q)h\left( d + \frac \Delta{2(1-q)}\right) \notag
\end{equation}
It follows that
\begin{equation}
{n \choose \lceil nq \rceil }^{-1} L_n(np + n\Delta, \lceil qn \rceil) \geq \frac {C} {\sqrt{n}} \exp\left\{-n g(\Delta)  \right\}
\end{equation}
whenever $L_n(k, \lceil qn \rceil)$ is nonzero, that is, whenever $\lceil nq \rceil - nd \leq k \leq \lceil nq \rceil + nd$,  and $g(\Delta) = 0$ otherwise.

Applying a Taylor series expansion in the vicinity of $\Delta = 0$ to $g(\Delta)$, we get
\begin{equation}
\label{eq_-2orderABMSg}
g(\Delta) = h(p) - h(d) + h^\prime(p) \Delta + \bigo{\Delta^2}
\end{equation}
Since $g(\Delta)$ is continuously differentiable with $g^\prime(0) = h^\prime(p) > 0$, there exist constants $\underline{b}, \bar b > 0$  such that $g(\Delta)$ is monotonically increasing on $(-\underline{b}, \bar b)$ and \eqref{eq_-ABMSrho} holds.
Let
\begin{align}
b_n &= \sqrt{\frac{p(1-p)}{n}} Q^{-1}\left( \epsilon_n\right) \label{eq_ABMSbn}\\
\epsilon_n &= \epsilon - \frac{2B_n}{\sqrt n} - \sqrt{\frac{V(d)}{2\pi n}}\frac{1}{\bar b} e^{-n \frac{{\bar b}^2}{2V(d)}}
 - \frac 1 {\sqrt n}\\
B_n &= 6 \frac{1 - 2p + 2p^2}{\sqrt{p(1-p)}}\\
R &= g(b_n)  + \frac 1 2 \frac {\log n}{n} + \frac 1 n \log \left( \frac {\log_e n}{2C}\right)\label{eq_ABMSRchosen}
\end{align}
Using \eqref{eq_-2orderABMSg} and applying a Taylor series expansion to $\Qinv{\cdot}$, it is easy to see that $R$ in \eqref{eq_ABMSRchosen} can be rewritten as the right side of \eqref{eq_2orderABMS}. Splitting the sum in \eqref{eq_ABMS''} into three sums and upper bounding each of them separately, we have
\begin{align}
\label{eq_ABMSsplit}
&~\sum_{k = 0}^n {  n \choose k } p^{k} (1 - p)^{n - k} e^{-{n \choose \lceil qn \rceil }^{-1}L_n(k, \lceil qn \rceil ) \exp\left(nR\right)} \notag\\
=&~  \sum_{k = 0}^{ \lfloor np - n\underline{b} \rfloor} +  \sum_{k =  \lfloor np - n\underline{b} \rfloor + 1}^{ \lfloor n p + n b_n  \rfloor } +   \sum_{k = \lfloor n p + n b_n  \rfloor + 1}^n\\
\leq&~ \Prob{\sum_{i = 1}^n X_i \leq np - n \underline{b}} \notag \\
 +&~ \sum_{k =  \lfloor np - n\underline{b} \rfloor + 1}^{ \lfloor n p + n b_n  \rfloor } {  n \choose k } p^{k} (1 - p)^{n - k}e^{ - \frac C {\sqrt n}\exp\left\{nR - n g\left( \frac k n - p\right)\right\}} \notag \\
 +&~ \Prob{\sum_{i = 1}^n X_i \geq np + n b_n}  \label{eq_-2orderABMS-a} \\
 \leq&~ \frac{B_n}{\sqrt n} + \sqrt{\frac{V(d)}{2\pi n}}\frac{1}{\bar b} e^{-n \frac{{\bar b}^2}{2V(d)}}
 + \frac 1 {\sqrt n}
 + \epsilon_n + \frac{B_n}{\sqrt n}\\
 =&~ \epsilon
\end{align}
 where $\{X_i\}$ are i.i.d. Bernoulli random variables with bias $p$. The first and third probabilities in the right side of \eqref{eq_-2orderABMS-a} are bounded using the Berry-Esseen bound \eqref{eq_BerryEsseen} and \eqref{eq_Qub}, while the second probability is bounded using the monotonicity of $g(\Delta)$ in $(-\underline b, b_n]$ for large enough $n$, in which case the minimum difference between $R$ and $g(\Delta)$ in $(-\underline b, b_n)$ is $\frac 1 2 \frac {\log n}{n} + \frac 1 n \log \left( \frac {\log_e n}{2C}\right)$.

\section{Proof of Theorem \ref{thm:2orderEDMS}}
\label{appx:2orderEDMS}
\begin{table*}[!b]
\normalsize
\setcounter{mytempeqncnt}{\value{equation}}
\setcounter{equation}{399}
\vspace*{4pt}
\hrulefill
\begin{align}
{n \choose \mathbf k} &\leq C_1 n^{-\frac{m-1}{2}} \exp ~ n\left\{ H(\mathsf X)+ \sum_{a = 1}^{m}\Delta_a \log\frac{1}{P_{\mathsf X}(a)} + \bigo{|\mathbf \Delta|^2} \right\}
\label{eq_multiszpankowski}\\
  {t_b^\star \choose \mathbf k_b } &\geq C_2 n^{-\frac {m-1}{2}} \exp n \left\{  P_{\mathsf Y}^\star(b) H\left( \mathsf X| \mathsf Y^\star = b \right)+ \sum_{a = 1}^m \delta(a,b) \log \frac 1 {P_{\mathsf X | \mathsf Y}^\star (a|b)}+ \bigo{|\mathbf \Delta|^2} \right\}  \label{eq_-multia}
\end{align}
\setcounter{equation}{\value{mytempeqncnt}}
\end{table*}
In order to study the asymptotics of \eqref{eq_CEDMS} and \eqref{eq_AEDMS}, we need to analyze the asymptotic behavior of $\hammingsum{n}{\lfloor nd \rfloor}$ which can be carried out similarly to the binary case. Recalling the inequality \eqref{eq_binoineq}, we have
\begin{align}
\hammingsum{n}{k} &= \sum_{j = 0}^{k} {n \choose j} (m-1)^{j}\\
&\leq {n \choose k} \sum_{j = 0}^{k} \left( \frac{k}{n-k}\right)^j (m-1)^{k-j}\\
&\leq {n \choose k}(m-1)^{k}\sum_{j = 0}^{\infty} \left( \frac{k}{(n-k)(m-1)}\right)^j\\
&= {n \choose k}(m-1)^{k} \frac{n-k}{n-k\frac m {m-1}} \label{eq_HammingSumUpper}
\end{align}
where \eqref{eq_HammingSumUpper} holds as long as the series converges, i.e. as long as $\frac {k}{n} < \frac {m-1}{m}$.
Using
\begin{align}
\hammingsum{n}{k} \geq {n \choose k} (m-1)^{k} \label{eq_HammingSumLower}
\end{align}
and applying Stirling's approximation \eqref{eq_StirlingLower} and \eqref{eq_StirlingUpper}, we have for $0 < d < \frac {m-1}{m}$
\begin{align}
\log \hammingsum{n}{\lfloor nd \rfloor} = \log {n \choose \lfloor nd \rfloor} + nd \log (m-1) + O(1) \\
= nh(d) + nd \log (m-1) - \frac 1 2 \log n+ O(1) \label{eq_HammingSumAsymptotics}
\end{align}

 Taking logarithms in \eqref{eq_CEDMS} and letting $\log M = nR$ for any $R \geq R(n, d, \epsilon)$, we obtain
\begin{align}
\log (1 - \epsilon) &\leq n(R - \log m) + \log \hammingsum{n}{\lfloor nd \rfloor}\\
&\leq n(R - \log m + h(d) + d \log (m-1)) \notag\\
& - \frac 1 2 \log n + O\left( 1\right) \label{eq_-2orderCEDMS}
\end{align}
Since \eqref{eq_-2orderCEDMS} holds for any $R \geq R(n, d, \epsilon)$, we conclude that
\begin{equation}
R(n, d, \epsilon) \geq R(d) + \frac 1 2 \frac{\log n}{n} + O\left( \frac 1 n\right)
\end{equation}
Similarly, Theorem \ref{thm:AEDMS} implies that there exists an $(\exp(nR), d , \epsilon)$ code with
\begin{align}
\log \epsilon &\leq \exp\left(nR\right) \log \left(1 - \frac{\hammingsum{n}{\lfloor nd \rfloor}}{m^n}\right) \\
&\leq - \exp\left(nR\right) \frac{\hammingsum{n}{\lfloor nd \rfloor}}{m^n } \log e\label{eq_-2orderAEDMS}
\end{align}
where we used $\log (1 + x) \leq x \log e$, $x > -1$. Taking the logarithm of the negative of both sides of \eqref{eq_-2orderAEDMS}, we have
\begin{align}
\log \log \frac{1}{\epsilon} &\geq n(R - \log m) + \log \hammingsum{n}{\lfloor nd \rfloor} + \log \log e\\
&= n(R - \log m + h(d)) - \frac 1 2 \log n  +  O\left( 1 \right)  \label{eq_-2orderAEDMS-a},
\end{align}
where \eqref{eq_-2orderAEDMS-a} follows from \eqref{eq_HammingSumAsymptotics}. Therefore,
\begin{equation}
R(n, d, \epsilon) \leq R(d) + \frac 1 2 \frac{\log n}{n} + O\left( \frac 1 n\right)
\end{equation}
The case $d = 0$ follows directly from \eqref{eq_2orderLossless0}, or can be obtained by observing that $\hammingsum{n}{0} = 1$ in the analysis above. 

\section{Gaussian approximation \\of the bound in Theorem \ref{thm:ADMS}}
\label{appx:2orderADMS}
Using Theorem \ref{thm:ADMS}, we show that
\begin{align}
R(n, d, \epsilon ) &\leq R(d) + \sqrt{\frac{V(d)}{n}} \Qinv{\epsilon}\label{eq_2orderADMS} \\
&+ \frac {(m-1)(m_\eta - 1)}2 \frac {\log n}{n} + \frac {\log \log n}{n} + \bigo{\frac 1 n} \notag
\end{align}
where $m_\eta$ is defined in \eqref{eq_DMSmeta}, and $V(d)$ is as in \eqref{eq_dispersionDMS}. Similar to the binary case, we express $L_n( \mathbf k, \mathbf t^\star)$ in terms of the rate-distortion function. Observe that whenever $L_n( \mathbf k, \mathbf t^\star)$ is nonzero,
\begin{align}
 {n \choose \mathbf t^\star}^{-1} L_n( \mathbf k, \mathbf t^\star) &=   {n \choose \mathbf t^\star}^{-1} \prod_{a = 1}^m {k_a \choose \mathbf t_a} \\ 
 &=  {n \choose \mathbf k}^{-1} \prod_{a = 1}^{m_\eta} {t_b^\star \choose \mathbf k_b}
\end{align}
where $\mathbf k_b = \left( t_{1, b}, \ldots, t_{m, b} \right)$. It can be shown \cite{szpankowski2009minimum} that for $n$ large enough, there exist positive constants $C_1, C_2$ such that \eqref{eq_multiszpankowski} and \eqref{eq_-multia} at the bottom of the page hold for small enough $|\mathbf \Delta|$, where $\mathbf \Delta = (\Delta_1, \ldots, \Delta_m)$. A simple calculation using $\sum_{a = 1}^m \Delta_a = 0$ reveals that
\addtocounter{equation}{2}
\begin{align}
 &~\sum_{a = 1}^m \sum_{b = 1}^{m_\eta} \delta(a, b) \log \frac 1{P_{\mathsf X| \mathsf Y}^\star (a|b)} \notag\\
 =&~  \sum_{a = 1}^{m_\eta} \Delta_a \log \frac 1 \eta + \sum_{a = m_\eta + 1}^m \Delta_a \log \frac 1 {P_{\mathsf X}(a)}
\end{align}
so invoking \eqref{eq_multiszpankowski} and \eqref{eq_-multia} one can write
\begin{equation}
{n \choose \mathbf k}^{-1} \prod_{a = 1}^{m_\eta} {t_b^\star \choose \mathbf k_b}  \geq  C n^{-\frac {(m-1)(m_\eta - 1)}{2}} \exp\left\{ -n g(\mathbf \Delta)\right\} \label{eq_-rhoDMS}
\end{equation}
where $C$ is a constant, and $g(\mathbf \Delta)$ is a twice differentiable function that satisfies
\begin{align}
g(\mathbf \Delta) &= R(d) +\sum_{a = 1}^{m} \Delta_a v(a) + \bigo{|\mathbf \Delta|^2}\label{eq_-2orderADMSg}\\
v(a) &= \min\left\{\imath_{X}(a), \log \frac 1 \eta\right\} 
\end{align}
Similar to the BMS case, $g(\mathbf \Delta)$ is monotonic in $\sum_{a = 1}^m \Delta_a v(a) \in (-\underline{b}, \bar b)$ for some constants $\underline{b}, \bar b > 0$ independent of $n$. Let
\begin{align}
b_n &= \sqrt{\frac{V(d)}{n}} \Qinv{\epsilon} \\
\epsilon_n&= \epsilon -  \frac {2B_n} {\sqrt n} - \frac 1 {\sqrt n} - \sqrt{\frac{V(d)}{2\pi n}}\frac{1}{\bar b} e^{-n \frac{{\bar b}^2}{2V(d)}} \\
R &= \max_{\substack{ \mathbf \Delta: \\ \sum_{a=1}^m \Delta_a v(a)  \in  (-\underline b, b_n]} } g(\mathbf \Delta) \notag\\
&+  \frac{(m-1)(m_\eta - 1)}{2}\frac{\log n}{n} + \frac 1 n \log \left( \frac {\log_e n}{2C}\right)  \label{eq_-RDMS}
\end{align}
where $B_n$ is the finite constant defined in \eqref{eq_BerryEsseenBn}. Using \eqref{eq_-2orderADMSg} and applying a Taylor series expansion to $\Qinv{\cdot}$, it is easy to see that $R$ in \eqref{eq_-RDMS} can be rewritten as the right side of \eqref{eq_2orderADMS}. Further, we use $nR = \log M$ and $(1 - x)^M \leq e^{-Mx}$ to weaken the right side of \eqref{eq_ADMS} to obtain
\begin{align}
& \sum_{\mathbf \Delta} {n \choose n(\mathbf p + \mathbf \Delta)} p^{n (\mathbf p + \mathbf \Delta)} e^{-  {n \choose \mathbf t^\star}^{-1} L_n( n(\mathbf p + \mathbf \Delta), \mathbf t^\star)  \exp\left(nR\right) }\notag \\
= & \sum_{\substack{ \mathbf \Delta: \\ \sum_{a=1}^m \Delta_a v(a) \leq -\underline b}}  + \sum_{\substack{ \mathbf \Delta: \\\sum_{a=1}^m \Delta_a v(a) \in (-\underline b, b_n)}}  + \sum_{\substack{ \mathbf \Delta: \\\sum_{a=1}^m \Delta_a v(a) \geq b_n}} \label{eq_-2orderADMS-a}\\
&\leq  \Prob{ \sum_{k = 1}^n v(X_k) \leq \E{v(\mathsf X)}- \underline b} \notag \\
&+\sup_{ \substack{ \mathbf \Delta: \\  \sum_{a=1}^m \Delta_a v(a)  \in  (-\underline b, b_n)}} e^{-  C n^{- \frac {(m-1)(m_\eta - 1)} 2 }\exp ~n\{ R -g(\mathbf \Delta) \} }\notag\\
&+ \Prob{ \sum_{k = 1}^n v(X_k) \geq \E{v(\mathsf X)}+ \underline b_n} \\
&\leq \frac {B_n} {\sqrt n} + \sqrt{\frac{V(d)}{2\pi n}}\frac{1}{\bar b} e^{-n \frac{{\bar b}^2}{2V(d)}} 
+ \frac 1 {\sqrt n} 
+  \epsilon_n +  \frac {B_n} {\sqrt n} 
\end{align}
where $P_{X_k}(a) = P_{\mathsf X}(a)$. The first and third probabilities in \eqref{eq_-2orderADMS-a} are bounded using the Berry-Esseen bound \eqref{eq_BerryEsseen} and \eqref{eq_Qub}. The middle probability is bounded by observing that the difference between $R$ and $g(\mathbf \Delta)$ in $\sum_{a=1}^m \Delta_a v(a) \in (-\underline b, b_n)$ is at least $\frac {(m -1)(m_\eta - 1)} 2 \frac {\log n}{n} + \frac 1 n \log \left( \frac {\log_e n}{2C}\right)$.

\section{Proof of Theorem \ref{thm:2orderBES} }
\label{appx:2orderBES}

\begin{proof}[Converse] The proof of the converse part follows the Gaussian approximation analysis of the converse bound in Theorem \ref{thm:CBES}. 
 Let $j = n\frac \delta 2 + n\Delta_1$ and $k = n\delta - n\Delta_2$. Using Stirling's approximation for the binomial sum \eqref{eq_BinosumAsymptotics}, after applying a Taylor series expansion we have
\begin{equation}
2^{-(n-k)}\binosum{n-k}{\lfloor nd - j \rfloor} = \frac {C(\mathbf \Delta)}{\sqrt n}\exp\left\{ - n\ g(\Delta_1, \Delta_2)\right\} \label{eq_-2orderBES}
\end{equation}
where $C(\mathbf \Delta)$ is such that there exist positive constants $\underline C$, $\bar C$, $\xi$ such that $\underline C \leq C(\mathbf \Delta) \leq \bar C$ for all $|\mathbf \Delta| \leq \xi$, and the twice differentiable function $g(\Delta_1, \Delta_2)$ can be written as
\begin{align}
g(\Delta_1, \Delta_2) &= R(d) + a_1\Delta_1  + a_2\Delta_2 + \bigo{|\mathbf \Delta|^2} \label{eq_-2orderBESg}\\
a_1 &= \log \frac{1 - d - \frac \delta 2 }{d - \frac \delta 2} = \lambda^\star\\
a_2 &= \log \frac {2\left(1 - d - \frac \delta 2\right)}{1 - \delta} = \log \frac 2 {1 + \exp( - \lambda^\star)}
\end{align}
It follows from \eqref{eq_-2orderBESg} that $g(\Delta_1, \Delta_2)$ is increasing in $a_1\Delta_1  + a_2\Delta_2 \in (-\underline{b},\bar b)$ for some constants $\underline{b}, \bar b >0$ (obviously, we can choose $\underline{b}, \bar b$ small enough in order for $\underline C \leq C(\mathbf \Delta) \leq \bar C$ to hold). In the sequel, we will represent the probabilities in the right side of \eqref{eq_CBES} via a sequence of i.i.d. random variables $Z_1, \ldots, Z_n$ with common distribution
\begin{equation}
\mathsf Z = \begin{cases}
a_1 & \text{ w.p. } \frac \delta 2\\
a_2 & \text{ w.p. } 1 - \delta\\
0 & \text{ otherwise}
\end{cases} \label{eq_-2orderBESZ}
\end{equation}
Note that
\begin{align}
\E{\mathsf Z} &= \frac{a_1 \delta}{2} + a_2(1-\delta)\\
\Var{\mathsf Z} &= \delta(1-\delta)\left(a_2 - \frac {a_1} 2\right)^2 + \frac{\delta a_1^2}{4} = V(d)
\end{align}
and the third central moment of $\mathsf Z$ is finite, so that $B_n$ in \eqref{eq_BerryEsseenBn} is a finite constant.
Let
\begin{align}
b_n &= \sqrt{\frac{V(d)}{n}}\Qinv{\epsilon_n}\\
\epsilon_n &= \left(1 - \frac {\bar C}{\sqrt n} \right)^{-1} \epsilon  + \frac{2 B_n}{\sqrt{n}} +  \sqrt{\frac{V(d)}{2\pi n}}\frac{1}{\bar b} e^{-n \frac{{\bar b}^2}{2V(d)}}  \label{eq_-2orderCBESepsn}\\
R &= \min_{\substack{\Delta_1, \ \Delta_2: \\ b_n \leq  a_1 \Delta_1 + a_2 \Delta_2 \leq \bar b}} g(\Delta_1, \Delta_2) \label{eq_-2orderCBES} \\
&= R(d) + b_n  + \bigo{b_n^2}
\end{align}
With $M = \exp\left(nR\right)$, since $R \leq g(\Delta_1, \Delta_2)$ for all $a_1\Delta_1  + a_2\Delta_2 \in [b_n, \bar b]$, for such $(\Delta_1, \Delta_2)$ it holds that
\begin{equation}
\left[1 - \frac {\bar C}{\sqrt n}M\exp\left\{ - n\ g(\Delta_1, \Delta_2)\right\} \right]^+ \geq 1 - \frac {\bar C}{\sqrt n} \label{eq_-2orderCBESa}
\end{equation}
Denoting the random variables
\begin{align}
N(x) &= \frac 1 n \sum_{i = 1}^n 1\{Z_i = x\}\\
G_n &= n \ g\left(N(a_1) -  \frac \delta 2, N(a_2) - 1 + \delta \right) \label{eq_-2orderCBESnotation}
\end{align}
and using \eqref{eq_-2orderBES} to express the probability in the right side of \eqref{eq_CBES} in terms of $Z_1, \ldots, Z_n$, we conclude that the excess-distortion probability is lower bounded by
\begin{align}
 &~ \E{\left( 1-\frac {\bar C}{\sqrt n}\exp\left\{ \log M - G_n \right\} \right)^+} \notag\\
\geq &~ \left(1 - \frac {\bar C}{\sqrt n} \right)\Prob{b_n \leq \sum_{i = 1}^n Z_i - n \E{\mathsf Z} < \bar b } \label{eq_-2orderCBESb} \\
\geq &~ \left(1 - \frac {\bar C}{\sqrt n} \right) \left( \epsilon_n  - \frac{2B_n}{\sqrt{n}} -  \sqrt{\frac{V(d)}{2\pi n}}\frac{1}{\bar b} e^{-n \frac{{\bar b}^2}{2V(d)}}\right) \label{eq_-2orderCBESc}\\
= &~  \epsilon \label{eq_2orderCBESd}
\end{align}
where \eqref{eq_-2orderCBESb} follows from \eqref{eq_-2orderCBESa}, and \eqref{eq_-2orderCBESc} follows from the Berry-Esseen inequality \eqref{eq_BerryEsseen} and \eqref{eq_Qub}, and \eqref{eq_2orderCBESd} is equivalent to \eqref{eq_-2orderCBESepsn}. 
\end{proof}

\begin{proof}[Achievability] We now proceed to the Gaussian approximation analysis of the achievability bound in Theorem \ref{thm:ABES}.
 Let
\begin{align}
b_n &= \sqrt{\frac{V(d)}{n}}\Qinv{\epsilon_n}\\
\epsilon_n &= \epsilon - \frac{2B_n}{\sqrt{n}} - \sqrt{\frac{V(d)}{2\pi n}}\frac{1}{\bar b} e^{-n \frac{{\bar b}^2}{2V(d)}} - \frac 1 {\sqrt n} \label{eq_-2orderABESepsn}\\
\log M  &= n \min_{\substack{\Delta_1, \ \Delta_2: \\ b_n \leq a_1 \Delta_1 + a_2 \Delta_2 \leq \bar b}} g(\Delta_1, \Delta_2) \notag\\
&+\frac 1 2 \log n + \log \left( \frac {\log_e n}{2\underline C}\right)\\
& = n R(d) + \sqrt{nV(d)} \Qinv{\epsilon}\notag \\
&+ \frac 1 2 \log n +  \log \log n  + \bigo{1} \label{eq_-2orderABESa}
\end{align}
where $g(\Delta_1, \Delta_2)$ is defined in \eqref{eq_-2orderBES}, and \eqref{eq_-2orderABESa} follows from \eqref{eq_-2orderBESg} and a Taylor series expansion of $\Qinv{\cdot}$.
Using \eqref{eq_-2orderBES} and $(1 - x)^M \leq e^{-Mx}$ to weaken the right side of \eqref{eq_ABES} and expressing the resulting probability in terms of i.i.d. random variables $Z_1, \ldots, Z_n$  with common distribution \eqref{eq_-2orderBESZ}, we conclude that the excess-distortion probability is upper bounded by (recall notation \eqref{eq_-2orderCBESnotation})
\begin{align}
&~\E{e^{-\frac {\underline C}{\sqrt n}\exp\left\{ \log M - G_n \right\}} } \notag\\
\leq &~ \Prob{\sum_{i = 1}^n Z_i \geq n\E{\mathsf Z} + n b_n} + \Prob{\sum_{i = 1}^n Z_i \leq n\E{\mathsf Z} - n \underline b} \notag\\
+&~ \E{e^{-\frac {\underline C}{\sqrt n}\exp\left\{ \log M - G_n \right\}}1\left\{n\underline b <\sum_{i = 1}^n Z_i  - n\E{\mathsf Z} <  n b_n\right\} } \\
\leq&~ \epsilon_n + \frac{B_n}{\sqrt{n}} + \frac{B_n}{\sqrt{n}} + \sqrt{\frac{V(d)}{2\pi n}}\frac{1}{\bar b} e^{-n \frac{{\bar b}^2}{2V(d)}} + \frac 1 {\sqrt n}\\
 =&~ \epsilon  \label{eq_-2orderABESb}
\end{align}
where the probabilities are upper bounded by the Berry-Esseen inequality \eqref{eq_BerryEsseen} and \eqref{eq_Qub}, and the expectation is bounded using the fact that in $\underline b < a_1 \Delta_1 + a_2 \Delta_2 < b_n$, the minimum difference between $\log M$ and $n \ g(\Delta_1, \Delta_2)$ is $\frac 1 2 \log n + \log \left( \frac {\log_e n}{2 \underline C}\right)$. Finally, \eqref{eq_-2orderABESb} is just \eqref{eq_-2orderABESepsn}. 
\end{proof}

\section{Gaussian approximation\\ of the bound in Theorem \ref{thm:AGMS}}
\label{appx:2orderAGMS}
Using Theorem \ref{thm:AGMS}, we show that $R(n,d,\epsilon)$ does not exceed the right-hand side of \eqref{eq_2orderGMS} with the remainder satisfying \eqref{eq_2orderAGMS}. Since the excess-distortion probability in \eqref{eq_AGMS} depends on $\sigma^2$ only through the ratio $\frac{d}{\sigma^2}$, for simplicity we let $\sigma^2 = 1$.
Using inequality $(1-x)^M \leq e^{-Mx}$, the right side of \eqref{eq_AGMS} can be upper bounded by
\begin{equation}
\int_{0}^{\infty} e^{-\rho(n,z)\exp\left(nR\right)} f_{\chi^2_n}\left(n z\right) n\ dz,
 \label{eq_AGMScor}
\end{equation}
From Stirling's approximation for the Gamma function
\begin{equation}
\Gamma \left( x \right) = \sqrt{\frac{2 \pi }{x}} \left( \frac{x}{e} \right)^x \left( 1 + \bigo{\frac 1 x}\right)
\end{equation}
it follows that
\begin{equation}
\frac{\Gamma \left( \frac n 2 + 1\right)}{\sqrt \pi n \Gamma \left( \frac{n-1}{2} + 1\right) } = \frac{1}{\sqrt{2 \pi n}}\left( 1 + O\left( \frac{1}{n}\right)\right),
\end{equation}
which is clearly lower bounded by $\frac{1}{2 \sqrt{\pi n}}$ when $n$ is large enough.
This implies that for all $a^2 \leq z \leq b^2$ and all $n$ large enough
\begin{align}
\label{eq_AGMScond}
\rho(n,z) &\geq \frac{1}{2 \sqrt{\pi n}} \exp\left\{  (n-1) \log \left( 1 - g(z) \right) ^{\frac 1 2} \right\}
\end{align}
where
\begin{equation}
g(z) = \frac{\left(1 + z - 2d\right)^2}{4 \left(1 - d\right) z}
\end{equation}
It is easy to check that $g(z)$ attains its global minimum at $z = [1 - 2d]^+$ and is monotonically increasing for $z > [1 - 2d]^+$. Let
\begin{align}
b_n &= \sqrt \frac{2}{n}\Qinv{\epsilon_n}\\
\epsilon_n &= \epsilon - \frac{2B_n}{\sqrt{n}} - \frac 1 {\sqrt n} - \frac{1}{4d \sqrt{\pi n}} e^{-{2d}^2 n}\\
R &=  -\frac 1 2 \log \left( 1 - g(1 +b_n) \right) + \frac 1 2 \frac{\log n}{n} + \frac 1 n \log \left( \sqrt \pi \log_e n\right)\label{eq_-AGMSRchosen}
\end{align}
where $B_n = 12 \sqrt 2$. Using a Taylor series expansion, it is not hard to check that $R$ in \eqref{eq_-AGMSRchosen} can be written as the right side of \eqref{eq_2orderGMS}. So, the theorem will be proven if we show that with $R$ in \eqref{eq_-AGMSRchosen}, \eqref{eq_AGMScor} is upper bounded by $\epsilon$ for $n$ sufficiently large.

Toward this end, we split the integral in \eqref{eq_AGMScor} into three integrals and upper bound each separately:
\begin{equation}
\int_{0}^{\infty} = \int_{0}^{[1 - 2d]^+} + \int_{[1-2d]^+}^{1 + b_n} + \int_{1 + b_n}^\infty\\
\end{equation}
The first and the third integrals can be upper bounded using the Berry-Esseen inequality \eqref{eq_BerryEsseen} and \eqref{eq_Qub}:
\begin{align}
\int_{0}^{[1 - 2d]^+} &\leq \mathbb P \left[ \sum_{i = 1}^n X_i^2 < n(1-2d)\right] \\
&\leq \frac{B_n}{\sqrt{n}} + \frac{1}{4d \sqrt{\pi n}} e^{-{2d}^2 n}\\
\int_{1 + b_n}^{\infty} &\leq \mathbb P \left[ \sum_{i = 1}^n X_i^2 > n(1+b_n)\right] \\
&\leq \epsilon_n + \frac{B_n}{\sqrt{n}}
\end{align}
Finally, the second integral is upper bounded by $\frac 1 {\sqrt n}$ because by the monotonicity of $g(z)$,
\begin{align}
 e^{-\rho(n,z)\exp\left(nR\right)} &\leq  e^{ -\frac{1}{2 \sqrt{\pi n}} \exp\left\{   \frac 1 2 \log n + \log \left( \sqrt \pi \log_e n\right)\right\}}\\
 &= \frac 1 {\sqrt n}
\end{align}
for all $[1-2d]^+ \leq z \leq 1 + b_n$.

\bibliographystyle{IEEEtran}
\bibliography{../../ratedistortion}

\begin{IEEEbiographynophoto}{Victoria Kostina}(S'12)
received the BachelorÕs degree with honors in applied mathematics
and physics from the Moscow Institute of Physics and Technology,
Russia, in 2004, where she was affiliated with the Institute for Information Transmission Problems of the Russian Academy of Sciences, and the MasterÕs degree in electrical engineering from the University
of Ottawa, Canada, in 2006.

She is currently pursuing a Ph.D. degree in electrical engineering at Princeton University. Her research interests lie in information theory, theory of random processes, coding, and wireless
communications.
\end{IEEEbiographynophoto}

\begin{IEEEbiographynophoto}{Sergio Verd\'{u}}(S'80--M'84--SM'88--F'93)
is on the faculty of the School of Engineering and Applied Science at Princeton University.

A member of the  National Academy of Engineering, Verd\'{u} 
is the recipient of the 2007 Claude E. Shannon Award 
and of the 2008 IEEE Richard W. Hamming Medal. 

In addition to the 28th Shannon Lecture, Verd\'{u} has given the
inaugural Nyquist Lecture at Yale University, 
the sixth Claude E. Shannon Memorial Lecture at the University of California, San Diego, 
and the tenth Viterbi Lecture at the University of Southern California.
\end{IEEEbiographynophoto}

\end{document}